\documentclass[aps,prd,twocolumn,english,preprintnumbers,amsmath,amssymb,nofootinbib,superscriptaddress,10pt]{revtex4-1}

\pdfoutput=1

\usepackage[usenames,dvipsnames]{color}

\usepackage[latin1]{inputenc}
\usepackage{graphicx}
\usepackage{color}
\usepackage{txfonts}
\usepackage{epstopdf}

\usepackage{bm}
\usepackage{bbm}
\usepackage{amsmath}
\usepackage{slashed}
\usepackage{esint}
\usepackage{amsfonts}
\usepackage{amssymb}
\usepackage{tabularx}

\usepackage{dsfont}

\newcommand{\dif}[1]{\ensuremath{\medspace \mbox{d} #1}}

\newcommand{\STr}{\mathrm{STr}}

\newcommand{\be}{\begin{eqnarray}}
\newcommand{\ee}{\end{eqnarray}}

\newcommand{\nn}{\nonumber }

\newcommand{\Nf}{N_{\text{f}}}
\newcommand{\Nfcr}{N_{\text{f,cr}}}
\newcommand{\Nfpc}{N_{\text{f,cr}}^{\text{qc}}{}}

\newcommand{\ksb}{k_{\text{SB}}}

\newcommand{\Eqref}[1]{Eq.~\eqref{#1}}

\usepackage{multirow}

\newcommand{\lfdif}[2]{\ensuremath{\frac{\overset{\rightarrow}{\delta} #1}{\delta #2}}}
\newcommand{\rfdif}[2]{\ensuremath{\frac{\overset{\leftarrow}{\delta} #1}{\delta #2}}}

\def\slash#1{\setbox0=\hbox{$#1$}               
   \dimen0=\wd0                                 
   \setbox1=\hbox{/} \dimen1=\wd1               
   \ifdim\dimen0>\dimen1                        
      \rlap{\hbox to \dimen0{\hfil/\hfil}}      
      #1                                        
   \else                            
      \rlap{\hbox to \dimen1{\hfil$#1$\hfil}}   
      /                                         
   \fi}                                         %

\newcommand{\chiral}{chiral }
\newcommand{\confcrit}{conformal-critical }
\newcommand{\Confcrit}{Conformal-critical }

\usepackage{babel}
\makeatother

\begin{document}

\title{On the Phase Structure of Many-Flavor QED${}_3$}
\author{Jens Braun} 
\affiliation{Institut f\"ur Kernphysik (Theoriezentrum), Technische Universit\"at Darmstadt, 
D-64289 Darmstadt, Germany}
\affiliation{ExtreMe Matter Institute EMMI, GSI, Planckstra{\ss}e 1, D-64291 Darmstadt, Germany}
\affiliation{Theoretisch-Physikalisches Institut,  Abbe Center of Photonics, Friedrich-Schiller-Universit\"at Jena, Max-Wien-Platz 1, D-07743 Jena, Germany}
\author{Holger Gies}
\affiliation{Theoretisch-Physikalisches Institut,  Abbe Center of Photonics, Friedrich-Schiller-Universit\"at Jena, Max-Wien-Platz 1, D-07743 Jena, Germany}
\affiliation{Helmholtz-Institut Jena, Fr\"obelstieg 3, D-07743 Jena, Germany}
\author{Lukas Janssen}
\affiliation{Department of Physics, Simon Fraser University, Burnaby, British Columbia, V5A 1S6, Canada}
\affiliation{Theoretisch-Physikalisches Institut,  Abbe Center of Photonics, Friedrich-Schiller-Universit\"at Jena, Max-Wien-Platz 1, D-07743 Jena, Germany}
\author{Dietrich Roscher}
\affiliation{Institut f\"ur Kernphysik (Theoriezentrum), Technische Universit\"at Darmstadt, 
D-64289 Darmstadt, Germany}
\affiliation{Theoretisch-Physikalisches Institut,  Abbe Center of Photonics, Friedrich-Schiller-Universit\"at Jena, Max-Wien-Platz 1, D-07743 Jena, Germany}

\begin{abstract}
We analyze the many-flavor phase diagram of quantum electrodynamics
(QED) in $2+1$ (Euclidean) space-time dimensions. We compute the
critical flavor number above which the theory is in the
quasi-conformal massless phase. For this, we study the renormalization
group fixed-point structure in the space of gauge {interactions} and pointlike
fermionic self-interactions{, the latter of} which are induced dynamically by
fermion-photon interactions. We find that a reliable estimate of the
critical flavor number crucially relies on {a careful treatment of the Fierz
ambiguity in the fermionic sector. Using a Fierz-complete basis,}
our results indicate that the phase transition towards a chirally-broken phase
occurring at small flavor numbers could be separated from the
quasi-conformal phase at larger flavor numbers, allowing for an
intermediate phase which is dominated by fluctuations in a vector
channel. If these interactions approach criticality, the intermediate
phase could be characterized by a Lorentz-breaking vector condensate.
\end{abstract}

\maketitle

\section{Introduction}

The competition between screening and anti-screening effects is at the
heart of the intriguing diversity of phases occurring in
asymptotically free theories. Not only thermal phase transitions
governed by parameters such as temperature or chemical potentials, but
also quantum phase transitions triggered by the number of active
degrees of freedom have recently been of central interest. Most
prominently, the number of light fermion degrees of freedom $\Nf$
often serves as a control parameter to tune the
screening--anti-screening competition. While chiral quantum phase
transitions of this type have attracted considerable attention in
4-dimensional non-abelian gauge theories because of their potential
relevance for embeddings of the Higgs sector in
beyond-standard-model scenarios~\cite{Weinberg:1979bn,Holdom:1981rm,Hong:2004td,
Sannino:2004qp,Dietrich:2005jn,Dietrich:2006cm,Sannino:2009za}, similar
theoretical mechanisms can be at work in the abelian theory of quantum
electrodynamics (QED) in $d=3$ (Euclidean) spacetime dimensions. Beyond the
predominantly conceptual interest, such studies gain significance from
layered condensed-matter systems for which $d=2+1$ dimensional QED
with four-component Dirac fermions can serve as an effective field
theory for low-energy excitations. Applications of this type have been
discussed, e.g., in the context of graphene, 
surface states of 3-dimensional topological insulators, and 
high-temperature cuprate superconductors.
For recent reviews on this rapidly evolving field, see, e.g.,
Refs.~\cite{Cortijo:2011aa, Vafek:2013mpa}.
{In particular, QED$_3$ has been proposed to model
the destruction of phase coherence in the
underdoped cuprates
\cite{Franz:2001zz,Franz:2002qy,Herbut:2002wd,Herbut:2002yq,Tesanovic:2002zz,
Mavromatos:2003ss,Herbut:2004ue}. Chiral symmetry breaking in
QED$_3$ then describes the zero-temperature transition from the
$d$-wave superconducting state into the antiferromagnetic state. The size of the
dynamically generated mass in the effective theory consequently determines the
band gap in the insulating phase of the underdoped cuprates.}

As the coupling constant of QED${}_3$ has a positive mass dimension,
the theory is asymptotically free for purely dimensional reasons: any
finite value of the coupling, if measured in terms of a
  reference scale, will become arbitrarily small if this reference
scale is pushed to asymptotically large energies or momenta. In turn,
one expects QED$_3$ to become more strongly coupled at low energies,
possibly generating fermion masses through a chiral phase
transition. By contrast, increasing the number of fermion flavors
enhances the screening properties of fermionic fluctuations. If
  this screening dominates, the coupling may remain small and the
theory can be expected to be in the disordered massless phase. More
precisely, the fluctuations can generate an infrared (IR) fixed point,
such that the theory remains quasi-conformal: it has a nontrivial RG
flow from the Gau\ss{}ian ultraviolet (UV) to the IR fixed point with
the transition scale set by the dimensionful gauge coupling. Scenarios
of this type have been suggested and analyzed in many works, and
  the critical flavor number $\Nfcr^{\chi}$ separating the chirally
  broken phase for small $\Nf$ from the {symmetric} for
  large $\Nf$ has been estimated by a variety of nonperturbative
  methods, see, e.g., Refs.%
~\cite{Appelquist:1988sr,Nash:1989xx,Pennington:1990bx,
    Atkinson:1989fp, Curtis:1992gm,
    Maris:1995ns,Ebihara:1995aa,Maris:1996zg,Aitchison:1997aa,Appelquist:1999hr,
    Gusynin:2003ww, Fischer:2004nq,Kaveh:2004qa,Kubota:2001kk,
    Hands:2002dv,Hands:2004bh, Mitra:2007aa,
    Strouthos:2008hs,Bashir:2008aa,Bashir:2009aa,Feng:2012aa,Grover:2012aa}.
  Predictions from self-consistent approximations of the
  Dyson-Schwinger equations (DSE) in their most advanced form yield
  results near $\Nfcr^\chi \approx 4$, see, e.g.,
  \cite{Fischer:2004nq}.
Recently, these studies have been extended to incorporate lattice
anisotropies as well as finite temperature in order to approach more
realistic
applications~\cite{Bonnet:2011hh,Bonnet:2011ds,Bonnet:2012az,Popovici:2013wfa}.
An early RG study found $\Nfcr^\chi \simeq
3.1$~\cite{Kubota:2001kk}. Based on a thermodynamic argument an inequality
$\Nfcr^\chi \leq 1.5$ has been conjectured~\cite{Appelquist:1999hr}, but was
challenged later~\cite{Mavromatos:2003ss}. Another upper bound $\Nfcr^\chi < 7$
has been claimed recently using an RG monotonicity argument~\cite{Grover:2012aa}.
On the other hand, lattice
simulations in QED$_3$ are difficult due to a large separation of scales;
however, they appear to agree at least on a lower bound $\Nfcr^\chi >
1$~\cite{Hands:2004bh, Strouthos:2008hs}.
{The actual value of $\Nfcr^\chi$ in QED$_3$ is in fact of profound
interest for the effective cuprate models, in which the number of four-component
Dirac flavors is $\Nf = 2$: If $\Nfcr^\chi > 2$, then the effective theory
predicts a direct transition from the $d$-wave superconducting into the
antiferromagnetic phase at $T=0$ as a function of the doping
\cite{Herbut:2002wd,
Herbut:2002yq}. Otherwise, a
small $\Nfcr^\chi < 2$ would leave the possibility of an unconventional
non-Fermi-liquid phase in the $T=0$ underdoped cuprates
\cite{Franz:2001zz,Franz:2002qy,Tesanovic:2002zz}.}

In the present work, we take a fresh look at the phase structure of
QED${}_3$ as a function of the fermion number. We pay particular
attention to all interaction channels allowed by the large U($2\Nf$)
flavor symmetry for Dirac fermions in the reducible
representation. Using the functional renormalization group (RG), we
find evidence for a more involved structure of the phase
diagram. Within our approach, we can straightforwardly identify the
{``conformal-}critical'' flavor number $\Nfpc$ above which the theory is in
the quasi-conformal phase. {A priori, $\Nfpc$ can be different from the
``chiral-critical'' flavor number $\Nfcr^\chi$ below which the theory is in the
chirally-broken phase.}
Our results suggest that $\Nfcr^\chi\lesssim\Nfpc$. This
includes the interesting possibility of a third intermediate phase
with $\Nf$ fermion flavors such that $\Nfcr^{\chi}<\Nf<\Nfpc$. Our
findings suggest that this phase is dominated by vector-channel
fluctuations. If they become critical, the model features a
Lorentz-breaking vector condensate and a correspondingly mixed
spectrum of photonlike massless Goldstone bosons and massive
excitations.

The present work is organized as follows: In Sec.~\ref{sec:sym}, we
discuss the symmetries and fermionic interaction channels of
QED${}_3$. Corresponding symmetry-breaking patterns are briefly
outlined in Sec.~\ref{sec:SBP}. In Sec.~\ref{sec:RG}, we introduce
and apply the functional RG as our central technical tool in order to
derive the RG flow equations for the interactions and wave-function
renormalizations. Section \ref{sec:fpana} is devoted to a fixed-point
analysis as a means to identify possible phase structures. An estimate
of the conformal-critical flavor number $\Nfpc$ marking the transition to
the disordered quasi-conformal phase is performed in Sec.~\ref{sec:quant}. After
illustrating the importance of Fierz completeness of {the}
fermionic interaction
channels in Sec.~\ref{sec:fierz}, we summarize our findings in the
form of a conjectured phase diagram in Sec.~\ref{sec:PS} and conclude
in Sec.~\ref{sec:conc}. Some technical details are summed up in the
Appendices.

\section{Symmetries and Fermionic Interaction Channels}
\label{sec:sym}
Let us first recapitulate the flavor symmetries of QED${}_3$ with many
flavors, paying attention to the diversity of interaction channels,
see~\cite{Gies:2010st,Janssen:2012oca} for an extended discussion.

The microscopic (classical) action of QED${}_3$ with $\Nf$ fermion
flavors in $d=3$ Euclidean space-time is given by
\be
\label{SQED3}
S = \int d^3{x}\left\{\bar{\psi}^a i\slashed{\partial}\psi^a +\bar{e}\bar{\psi}^a\slashed{A}\psi^a+ \frac{1}{4}F_{\mu\nu}F^{\mu\nu}\right\}\,,
\ee
where~$\bar{e}$ denotes the {bare {dimensionful} gauge coupling and summation over flavor indices $a$ is tacitly assumed.} The fermions $\psi,
\bar\psi$ are considered to be four-component Dirac spinors, naturally occurring,
e.g., in effective theories for electrons on a honeycomb lattice
\cite{Semenoff:1984dq,Hands:2008id,Herbut:2009qb,
  Drut:2009aj,Armour:2009vj, Gusynin:2007ix,Smith:2014tha} or in
cuprates \cite{Franz:2001zz, Franz:2002qy, Herbut:2002wd,
  Herbut:2002yq, Tesanovic:2002zz, Herbut:2004ue, Mavromatos:2003ss}. They transform
under a reducible representation of the Dirac algebra
$\{\gamma_\mu,\gamma_\nu\} = 2\delta_{\mu\nu}$ in terms of $4\times 4$
Dirac matrices
\begin{equation} \label{eq:RedRep4a}
\gamma_\mu= 
\begin{pmatrix}
0 & -i \sigma_\mu\\
i \sigma_\mu & 0\\
\end{pmatrix}, \qquad \mu=1,2,3,
\end{equation}
where $\{\sigma_\mu\}_{\mu=1,2,3}$ denote the standard Pauli matrices.  
The Clifford algebra can be spanned with the aid of two further $4\times 4$ matrices
\begin{equation} \label{eq:RedRep4b}
\gamma_4=
\begin{pmatrix}
0 & \mathbbm{1}_2\\
\mathbbm{1}_2 & 0\\
\end{pmatrix} \quad \text{and} \quad
\gamma_5=\gamma_1\gamma_2\gamma_3\gamma_4=
\begin{pmatrix}
\mathbbm{1}_2 & 0\\
0 & -\mathbbm{1}_2\\
\end{pmatrix},
\end{equation}
which anticommute with each other as well as with all $\gamma_\mu$. A
complete Clifford basis is given by
\begin{equation} \label{eq:CliffordBasis}
\left\{\gamma_A\right\}_{A=1,\dots,16} =
\left\{\mathbbm{1}_4, \gamma_\mu, \gamma_4, \gamma_{\mu\nu}, i\gamma_\mu\gamma_4,
i\gamma_\mu\gamma_5, \gamma_{45}, \gamma_5\right\}, 
\end{equation}
where $\gamma_{45}= i\gamma_4\gamma_5$ and $\gamma_{\mu\nu}=
\frac{i}{2}[\gamma_\mu,\gamma_\nu]$ (in \Eqref{eq:CliffordBasis}, only
those $\gamma_{\mu\nu}$ with $\mu<\nu$ are counted as independent).

The obvious U($\Nf$) flavor symmetry of \Eqref{SQED3} together with rotations
in the space of irreducible subcomponents of the Dirac spinors leads to an
enhanced U(2$\Nf$) flavor (or ``chiral'') symmetry of QED${}_3$, see
App.~\ref{app:irr} for details.

From a renormalization group perspective, it is convenient to view the
approach from the microscopic theory towards possible symmetry-broken
regimes as a two-stage process: first, fluctuations involving
gauge-fermion interactions induce effective fermionic
self-interactions. Second, further fluctuations may lead to a rapid
growth of the fermionic interactions driving the system to criticality
and giving rise to possible condensation phenomena. 

In the present work, we study the fermionic self-interactions in the
pointlike (i.e., the zero-momentum) limit. 
To this end, we first classify all possible fermionic self-interactions which
are compatible
with the U(2$\Nf$) flavor symmetry as well as with the discrete $C$,
$P$, and $T$ symmetries of the model. Following \cite{Kubota:2001kk,
  Kaveh:2004qa, Herbut:2009qb, Gies:2010st, Janssen:2012oca}, these
interactions are given by the flavor-singlet channels
\begin{align}
(V)^2 & = \left(\bar{\psi}^a \gamma_\mu\psi^a\right)^2, &
(P)^2 & = \left(\bar{\psi}^a \gamma_{45}\psi^a\right)^2,
\label{eq:SingChan}
\end{align}
and the flavor-nonsinglet channels
\begin{align}
(S)^2 & =
\left(\bar{\psi}^a\psi^b\right)^2-\left(\bar{\psi}^a\gamma_4\psi^b\right)^2-
\left(\bar{\psi}^a\gamma_5\psi^b\right)^2
\nonumber \\ &\quad 
+\left(\bar{\psi}^a\gamma_{45}\psi^b\right)^2, \label{eq:nonsingchan}\\
(A)^2 & =
\left(\bar{\psi}^a\gamma_\mu\psi^b\right)^2+
\frac{1}{2}\left(\bar{\psi}^a\gamma_{\mu\nu}\psi^b\right)^2-
\left(\bar{\psi}^a i\gamma_\mu\gamma_4\psi^b\right)^2
\nonumber \\ &\quad
-\left(\bar{\psi}^a i\gamma_\mu\gamma_5\psi^b\right)^2.
\end{align}
Here, we have used the convention $(\bar{\psi}^a\psi^b)^2 \equiv
\bar{\psi}^a\psi^b\bar{\psi}^b\psi^a$, etc. The corresponding 4-point
correlation functions of these fermion interactions can develop
largely independent structures in momentum space. By contrast, in the
zero-momentum (pointlike) limit, these four-fermion interactions are
connected {due to} Fierz identities,
\begin{align} \label{eq:FierzIdentities}
(V)^2 + (S)^2 + (P)^2 & = 0, &
-4 (V)^2 - 3 (S)^2 + (A)^2 & = 0.
\end{align}
In this limit, only two four-fermion terms are linearly independent. 
We choose to work with the flavor singlets
and parametrize the corresponding part
of the (effective) Lagrangian {as} 
\begin{eqnarray} 
\label{eq:LPsiInt}
\mathcal{L}_{\psi,\text{int}}  &=&
 \frac{\bar{g}}{2\Nf}
(V)^2
+ \frac{\bar{\tilde{g}}}{2\Nf}
(P)^2 \nonumber\\
&=&
 \frac{\bar{g}}{2\Nf}
(\bar{\psi}^a \gamma_\mu \psi^a)^2
+ \frac{\bar{\tilde{g}}}{2\Nf}
(\bar{\psi}^a \gamma_{45}\psi^a)^2,
\end{eqnarray}
with the bare couplings $\bar{g},\bar{\tilde{g}}$.
In our RG study below, $\bar g$ and $\bar{\tilde g}$ are set to zero at the
initial scale. However, they can be
generated dynamically during the RG flow. In any case,
the first term $\sim \bar{g}$ corresponds to
the interaction known from the Thirring model, whereas the second {one} $\sim
\bar{\tilde{g}}$ is similar to a Gross-Neveu interaction\footnote{If expressed
  in terms of two-component Weyl spinors, this interaction is indeed identical
  to the Gross-Neveu interaction, cf. App.~\ref{app:irr}.}.

{For $\Nf > 1$, another Fierz basis may be of interest from a physical
point of view:}
\begin{align} \label{eq:LPsiInt2}
\mathcal{L}_{\psi,\text{int}} & =
- \frac{\bar{g}_V}{2\Nf}
(V)^2
+ \frac{\bar{g}_\phi}{4\Nf}
(S)^2,
\end{align}
where the couplings are related to those of \Eqref{eq:LPsiInt} by
\begin{align}
\bar g_V & =  \bar{\tilde g} - \bar g, \nonumber\\
\bar g_\phi & = -2 \bar{\tilde g}.\label{eq:coupFierz}
\end{align}
In addition to the vector (Thirring) channel $\sim(V)^2$, we encounter
the nonsinglet channel $\sim(S)^2$ of \Eqref{eq:nonsingchan} {reminiscent to the
  Nambu--Jona-Lasinio (NJL) model}. We emphasize that the description
of the system in terms of \Eqref{eq:LPsiInt} is completely equivalent
to that of \Eqref{eq:LPsiInt2} in the pointlike limit. The same is
true for any other combination of two linearly independent
(``Fierz-complete'') interactions out of the four channels $(V)^2$,
$(P)^2$, $(S)^2$, or $(A)^2$.

We conclude this section by critically assessing the pointlike limit: from a
more general viewpoint, pointlike interactions are
only a special limit of  fermionic correlation functions $\Gamma^{(n)}$, i.e., 
\begin{eqnarray}
&&g_{\mathcal{O}} (\bar\psi \mathcal{O} \psi)^2 \label{eq:CorFunc}\\
&&\,\,\,  = \lim_{p_i\to 0}
\bar\psi^a(p_1) \bar\psi^b(p_2)
\Gamma^{(4), abcd}_{\mathcal{O}} (p_1,p_2,p_3,p_4) \psi^c(p_3) \psi^d(p_4).
\nonumber
\end{eqnarray}
A priori, the pointlike limit hence ignores a substantial amount of
momentum-dependent information\footnote{The functional renormalization group
  approach used below actually
  reinstates part of the momentum-dependent information in an effective
  manner.}. Most importantly, since bound-state formation is encoded in the
momentum structure of correlation functions (e.g. as $s$-channel poles in
Minkowski space), we cannot expect to obtain reliable information about the
mass spectrum of the system. Moreover, the formation of a condensate goes
along with a singularity in the fermionic four-point function, such that the
fermionic pointlike description cannot access the symmetry-broken regime.

In turn, this implies that the pointlike limit can only be used to
study the system within the symmetric regime. In fact, it is adequate
to address the large-$\Nf$ limit which is expected to lie in the
symmetric phase. By lowering the flavor number $\Nf$, we can therefore
study the approach to the symmetry-broken phase of the theory, as
symmetry-breaking inevitably goes along with a break-down of the
pointlike description. In this manner, we can determine a
\confcrit flavor number $\Nfpc$ below which the pointlike
description breaks down, possibly indicating condensate and
bound-state formation. In the case that the approach to $\Nfpc$ from
above exhibits a clear signature for condensation in a particular
channel, the \confcrit flavor number can agree with {a specific}
critical flavor number $\Nfcr$ below which the system is in a particular
symmetry-broken phase. This reasoning has been used in
\cite{Gies:2005as,Braun:2005uj,Braun:2006jd} to determine the
many-flavor phase diagram of QCD.

{However,} because of the diversity of possible symmetry-breaking patterns
as discussed below, the meaning of $\Nfpc$ {in QED$_3$} is less obvious.
{In fact, our results indicate that there may exist more than one critical
flavor number corresponding to different symmetry-broken phases. The \confcrit
flavor number $\Nfpc$, which we aim to estimate in the present work, provides an
upper bound on the potentially existing critical flavor numbers for all kinds of
broken phases.

\section{Symmetry breaking patterns}
\label{sec:SBP}

Let us discuss the various symmetry-breaking patterns that can arise if
the fermion
self-interactions become critical. Symmetry breaking can give rise to two
fundamentally
different fermion mass terms: $i m \bar\psi \psi$ and $i \tilde{m} \bar\psi
\gamma_{45} \psi$. Further fermion bilinears involving $\gamma_4$ and
$\gamma_5$ are U($2\Nf$) equivalent to these mass terms. 

The relation between fermion mass generation and symmetry breaking becomes
transparent by means of a Hubbard-Stratonovich transformation
\cite{Hubbard:1959ub,Stratonovich}. This partial bosonization allows us to
treat composites of two fermions in terms of effective bosons,
schematically, $\phi \sim \bar{\psi}\psi$.  More formally, such a
transformation allows us to trade in the four-fermion interaction term for a
corresponding term bilinear in bosonic fields and a Yukawa-type interaction
term on the level of the path integral:
\be
\bar{g}_{\mathcal O}(\bar{\psi}{\mathcal O}\psi)^2 \quad\longrightarrow \quad {\bar{g}_{\mathcal O}}^{-1}\phi_{\mathcal O}^2 + \bar{\psi}\bar{h}_{\mathcal O}\phi_{\mathcal O}\psi\,,\label{eq:bossektch}
\ee
{where the Yukawa-type coupling $\bar{h}_{\mathcal O}$ can possibly be flavor- or Dirac-matrix-valued.}
The quantum numbers and transformation properties
of the new bosonic field~$\phi_{\mathcal O}$ depend on the exact definition of
the four-fermion interaction associated with the operator~$\mathcal O$.
The Yukawa coupling is normalized such that the four-fermion coupling is
reproduced upon integrating out the bosonic field. 

From \Eqref{eq:bossektch}, we deduce that the four-fermion couplings are
inversely proportional to the mass term
$\sim \phi_{\mathcal O}^2$ of the
bosonic field. Upon fluctuations, we expect that a full Ginzburg-Landau-type
effective potential is generated for the boson field. Therefore, a singularity
of the pointlike fermionic {coupling goes along} with the effective potential
developing a nontrivial minimum. If so, the expectation value of
$\phi_{\mathcal{O}}$ serves as an order parameter for symmetry breaking. Vice
versa, if we observe a divergence of the fermionic self-interactions at a
finite RG scale~$\ksb$ in the purely fermionic language, this serves as an
indication for the possible onset of spontaneous symmetry breaking.

Whereas Fierz completeness can be fully preserved by choosing a suitable basis
in the {purely} fermionic language, simple approximations on the
 {partially bosonized} side can
actually violate this property. For instance, in mean-field approximations
this is known as the ``Fierz ambiguity'' or ``mean-field
ambiguity''~\cite{Baier:2000yc}, the resolution of which requires dynamical
bosonization techniques on the bosonic side
\cite{Gies:2001nw,Jaeckel:2002rm,Janssen:2012pq}.

In the present work, we anyway study the system by approaching the phase
boundary from the symmetric phase, hence the quantitative details of
bosonization are not important for our purpose. In order to get a first
picture of possible symmetry-breaking patterns, let us take a closer look at
the partially bosonized version of \Eqref{eq:LPsiInt2} that uses the $(V)^2$
and $(S)^2$ channels, which are considered to be the relevant channels also in
the Thirring model \cite{Janssen:2012pq}. Using the irreducible representation
in terms of two-component fermions $\chi$, see App.~\ref{app:irr}, we get for
the vector channel
\begin{equation}
- \frac{\bar{g}_V}{2\Nf} (V)^2 \to \frac{1}{2} \bar{m}_V^2 V_\mu V_\mu -  \bar
h_V V_\mu \bar\chi^i \sigma_\mu \chi^i, {\quad i=1,\dots, 2\Nf,}
\end{equation}
where $V_\mu$ denotes a real vector boson, and the $(S)^2$ channel yields
\begin{equation}
 \frac{\bar{g}_\phi}{4\Nf} (S)^2 \to \frac{1}{2} \bar m_\phi^2
\phi^{ij}\phi^{ji} + i \bar h_\phi
\bar\chi^i \phi^{ij} \chi^j,
\end{equation}
where $\phi^\dagger=\phi$ denotes a scalar field represented by
a hermitean
$2\Nf\times 2\Nf$ matrix. The equivalence with the fermionic action holds also
on the path integral level, if the bare couplings satisfy the constraint
\begin{equation}
\frac{\bar h_\phi^2}{2\bar m_\phi^2}  = \frac{\bar g_\phi}{2\Nf},\quad
\frac{\bar h_V^2}{2\bar m_V^2}  = \frac{\bar g_V}{2\Nf}.
\label{eq:Constraint}
\end{equation}
Whereas the vector field $V_\mu$ is invariant under U($2\Nf$) transformations,
the scalar field transforms according to the bifundamental
representation. Different symmetry-breaking patterns arise depending on which
bosonic field component eventually develops a {finite} vacuum expectation value. For
instance, if $\phi^{ij}$ acquires an expectation value $\sim \delta^{ij}$, a
fermion mass term $\sim i \tilde{m} \bar\psi^a \gamma_{45} \psi^a$ is
generated. As is obvious from the form of the expectation value, this mass
term does not break the U$(2\Nf)$ symmetry. It breaks
parity {and time-reversal symmetry \cite{Janssen:2012oca}}. By
contrast, an expectation value of the form 
\begin{equation}
\phi^{ij} \sim
\left( \begin{array}{cc} \mathbbm{1} & 0 \\ 0 & {-\mathbbm{1}}
  \end{array} \right)
\end{equation}
gives rise to a mass term $i m( \bar\chi^a \chi^a- \bar\chi^{a+\Nf}
\chi^{a+\Nf}) = i m \bar\psi^a \psi^a$ which corresponds to a
symmetry-breaking pattern of the form
\begin{equation} \label{eq:BreakPatt}
\mathrm{U}(2\Nf) \rightarrow \mathrm{U}(\Nf) \otimes
\mathrm{U}(\Nf) .
\end{equation}
This is the pattern expected to occur for small flavor numbers in
QED${}_3$. {For $\Nf > 2$}, more breaking patterns arising
from the scalar sector
are in principle conceivable, but have not been considered
in the literature so far and will also be ignored in this work.

Another option is that the vector field $V_\mu$ develops an
expectation value. This would leave the U($2\Nf$) flavor symmetry
intact, but would break Lorentz invariance. Breaking patterns of this
type have already been considered during the heyday of the NJL model
and the development of the Higgs
mechanism~\cite{Bjorken:1963vg,BialynickiBirula:1963zz,Guralnik:1964zz}.
For instance, if the expectation value of $V_\mu$ was time-like, the
corresponding Goldstone bosons {may resemble in some aspects} a
photon field in temporal gauge. In the present case of QED${}_3$,
these Goldstone bosons could mix with the photon. In addition, a
massive {bosonic excitation and Lorentz violating features in
  correlation functions} {could} be expected to occur. However,
  {the number of non-perturbative studies of this symmetry breaking
  scenario {and} the nature of the transition is limited, see,
  e.g., \cite{Banks:1980rh,oai:arXiv.org:hep-th/0203221}}.
 
\section{Renormalization Group Flow of QED${}_3$}
\label{sec:RG}

The preceding sections already anticipated an RG viewpoint on the model. In
fact, our quantitative analysis will be based on the functional
RG formulated in terms of the Wetterich 
equation~\cite{Wetterich:1992yh} which is a flow equation for the
coarse-grained quantum effective action~$\Gamma{_k}$:
\be
\partial _t \Gamma_k = \frac{1}{2}{\rm STr}\left[(\partial_t R_k)\cdot \left[\Gamma_k^{(2)} + R_k\right]^{-1}\right]\,.\label{eq:wetterich}
\ee
Here, $\Gamma^{(2)}_k$ is the second functional derivative
of~$\Gamma_k$ with respect to the fields, $t=\ln(k/\Lambda)$, {and
$k$ is a flowing IR cutoff scale} which is used to set up the RG flow of the quantum
effective action. The regularization is implemented with the aid of
the regulator function $R_k$ specifying the details of the Wilsonian
momentum shell integrations. In the long-range limit, $k\to 0$, $R_k$
also vanishes such that all quantum fluctuations have been integrated
out. The initial condition of the RG flow is determined by the
classical action~$S$ in the limit $k\to\Lambda$:
$\Gamma_{k\to\Lambda\to\infty}\to S$. {In an exact solution to
  Eq.~\eqref{eq:wetterich},} the results for physical observables to
be read off for $k\to 0$ should not depend on our specific choice for
the regularization scheme, i.e., the function~$R_k$ in our case. In
this work, we exploit a variation of the scheme to test the predictive
power of our approximations, see Sec.~\ref{sec:quant}.

Solving the Wetterich equation yields an RG trajectory in theory space, i.e.,
the space of all action functionals parametrized for instance by all possible
field operators compatible with the symmetries of the theory. In the present
work, we confine ourselves to an investigation of the RG flow within a
hypersurface of theory space, parametrized by the ansatz
\be
\label{eq:GkQED3}
&&\Gamma_k[\bar{\psi},\psi,A]= \!\int\!d^3x\bigg\{\bar{\psi}\left( {\rm
    i}Z_\psi\slashed{\partial} \!+\!
  Z_{\bar{\psi}A\psi}\bar{e}\slashed{A}\right)\psi  \nonumber\\
&& \quad\;
\!+\! \frac{1}{2}A_\mu {Z_A} (-(\partial^2) \delta_{\mu\nu}
+ \partial_\mu\partial_\nu) A_{\nu} 
+ \frac{1}{2\xi}A_\mu Z_\xi \partial_\mu \partial_\nu A_\nu
\nn\\ 
&& \quad\;
+ \frac{\bar{\tilde{g}}}{2N_{\rm{f}}}(\bar{\psi}\gamma_{45}\psi)^2
+\frac{\bar{g}}{2N_{\rm{f}}}(\bar{\psi}\gamma_\mu\psi)^2 \bigg\}\,, 
\ee
where the couplings $\bar{g}$, $\tilde{\bar{g}}$, the wave-function
renormalizations~$Z_{\psi}$, $Z_A$, and the vertex
renormalization~$Z_{\bar{\psi}A\psi}$, governing the renormalization
of~$\bar{e}$, are assumed to be functions of the RG scale $k$. As discussed
above, we consider the four-fermion couplings $\bar g$ and $\bar{\tilde{g}}$
in the pointlike limit. In addition, also the couling $Z_{\bar{\psi}A\psi}
\bar{e}$, parametrizing the photon-electron vertex, and the
{fermionic} wave-function
renormalization $Z_\psi$ will be considered in the zero-momentum limit. In
fact, as the flow equation is local in momentum space, receiving its dominant
contributions from momenta $p\simeq k$ for a given scale $k$, the $k$ dependence
of all these couplings can be viewed as an effective momentum
dependence of the corresponding vertices and propagators, see also our discussion below.

Within the functional RG approach,
the restriction to the pointlike limit is therefore less severe as it
may seem: only highly asymmetric
momentum
dependencies of the vertices are neglected, whereas an overall momentum
dependence is effectively parametrized by the $k$ dependence of the
couplings. 

The situation is slightly but decisively different for the photon
wave-function renormalization, which we a priori consider to be a
function of momentum $Z_A=Z_A(p^2)$. While all qualitative features
could still be extracted from the zero-momentum limit, the
quantitative description of QED${}_3$ depends rather strongly on the
precise form of the momentum dependence of the photon propagator. The
reason for this is the qualitative change of the momentum dependence
of the polarization tensor~$\Pi_{\mu\nu}$,
\be
\Pi_{\mu\nu}(p)=(p^2\delta_{\mu\nu}-p^{\mu}p^{\nu})\Pi(p)\,,
\ee
across the scale set by the dimensionful QED coupling $\bar{e}^2$ in
three dimensions.\footnote{This is a peculiarity of three-dimensional
  theories and occurs generically for bosonic propagators dressed by
  fermion loops, see, e.g., Ref.~\cite{Strack:2009ia} for further
  examples. By contrast, no such severe momentum-dependence is known
  in four-dimensional theories: In studies of QED$_4$
    and QCD$_4$, for example, RG flows using the
  background-field
  method~\cite{Abbott:1980hw,Abbott:1981ke,Dittrich:1985tr} to
    compute $Z_A$ have been quite successful, see, e.g.,
  Refs.~\cite{Reuter:1993kw,Reuter:1997gx,Gies:2002af,Gies:2004hy,Gies:2006wv,Braun:2006jd}.}
For instance, in the large-$\Nf$ limit, the dressing function of
  the polarization tensor is known to behave as
\cite{Pisarski:1984dj}
\be
\Pi(p)\sim \frac{1}{p}\,,
\ee
which can have a rather strong effect on the photon wave-function renormalization~$Z_{A}$,
\be
Z_{A}(p) = 1+\Pi(p).
\ee
We need $Z_A(p^2)$ mainly in order to extract the running of the gauge
coupling. Since the momentum dependence of $Z_A(p^2)$ is expected to be
sensitive to the value of the gauge coupling, it appears quantitatively
mandatory to resolve the momentum dependence of $Z_A(p^2)$ in QED${}_3$ as
accurately as possible.

In addition to the kinetic term of the photon, the gauge sector also
comes with a gauge fixing term with gauge parameter $\xi$ and a
corresponding wave-function renormalization $Z_\xi$. In the present
work, we work in the Landau gauge $\xi\to 0$ which is known to be a
fixed point of the RG flow
\cite{Ellwanger:1995qf,Ellwanger:1996wy,Litim:1998qi,Lauscher:2001ya}. This
suggest to choose $Z_\xi=Z_A$ for simplicity.

With these prerequisites, it is in principle straightforward to derive
the flow of general action functionals spanned by the
ansatz~\eqref{eq:GkQED3}. In order to make proper contact with
QED${}_3$, we have to provide initial conditions for the flow
parameters in \Eqref{eq:GkQED3}. With regard to the classical action
\Eqref{SQED3}, these initial conditions are given at the microscopic
UV scale $\Lambda$ by
\begin{align}
Z_{\psi}\big|_{\Lambda\to\infty} & \to 1\,, &
Z_{A} \big|_{\Lambda\to\infty} & \to 1\,, &
Z_{\bar{\psi}A\psi}\big|_{\Lambda\to\infty} & \to 1\,, \nn\\
\bar{e}^2\big|_{\Lambda\to\infty} & > 0\,, &
\bar{g}\big|_{\Lambda\to\infty} & \to 0 \,, &
\bar{\tilde{g}}\big|_{\Lambda\to\infty} & \to 0\,. \label{eq:initcond}
\end{align}
Note that in particular the four-fermion self-interactions are not considered
to be independent parameters. If they appear in the RG flow, they are solely
generated by quantum fluctuations. 

The RG flows for the couplings can conveniently be formulated for the
dimensionless renormalized couplings. For the fermionic interactions, these
are given by
\be
\tilde{g}=Z_{\psi}^{-2}k\bar{\tilde{g}}\quad\text{and}\quad
{g}=Z_{\psi}^{-2}k\bar{{g}}\,. \label{eq:defg}
\ee
The running of the fermionic wave-function renormalization in turn can be parametrized in
terms of the fermionic anomalous dimension
\be
\eta_{\psi}=-\partial_t \ln Z_{\psi}\,. \label{eq:defetapsi}
\ee
The calculation of the corresponding {fermionic} flows is straightforward
with standard techniques, see Ref.~\cite{Braun:2011pp}, {and} the results will
be summarized below. 

The RG flow of the gauge sector requires a more careful discussion. The
corresponding definition of the dimensionless gauge coupling is
\be
e^2=\frac{\bar{e}^2 Z_{\bar{\psi}A\psi}^2}{Z_AZ_{\psi}^2k}\,.\label{eq:defeq}
\ee
In ordinary perturbation theory, the Ward indentity for the
photon-electron vertex} enforces 
$Z_{\bar{\psi}A\psi}=Z_{\psi}$ to hold at each order {in a coupling}
expansion, see, e.g.\ \cite{Pokorski:1987ed}. In the Wetterich
formulation of the functional RG, the regulator, being introduced as a
momentum-dependent mass term, also contributes to the breaking of the
gauge symmetry similar to the gauge-fixing procedure. This also
affects the Ward identities which are accordingly modified by
regulator-dependent terms
\cite{Reuter:1993kw,Ellwanger:1994iz,Ellwanger:1995qf,Reuter:1997gx,Litim:1998nf,Litim:1998qi,Litim:2002ce,Pawlowski:2005xe,Gies:2006wv}. For
our case, these terms can be worked out explicitly along the lines of
\cite{Gies:2003dp,Gies:2004hy}, yielding the modified relation
\be
Z_{\bar{\psi}A\psi}=Z_{\psi}\left( 1 - C_{g} g - C_{\tilde{g}}\tilde{g}\right)\,,
\ee
where $C_{g}$ and~$C_{\tilde{g}}$ are constants depending on the number of
fermion flavors as well as the regularization scheme. 

At this point, let us
schematically define the photon anomalous dimension analogously to
\Eqref{eq:defetapsi} as $\eta_A=- \partial_t \ln Z_{A}$ (a more precise
definition also accounting for the momentum dependence of $Z_A$ will be given
below). Then, the flow equation for the gauge coupling \eqref{eq:defeq} reads
\be
\partial_t e^2 =  (\eta_{A} - 1)e^2 
 - 2 \frac{\left( C_{g} (\partial _t g) \!+\! C_{\tilde{g}}(\partial_t
     \tilde{g})\right)}{\left( 1 \!-\! C_{g} g\! -\!
     C_{\tilde{g}}\tilde{g}\right)}.
\label{eq:betaeq}
\ee
In addition to the first term expected from perturbation theory, we encounter
additional terms proportional to the flows of the fermion couplings which
diagrammatically correspond to a resummation of a large class of
diagrams. Below, we will investigate the approach to possible phase
transitions as a function of $\Nf$ by means of a fixed-point analysis. As
fixed points are defined as points in theory space where the RG flow vanishes,
i.e., $\partial_t g = \partial_t \tilde{g}=0$, the additional 
terms in \Eqref{eq:betaeq} vanish identically at the fermionic fixed points
and thus are irrelevant for the determination of the fixed point of the full
system. For our fixed-point analysis presented below, these additional terms can
therefore be ignored. 
\begin{figure}
\begin{center}
\includegraphics[clip=true,width=0.8\columnwidth]{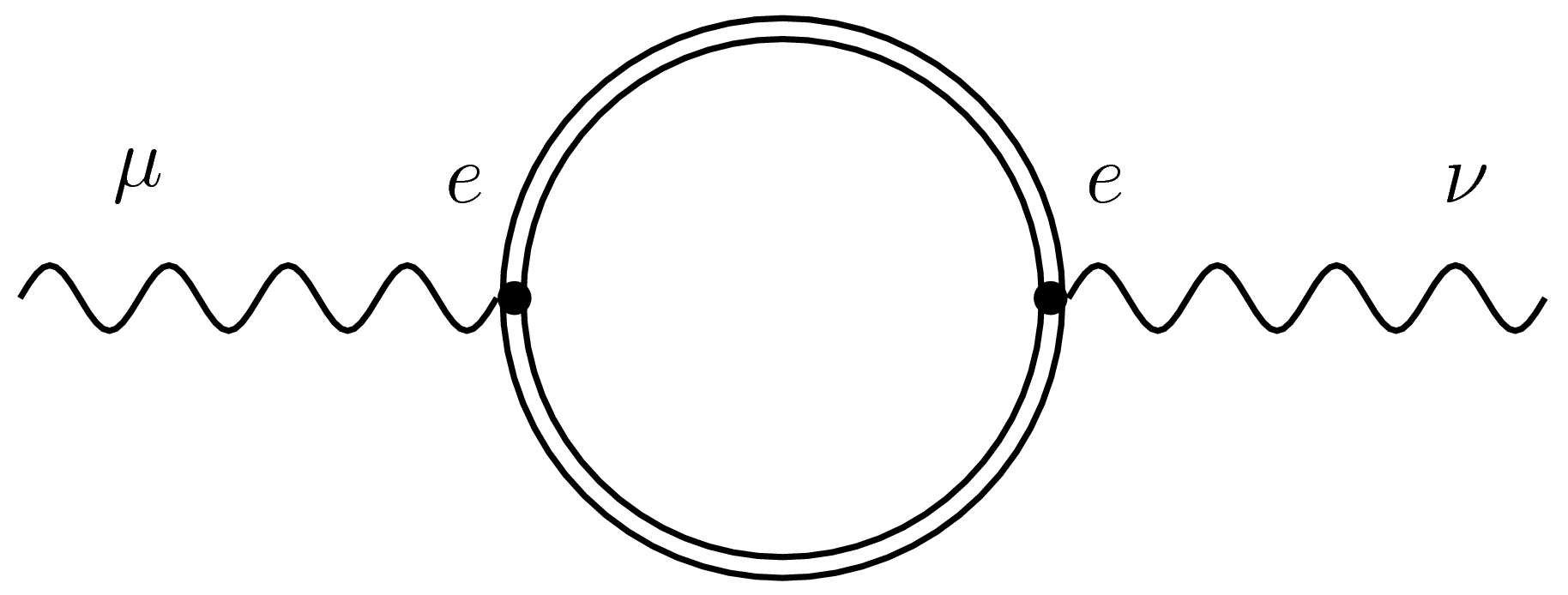}
\end{center}
\caption{1PI diagram contributing to the vacuum polarization tensor~$\Pi_{\mu\nu}$: the double lines represent
(full) scale-dependent regularized fermion propagators. The flow of the photon
wave-function renormalization is driven by the scale-derivative of this
diagram with respect to the regulator.}
\label{fig:pp}
\end{figure}

Finally, we have to give a precise definition of the photon anomalous
dimension in order to complete our set of flow equations for our
truncation. The evaluation of the photon polarization tensor, corresponding to
the diagram in Fig.~\ref{fig:pp},
yields a fully momentum dependent wave-function renormalization $Z_A(p^2)$. 
Since the integrand of the momentum trace in the flow equation by construction is peaked
for loop-momenta $q$ near the regulator scale, $q^2\simeq k^2$, it is crucial to 
obtain
a reliable estimate of the gauge coupling that parametrizes the
photon-fermion interaction strength of the modes interacting at momentum
transfer of the order of the scale $k$. As the running of the gauge coupling
is dominated by the photon anomalous dimension (at least near fermionic fixed
points), we define $\eta_A$ with the aid of the scale derivative of $Z_A(p^2)$
at a momentum scale $p^2$ evaluated near $k^2$.
To be more specific, we define
\begin{equation}
\eta_A=- \partial_t \ln Z_A(p^2=\zeta^2k^2)
\label{def:etaA}
\end{equation}
where $\zeta$ serves as a control parameter that can be used to
estimate the dependence of our final results on the details of the
definition of~$\eta_{A}$ and thus on the definition of the gauge
coupling. The parameter $\zeta$ fixes the momentum scale $p$
  serving as the (re-)normalization point of the photon field
  amplitude relative to the Wilsonian momentum shell $k$. Large values
  of $\zeta\gg 1$ therefore appear to be artificial, since the
  physically relevant momenta would then lie far beyond the Wilsonian
  momentum shell. As a consequence, we expect $\eta_A$ to be a
  decreasing function of $\zeta$ for large $\zeta$ for purely
  kinematical reasons. The natural range of physically relevant
  $\zeta$ values hence is $0\leq \zeta \lesssim 1$, with $\zeta\to 0$
  corresponding to the pointlike limit. For a more adapted resolution
  of nontrivial momentum-dependencies of $Z_A(p^2)$, the choice
  $\zeta=1$ appears a priori preferrable.

In the determination of $Z_A(p^2)$ via the polarization tensor,
another subtlety is hidden: the standard Ward identity for the
polarization tensor $p_\mu\Pi(p)_{\mu\nu}=0$ is also affected by the
presence of the regulator, yielding a nonzero regulator-dependent term
on the right-hand side that vanishes in the limit $k\to0$. 
This is a known peculiarity of the present Wilsonian-type of RG flow,
see, e.g., Refs.~\cite{Reuter:1993kw,Reuter:1997gx,Freire:2000bq,
Litim:2002ce,Fischer:2004uk,Pawlowski:2005xe,Gies:2006wv,herbut2007modern} for a
more detailed discussion of this issue.
In order to
avoid a contamination of our gauge coupling definition with these
artificial regulator-dependent terms, we subtract the $p\to0$ limit of
$\Pi_{\mu\nu}$ for finite $k$ in the determination of $Z_A(p^2)$.
This guarantees that the information entering the anomalous dimension
$\eta_A$ is not contaminated by contributions that arise in the
RG flow {only} in order to satisfy the regulator-dependent constraint on
the (unphysical) longitudinal modes. The technical details of the
construction of $\eta_A$ are summarized in Appendix~\ref{app:za}. In
any case, the result for $\eta_A$ has a comparatively simple form,
\be
\eta_A = 8 v_3 N_{\rm{f}} e^2 {\mathcal L}^{\rm
  (F)}_1(\eta_\psi;\zeta)\,,\label{eq:etaAFlow} 
\ee
{where}~$v_3=1/(8\pi)^2$, and ${\mathcal L}^{(\text F)}_1$ denotes a
threshold function that corresponds to the regularized one-particle
irreducible (1PI) Feynman diagram shown in Fig.~\ref{fig:pp}. It
depends on the choice of the regulator, thus encoding the RG-scheme
dependence, and also on the control parameter $\zeta$ introduced
above. The dependence on the fermion anomalous dimension $\eta_\psi$
signals the ``RG-improvement'' inherent in the functional
RG. The explicit integral representation of ${\mathcal L}^{\rm
  (F)}_1(\eta_\psi;\zeta)$ can be found in \Eqref{eq:AppL}.

We conclude this section by listing the fermion anomalous dimension,
\be
\!\!\!\!\!\eta_\psi = \frac{16}{3}v_3 e^2 \left(m^{\rm (F,B)}_{2,1}(\eta_\psi,\eta_A) - \tilde{m}^{\rm (F,B)}_{1,1}(\eta_\psi,\eta_A)\right),\label{eq:etapsiFlow}
\ee
with the regulator-dependent threshold {functions $m^{\rm (F,B)}_{2,1}$ and
$\tilde{m}^{\rm (F,B)}_{1,1}$, as defined, e.g., in Refs.~\cite{Berges:2000ew,Gies:2002hq,Braun:2011pp}.}
As the threshold functions are linear in
the anomalous dimensions, Eqs.~\eqref{eq:etaAFlow} and \eqref{eq:etapsiFlow}
can unambiguously be solved for $\eta_\psi$ and $\eta_A$ as functions of the
gauge coupling.

The RG $\beta$ functions for the fermion sector read
\be
\partial_t \tilde{g} &=&(1\!+\!2\eta_\psi) \tilde{g} -8v_3\left(\frac{2N_{\rm{f}}-1}{N_{\rm{f}}}\tilde{g}^2\!-\!\frac{3}{N_{\rm{f}}}\tilde{g}g\!-\!\frac{2}{N_{\rm{f}}}g^2\right) l^{\rm (F)}_1\nn\\
&& \quad -8v_3\left(2\tilde{g}e^2\!+\!4ge^2\right)l^{\rm (F,B)}_{1,1}\!+\! 16v_3N_{\rm{f}}e^4l^{\rm (F,B)}_{2,1}\,,\label{eq:sFlow}\\
\partial_t g &=& g(1\!+\!2\eta_\psi)+ 8v_3\left(\frac{1}{N_{\rm{f}}}\tilde{g}g+\frac{2N_{\rm{f}}+1}{3N_{\rm{f}}}g^2\right)l^{\rm (F)}_1\nn\\
&& \quad -\frac{8}{3}v_3\left(4\tilde{g}e^2-2ge^2\right)l^{\rm (F,B)}_{1,1}\,, \label{eq:gFlow}
\ee
where the threshold functions $l$ again carry the regulator dependence
and depend linearly on $\eta_\psi$ via $l^{(\text F)}_1$. For the
evaluation of the photon exchange diagrams, we neglect the full
momentum dependence of the photon propagator, but take the photon
field renormalization at the renormalization point
$Z_A(p^2)=Z_A(\zeta^2k^2)$ into account. Hence, the threshold
functions $l^{(\text{F,B})}_{1,1}$ and $l^{(\text{F,B})}_{2,1}$ depend
also on $\eta_A$. For the so-called sharp-cutoff,
Eqs.~\eqref{eq:sFlow}--\eqref{eq:gFlow} are equivalent to the results
reported in Ref.~\cite{Kubota:2001kk}. In the limit of large flavor
number $\Nf$, they also reduce to the large-$\Nf$ flow equations found
previously within the conventional Wilsonian RG approach
\cite{Kaveh:2004qa}. We would like to add that the sharp-cutoff
  regulator has to be handled with some care. Whereas this type of
  regulator can be used to compute the flow equations for the
  pointlike four-fermion couplings without any difficulty, the
  computation of the flow equations for the wave-function
  renormalizations suffers from ambiguities which can be traced back
  to the fact that there is no unique definition for this regulator,
  see Appendix~\ref{app:reg}.  Since the photon wave-function
  renormalization plays a prominent role in our study of the
  many-flavor phase structure, we refrain from using this regulator in
  the following. Instead, we only consider a smeared-out version of
  this regulator which is free of these
  difficulties.\footnote{This amounts to using a finite value
      for the parameter~$b$ in our definition of the sharp-cutoff
      regulator, see Eq.~\eqref{eq:defscreg}.} For the latter we have
  found that it yields results for the phase structure that are in
  accordance with those reported in Sect.~\ref{sec:quant} below.
 
{For vanishing gauge coupling $e^2=0$, we observe that
the fermionic $\beta$ functions~\eqref{eq:sFlow} and~\eqref{eq:gFlow} vanish} identically if $g,\tilde{g}$
are zero {at a particular scale} (as, e.g., required by the initial
conditions \eqref{eq:initcond}). This obvious fixed point of the flow
corresponds to the non-interacting Gau\ss{}ian fixed point of the
theory. For $e^2\neq0$, the point of vanishing fermionic couplings is
no longer a fixed point due to the last term $\sim e^4$ in
\Eqref{eq:sFlow}. 

Finally, the flow of the gauge coupling is given by
\Eqref{eq:betaeq} upon insertion of the anomalous dimension $\eta_A$
and the fermionic flows. Near fixed points of the fermionic flow,
where $\partial_t g,\partial_t \tilde{g} \simeq 0$, the
$\beta$ function of the gauge coupling simplifies to
\be
\beta_{e^2}\equiv\partial_t e^2 &=& (\eta_{A} - 1)e^2.\label{eq:eFlowsimple}
\ee
For the fixed-point analysis carried out in the present work, we
consider this simplified flow.

We close this section with a few comments on the reliability of the
approximations involved in our truncation. In our numerical studies, 
we indeed find that~$|\eta_{\psi}| \lesssim 1$ in the symmetric
large-$\Nf$ regime where the RG flow is governed by the presence of a fixed point, see also
our discussion in the subsequent section. This is a strong support for our implicit
assertion that momentum dependencies in the fermion sector are less
important, such that higher derivative terms of fermionic operators
can safely be dropped in this regime.  Moreover, it is worthwhile to
point out that in the pointlike limit the RG flow of a Fierz-complete
set of four-fermion couplings is completely decoupled from the RG flow
of fermionic $n$-point functions of higher order. In particular,
$8$-fermion interactions do not contribute to the flow of the
four-fermion interactions in this limit. This observation corroborates
the truncation on the four-fermion level.  Further tests of the
truncation -- particularly of the gauge sector -- will actively be
pursuit in the following sections by studying the amount of artificial
regularization-scheme dependence of observables.

\section{Fixed-Point Analysis}\label{sec:fpana}

The RG fixed-point structure of a theory is intimately related to the
phase diagram. Fixed points are defined as common zeros of all
$\beta$ functions, in our case by the requirement
\begin{equation}
 \partial_t e^2|_{e^2_\ast,g_\ast,\tilde{g}_\ast}  = \partial_t g|_{e^2_\ast,g_\ast,\tilde{g}_\ast} = \partial_t
 \tilde{g}|_{e^2_\ast,g_\ast,\tilde{g}_\ast} =0,
\label{eq:fixedpointcond}
\end{equation}
where $e^2_\ast,g_\ast,\tilde{g}_\ast$ denote the values of the
dimensionless couplings at the fixed point. Whereas the fixed-point
values {themselves} are non-universal, i.e., depend on the choice of the
regularization scheme, the critical exponents as well as the anomalous
dimensions $\eta_{\psi,\ast}$ and $\eta_{A,\ast}$ at a fixed point are
universal. Summarizing all couplings in
$\mathbf{G}=(e^2,g,\tilde{g})$, the critical exponents $\theta_I$ are
defined in terms of (minus) the eigenvalues of the stability matrix
$B_i{}^j$,
\begin{equation}
\partial_t G_i = \beta_i(\mathbf{G}), \quad B_i{}^j =
\left. \frac{\partial{\beta_i}}{\partial G_j}
\right|_{\mathbf{G}=\mathbf{G}_\ast}, \label{eq:stabmat}
\end{equation}
with $-\theta_I$ labeling the eigenvalues of $B_i{}^j$, and $I$ running from 1
to the number of couplings considered ($I=1,2,3$ in our case). For instance,
at the Gau\ss{}ian fixed point, $\mathbf{G}=0$, we have
$\theta_I=\{+1,-1,-1\}$, with the positive exponent $+1$ related to the RG
relevant gauge coupling. The negative exponents $-1$ correspond to the RG
irrelevant fermionic couplings in QED${}_3$. At the Gau\ss{}ian fixed point,
the critical exponents simply correspond to the power-counting dimension of
the couplings. 

In order to illustrate the fixed-point structure of the theory, let us
start with the flow of the gauge coupling. Assuming that the fixed-point
conditions for the fermion {couplings are satisfied, we can} use~\Eqref{eq:eFlowsimple}. In
addition to the Gau\ss{}ian fixed point, a non-Gau\ss{}ian, i.e.,
interacting, fixed-point exists for
\be
\eta_{A,\ast}=1, \quad e^2_{\ast}=\frac{1}{8v_3\Nf{\mathcal
      L}^{(\text{F})}_1(\eta_{\psi,\ast};\zeta)}\,, \label{eq:IRFPeq}
\ee
where the threshold function ${\mathcal L}^{\rm (F)}_1$ with $\eta_{\psi}$
evaluated at the IR fixed point is a regulator-dependent but 
real-valued positive number.\footnote{Negative values could only occur for very large
$\eta_{\psi,\ast}$ which would indicate the breakdown of our truncation
anyway. For all flows studied in this work, $\eta_\psi$ generically remains
rather small, $|\eta_\psi| \lesssim 1$, provided that the dynamics is governed by
a fixed point. If, on the other hand, the IR fixed point of the gauge coupling is destabilized
by, e.g., spontaneous (chiral) symmetry breaking, then~$\eta_{\psi}$ may grow 
rapidly as well. However, a detailed analysis of this scenario is beyond the scope of our 
present work.} The crucial observation is that the
value of the fixed point scales with the flavor number $\Nf$ as
$e^2_{\ast}\sim 1/\Nf$. 

\begin{figure}
\begin{center}
\includegraphics[clip=true,width=1\columnwidth]{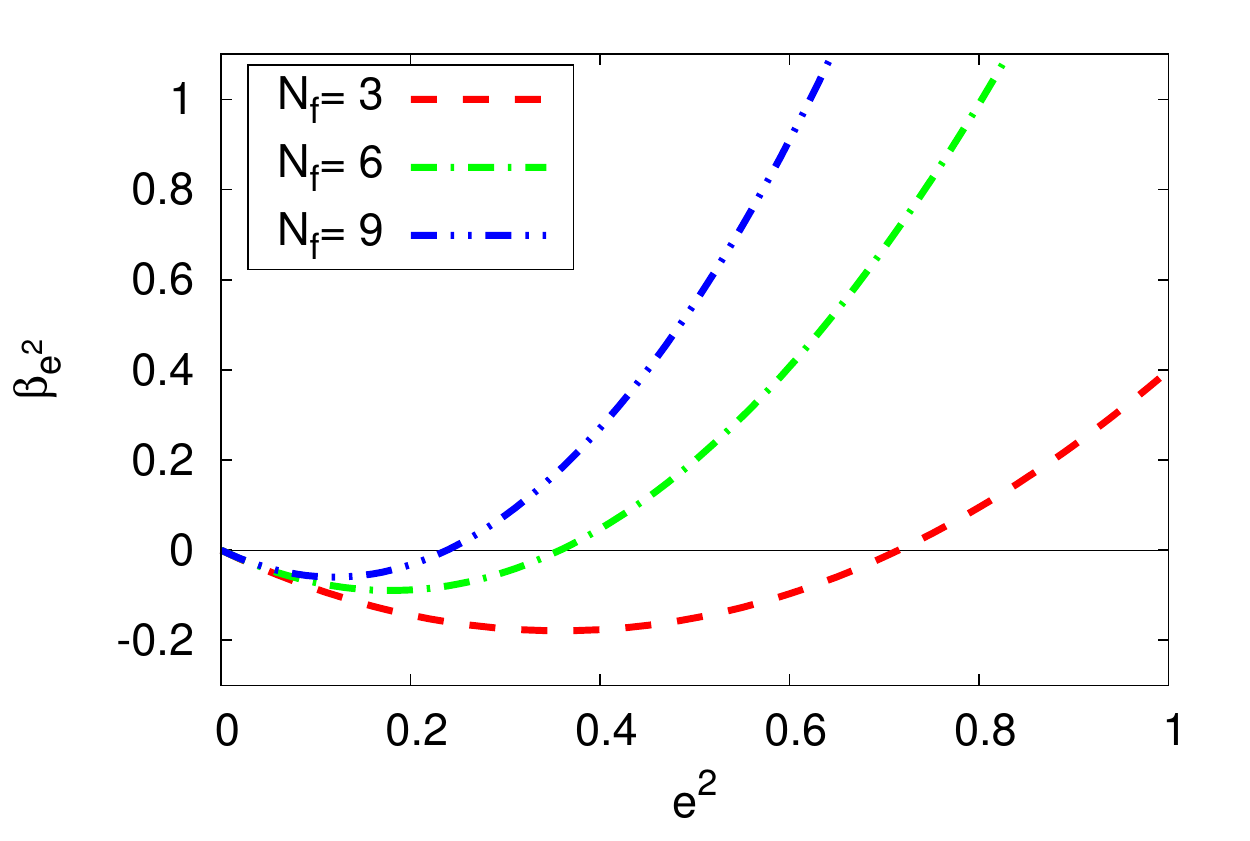}
\end{center}
\caption{(color online) $\beta_{e^2}$ function for three different
  values of~$\Nf$ as obtained from the linear regulator evaluated
  for~{$\zeta=1$ and} $\eta_{\psi}=0$ for simplicity. }
\label{fig:e2beta}
\end{figure}

Starting the RG flow near the Gau\ss{}ian fixed point at~$e^2\ll 1$,
the $\beta$ function $\partial_t e^2$ is negative, implying that the
coupling is asymptotically free towards the UV and increases towards
the IR. Hence, the gauge coupling is expected to approach the
non-Gau\ss ian fixed-point in the long-range limit, see
Fig.~\ref{fig:e2beta}.  As long as no fermion-mass generating phase
transition occurs in which case the dynamics of the theory would be
governed by a different sector of the theory, the whole system remains
massless and the IR fixed point [\Eqref{eq:IRFPeq}] is reached 
asymptotically {at} small momentum scales. In that case, the theory is
quasi-conformal, i.e., near-conformal in the UV near the Gau\ss{}ian
fixed point as well as near-conformal in the IR near the
non-Gau\ss{}ian fixed point. The two near-conformal regimes are
smoothly connected by a crossover occurring at momentum scales near
the scale approximately set by the bare coupling $\bar{e}^2$. Note
that the maximum coupling strength of the dimensionless coupling is
set by the IR fixed-point value, see~\Eqref{eq:IRFPeq}. In particular,
the maximum coupling strength is smaller for larger flavor numbers.

\begin{figure}
\includegraphics[clip=true,width=\columnwidth]{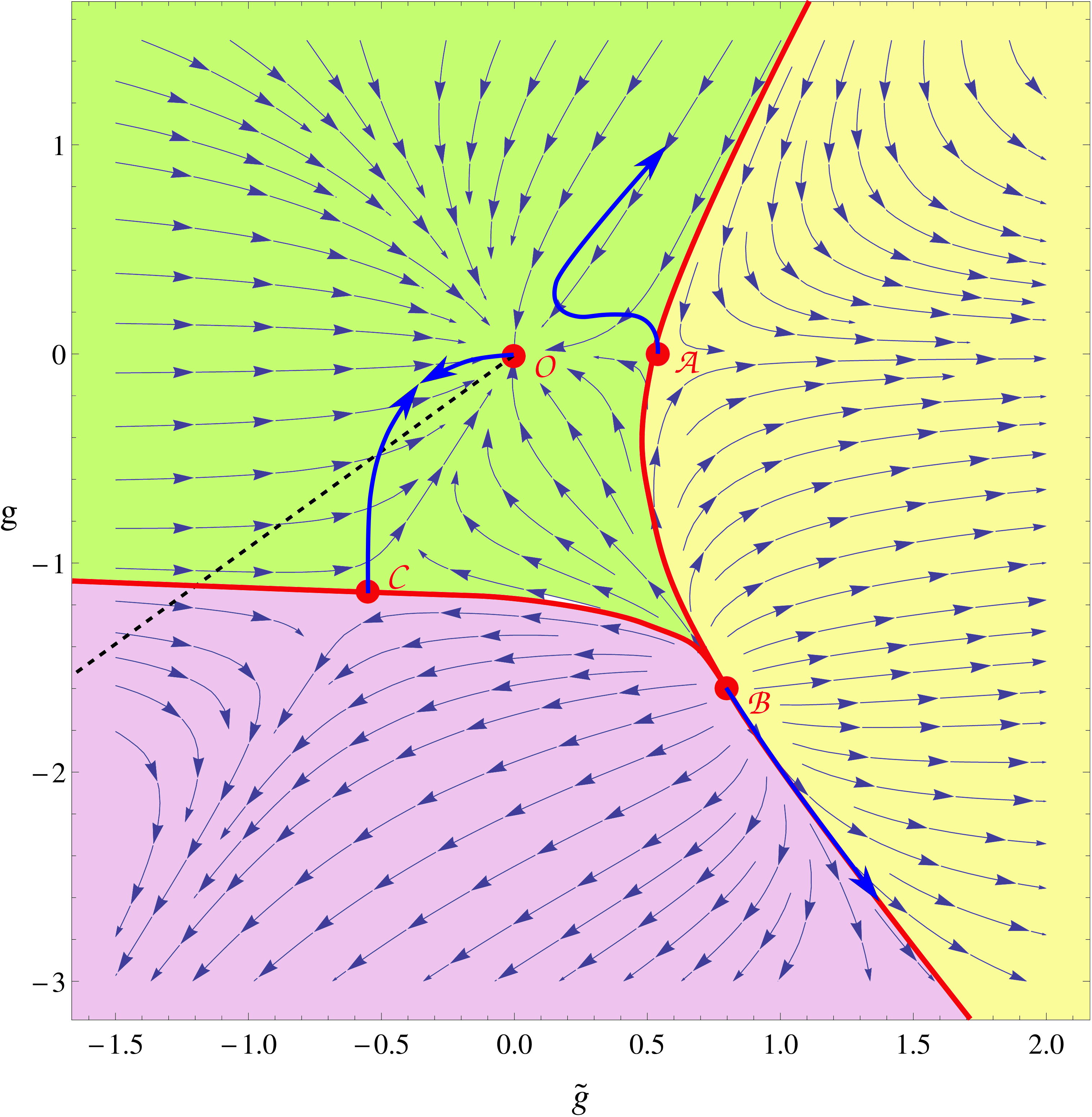}
\caption{{(color online)} RG trajectories in the plane spanned by the
four-fermion couplings~$\tilde g$ and~$g$ for $\Nf=4$ and~$e^2=0$ {using
    the linear regulator}. The fixed points
  are depicted by the red dots, where~$\mathcal O$ is the IR stable
  {Gau\ss{}ian} fixed point, $\mathcal A$ and~$\mathcal B$ are
  fixed points with one IR attractive and one IR repulsive direction,
  and~$\mathcal B$ is an unstable fixed point with two IR
  repulsive
  directions. The thin arrows indicate the RG flow towards the IR
  regime. The dashed line ($g=\tilde{g}$) corresponds to the
  \chiral channel (where $g_V=0$ and $g_\phi$ is nonzero), potentially
  associated with chiral symmetry breaking, see also
  Eq.~\eqref{eq:coupFierz}. This channel is typically chosen in
  Fierz-incomplete studies. The blue/{bold} 
  {arrows} attached to the
  four fixed points indicate the shift of the fixed
  points induced by an increase of the gauge coupling {$e^2 > 0$}.  }
\label{fig:fpnf3}
\end{figure}

Let us now turn to the fermionic sector with the corresponding flows
given in Eqs.~\eqref{eq:sFlow} and \eqref{eq:gFlow}, treating the
gauge coupling as an external parameter for the moment. As the
fixed-point conditions for $g$ and $\tilde{g}$
[\Eqref{eq:fixedpointcond}] correspond to two coupled quadratic
equations, we generically expect up to four distinct fixed-point
solutions. Provided that the gauge
coupling is sufficiently small, we find four distinct real solutions
  which thus represent candidates for physically relevant
fixed points.  For finite $e^2 > 0$, these points in
  coupling space are no longer fixed points of the total system, as
  their positions change with the gauge coupling $e^2$. In a slight abuse of
language, we still call them fixed points, as for a given value of $e^2$ they 
govern the flow in the fermionic sector. 
In the limit $e^2 \to 0$, one of the four fixed points is continuously
connected to the (true) Gau\ss{}ian fixed point at $\mathbf G
  = 0$.  For small but finite $e^2$, this fixed point is slightly
shifted to nonzero couplings $\tilde g_\ast,g_{\ast}$ but continues to
have two RG irrelevant directions. This fixed point, named $\mathcal
O$ in Fig.~\ref{fig:fpnf3}, is {thus} IR attractive in the $(\tilde
g,g)$ plane.  Two further fixed points $\mathcal A$ and $\mathcal C$
have one IR attractive (RG irrelevant) and one IR repulsive (RG
relevant) direction, and the fixed point $\mathcal B$ exhibits two IR
repulsive directions, see. Fig.~\ref{fig:fpnf3}.

For vanishing gauge coupling, {$e^2=0$}, the Gau\ss{}ian {fixed point
  $\mathcal{O}$} describes a free theory of non-interacting
fermions. The fixed point $\mathcal C$ has been extensively studied in
\cite{Gies:2010st,Janssen:2012pq,Janssen:2012oca}. It can be
associated with the asymptotically safe three-dimensional Thirring
model. For sufficiently small flavor numbers $\Nf<\Nfcr^{\chi,
  \text{Thirring}}$, the fixed point controls a second-order quantum
phase transition, separating the massless phase from the phase of
chiral symmetry {breaking, see, e.g., \cite{oai:arXiv.org:1207.4054}
  for a study of the $\Nf=1$ model}. In
Refs.~\cite{Janssen:2012pq,Janssen:2012oca}, the critical flavor
number of the Thirring model has been estimated as $\Nfcr^{\chi,
  \text{Thirring}}\simeq 5.1$.  {Lattice studies of the
  Thirrig model with a different realization of the chiral symmetry
  using staggered fermions found ${\Nfcr^{\chi,\text{Thirring}}}
\simeq 6.6$
  \cite{oai:arXiv.org:hep-lat/0701016}.}
\footnote{{In the literature, estimates for the critical flavor number
    of the Thirring model span a wide range of values
    \cite{Gomes:1991aa,Hong:1993qk,oai:arXiv.org:hep-th/9411201,oai:arXiv.org:hep-th/9611198,oai:arXiv.org:hep-th/9502070,
      oai:arXiv.org:hep-lat/9605021,oai:arXiv.org:hep-lat/9701016,oai:arXiv.org:hep-lat/9906008,oai:arXiv.org:hep-lat/0701016}. Many
    of the analytical estimates show a strong similarity to the
    corresponding QED${}_3$ results.}} 

The fixed point $\mathcal A$ corresponds to a variant of the
three-dimensional Gross-Neveu model. Different versions of this
  model exist in $d=3$, all of which are asymptotically safe because of
  such a non-Gau\ss{}ian fixed point
  \cite{Rosa:2000ju,Hofling:2002hj,Braun:2010tt}. 
  This fixed point governs the second-order quantum phase transition of a
  discrete $\mathbbm{Z}_2$ symmetry (parity symmetry in this case)
  which is known to occur for any $\Nf$. By contrast, the fixed point
  $\mathcal B$ has less well been studied, but could equivalently give
  rise to an asymptotically safe fermionic model potentially
  exhibiting first-order phase transitions to various phases in the
  IR.

Returning now back to QED${}_3$, the initial conditions~\eqref{eq:initcond} put the system into the vicinity of the
Gau\ss{}ian fixed point $\mathcal O$ at the microscopic scale $k\to\Lambda$,
leaving us with one RG relevant parameter, namely the gauge coupling,
as it should be. Towards the UV, the full system is asymptotically
free. Towards the IR, the gauge coupling increases, shifting the
Gau\ss ian fixed point $\mathcal O$ slightly in the $(\tilde{g},g)$ plane, see
blue/{bold} arrows in Fig.~\ref{fig:fpnf3}. Since $\mathcal O$
remains IR
attractive in the fermionic directions, the flow of $\tilde{g},g$
follows this IR attractive fixed point.

If the gauge coupling approaches a critical value~$e^2_{\rm cr}$, the
fixed points~$\mathcal C$ and~$\mathcal O$ annihilate, see Fig.~\ref{fig:fpnf3}. If we increase the
gauge coupling even further, then the flow of the four-fermion couplings is no
longer bounded by the existence of an IR attractive fixed point. On the
contrary, the four-fermion interactions start to grow rapidly and diverge at a
finite RG scale~$\ksb$, potentially indicating dynamical symmetry breaking, as
discussed above.  

From the fixed-point analysis itself, we do not gain immediate insight
into the exact type of spontaneous symmetry breaking, as this is a
result of the full RG flow towards the IR. Nevertheless, the fixed-point
analysis provides for a criterion for symmetry breaking to be
possible at all: as long as the fixed point $\mathcal O$ exists, being IR
attractive for the fermionic couplings, no approach to criticality in
the fermion sector can occur. Thus, monitoring the existence of this
fixed point as a function of $\Nf$ provides first information about
the structure of the phase diagram as a function of $\Nf$. 

\section{\Confcrit Flavor Number}\label{sec:quant}

\begin{figure}
\includegraphics[clip=true,width=\columnwidth]{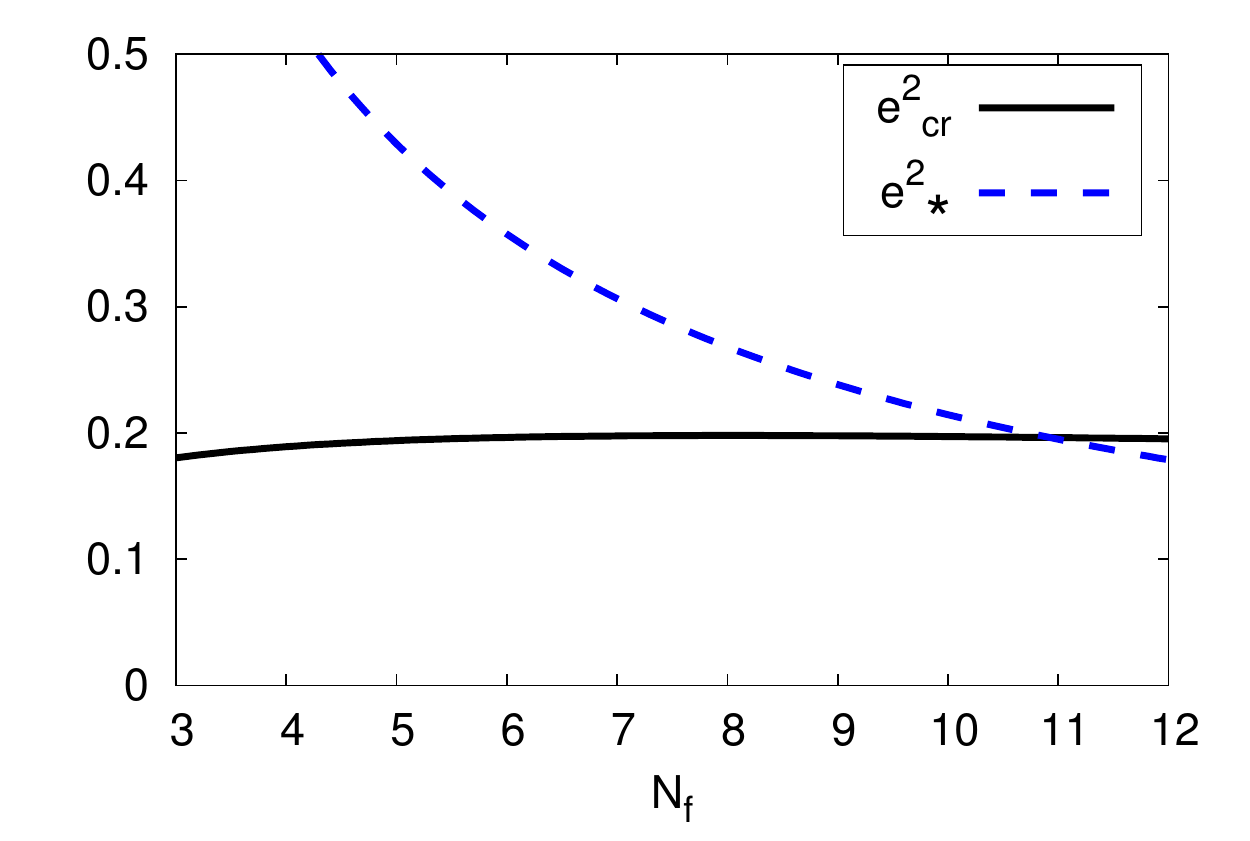}
\caption{(color online) The critical value $e^2_{\rm cr}$ for the
  gauge coupling and the value~$e^2_{\ast}$ of the IR fixed point as a
  function of~$\Nf$ as obtained from the linear regulator
  for~{$\zeta=1$ and} $\eta_{\psi}=0$ for
  simplicity.  The intersection point of both lines determines the
  \confcrit flavor number~$\Nfpc$, see Eq.~\eqref{eq:critcond}.  Note
  that the depicted $\Nf$ dependence of~$e^2_{\ast}$ has been computed
  with the aid of~\Eqref{eq:eFlowsimple}.  However, the associated IR
  fixed-point~$e^2_{\ast}$ is only approached for $\Nf\geq
  \Nfpc$. For~$\Nf<\Nfpc$, this fixed point is destabilized due to
  spontaneous symmetry breaking.
}
\label{fig:efpecr}
\end{figure}

From the preceding discussion, we expect the system to be quasi-conformal as
long as the {fixed point $\mathcal{O}$} in the fermion sector persists and remains IR
attractive in the fermionic couplings. The {fixed point $\mathcal{O}$} vanishes if the
gauge coupling exceeds a critical coupling strength $e^2_{\rm cr}$. In the
quasi-conformal phase, the IR fixed point $e^2_{\ast}$ as given in
\Eqref{eq:IRFPeq} is a measure for the maximum coupling strength. Since
$e^2_{\ast}$ is small for large $\Nf$, the quasi-conformal phase occurs at
large $\Nf$ extending to $\Nf\to \infty$. Lowering $\Nf$, the annihilation of
the fixed points $\mathcal O$ and $\mathcal C$ indicate the boundary of the quasi-conformal
phase and a possible onset of a different phase. The corresponding value of
$\Nf$ defines the \confcrit flavor number $\Nfpc$ which is defined by
the criticality condition 
\be
e^2_{\ast}(\Nfpc)\stackrel{!}{=}e^2_{\rm cr}(\Nfpc)\,,\label{eq:critcond}
\ee
see also Fig.~\ref{fig:efpecr}.  Whereas both {$e^2_\ast$ and
$e^2_\mathrm{cr}$} are non-universal
and depend on the choice of the regularization scheme, the
\confcrit flavor number $\Nfpc$ is expected to be
universal.\footnote{Since $\Nfpc$ presumably is not an integer, its
  value might depend on the manner, how theories with non-integer
  flavor numbers are constructed. Nevertheless, the result that
  systems with integer $\Nf>\Nfpc$ have long-range properties
  substantially different from those with integer $\Nf<\Nfpc$ is in
  principle a universal and observable phenomenon.} 
However, the fact that we
consider an approximation of the exact RG flow 
implies that also the universality of $\Nfpc$ holds only approximately.

In Tab.~\ref{tab:Nfpc}, we list our results for~$\Nfpc$ as obtained
from our computations with three different regulator functions,
see App.~\ref{app:reg} for the definitions of these
functions. We also consider two different values of the control
  parameter $\zeta$ which parametrizes the external photon momentum
of the vacuum polarization diagram relative to the cutoff scale,
cf.\ Eq.~\eqref{def:etaA}. Whereas the choice $\zeta=1$ appears
  more adapted to resolve the momentum dependence of the photon wave
  function, the choice $\zeta=0$ conforms with the pointlike
  approximation in the fermion sector. In either case, we obtain the
  smallest value of $\Nfpc$ for the Callan-Symanzik regulator. Since
  the latter is equivalent to a mass term $\sim k$ without any
  momentum dependence, it does not entail a UV suppression and
  therefore is likely to give rise to stronger truncation artifacts,
  as is also known from many other RG studies. The two other
  regulators, the exponential and the linear regulator, cf.,
  App.~\ref{app:reg} for details, provide for both a UV and IR
  regularization and are thus considered as quantitatively more
  reliable. These two regulators span the range of estimates for
  $\Nfpc$ of $\Nfpc\simeq 8, \dots, 10$ for $\zeta=1$ and $\Nfpc\simeq
  4, \dots, 5.7$ for $\zeta=0$ with the largest $\Nfpc$ value arising
  from the linear regulator, respectively. Intermediate values of
  $\zeta$ yield ranges that interpolate between the $\zeta=0$ and
  $\zeta=1$ case.\footnote{Incidentally, a smeared version of
      the sharp-cutoff with smearing parameter $b\simeq2$ (see
      App.~\ref{app:reg}) yields values for $\Nfpc$ within the ranges
      spanned by the exponential and the linear regulator.} We
  observe that the variation with respect to the control parameter
  $\zeta$ is even larger than the regulator dependence. We interpret
  this as a signature for the importance of the precise resolution of
  the momentum dependencies of the correlation functions.

In general, these uncertainties indicate a systematic error to
be associated with the employed truncation. For example, the inclusion
of the full momentum dependence especially of the photon-propagator
and the fermion-photon vertex may be required to determine~$\Nfpc$
more precisely.
\begin{table}[t]
\centering
\begin{tabularx}{\columnwidth}{@{\extracolsep{\fill}} cccc}
\hline\hline
\noalign{\smallskip}
regulator & $R_{\rm{CS}}$& $R_{\rm{exp}}$ & $R_{\rm{lin}}$ \\
 \noalign{\smallskip}\hline\noalign{\smallskip}
$\Nfpc (\zeta=1)$ & 7.5 & 8.1 & 10.0 \\ 
$\Nfpc (\zeta=0)$ & 3.7 & 4.1 & 5.7 \\ 
\noalign{\smallskip}\hline\hline
\end{tabularx}
\caption{\Confcrit flavor number for different regulator functions,
  Callan-Symanzik regulator (CS), exponential regulator (exp), linear
  regulator (lin), and for different choices of the control parameter $\zeta=1$ and
  $\zeta=0$}.
\label{tab:Nfpc}
\end{table}
In order to assess the stability of our results for the
  \confcrit flavor number, let us discuss the variations of the
  regulator and the control parameter in more detail: First, the
  dependence on the regulator is a natural consequence of truncated
  flows. This dependence can be lifted by identifying ``optimized''
  regularization schemes satisfying a-priori-criteria that can be
  argued to be closest to the exact results within a given truncation
  \cite{Litim:2000ci,Litim:2001up,Litim:2001fd,Pawlowski:2005xe}. The
  linear regulator is such an optimized regulator for the pointlike
  limit and with $\zeta=0$. For $\zeta=1$, none of our regulators is
  optimized in a similar sense. Different values of $\zeta$ should
  therefore be considered as different truncations. 

\begin{figure}
\begin{center}
\includegraphics[clip=true,width=0.8\columnwidth]{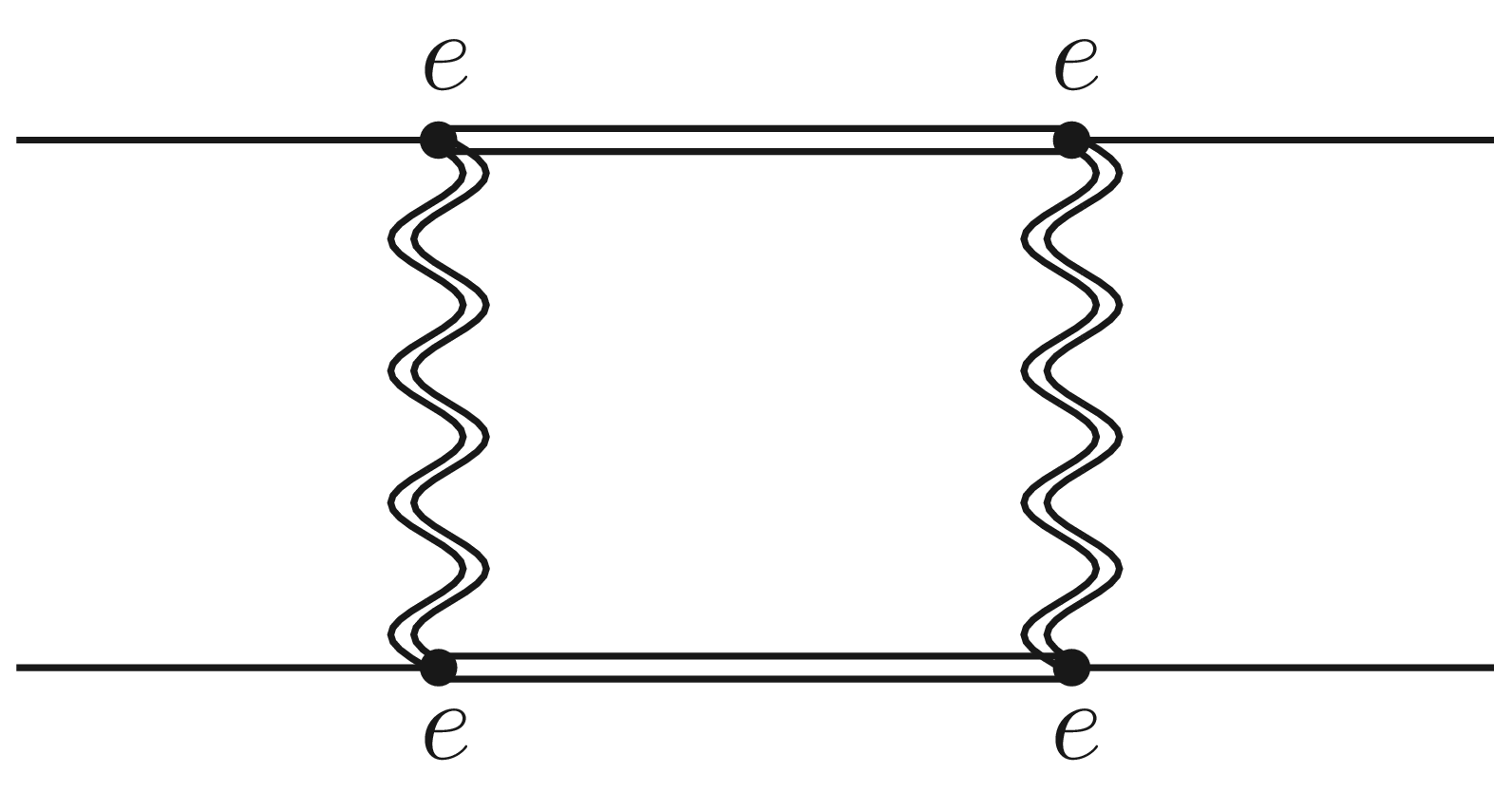}
\end{center}
\caption{1PI diagram contributing to the RG flow of the four-fermion couplings: the double lines represent
(full) scale-dependent regularized fermion and photon propagators.}
\label{fig:box}
\end{figure}

According to its definition $\zeta= |p|/k$, the control
parameter measures the relation between the incoming photon momentum
and the regularization scale of the internal fermion loop of the
vacuum polarization diagram, see Fig.~\ref{fig:pp}. For a
  reconstruction of the full momentum dependence of the photon wave
  function $Z_A(p^2)$ via the anomalous dimension formula
  \Eqref{def:etaA}, we hence consider the choice $\zeta=1$ more
  reliable. On the other hand, the vacuum polarization diagram is only used to
estimate the running coupling, which in turn enters the fermion box
diagrams as an estimate for the fermion-photon vertex, see also
Fig.~\ref{fig:box}. This estimate can be afflicted with the
  following problem: As we evaluate the box diagrams in the pointlike
limit, i.e., in the limit of zero external momentum, the vertex
  enters the flow equations at an asymmetric point, since the internal
  lines of the diagram carry an in general finite loop
  momentum. Therefore, {potentially asymmetric} structures of the vertices
  are neglected by our approximation. The intrinsic tension between
  such structures and our estimate for the running coupling could even
  be amplified by choosing a nonzero $\zeta$.

With this analysis of the regulator and $\zeta$ dependence, we
can now summarize our estimates for the location of the \confcrit
flavor number $\Nfpc$. From a conservative perspective, we have not
been able to find estimates of $\Nfpc$ with values smaller than
$\Nf\simeq 3.7$ or larger than $\Nf\simeq 10.0$ also including  extreme regulator
choices such as the Callan-Symanzik regulator. We hence conclude $\Nfpc$ to lie within this
interval. Excluding the Callan-Symanzik regulator in order to avoid
regulator artifacts, our results span a smaller region. The regulator
and $\zeta$ dependence analysis given above suggest the \confcrit
flavor number of QED${}_3$ to lie in the region
\be \Nfpc \approx 4.1 \dots 10.0\,.  \label{eq:NfpcEst}
\ee
We emphasize, however, that the upper and lower end of this interval
should not be viewed as a strict boundary, but may change upon
improvements of the approximation. Despite these uncertainties, this
estimate represents one of the main results of our study.

\section{Fierz Completeness}
\label{sec:fierz}

The above given estimate for the \confcrit flavor number $\Nfpc$ --
though coming with a large uncertainty -- appears to include values
significantly larger than
many results for the critical flavor number for
chiral symmetry breaking reviewed in the introduction. While there are
many sources that can take a strong influence on the final result
(e.g., large finite volume effects in finite-volume studies
\cite{Goecke:2008zh,Bonnet:2011hh}), we
emphasize in this work two issues that have not yet received sufficient
attention. 

First, we have determined the \confcrit flavor number $\Nfpc$
above which the system is quasi conformal. While this value is likely
to mark a region in the many-flavor phase diagram where a crossover or
a phase transition is expected to occur, it 
{does not necessarily have to} 
agree with the
critical flavor number for the chiral phase transition {$\Nfcr^\chi$}.
As we can only detect the quasi-conformal regime with our pointlike
approximation, we can only conclude so far that $\Nfcr^\chi\leq\Nfpc$,
cf.\ also next section for a discussion. Hence, there is no immediate
disagreement with the literature in this respect.

Second, we have emphasized that our ansatz for the effective action is
Fierz complete in the sense that it includes all pointlike four-fermion
interactions compatible with the symmetries of the model. The
significance of Fierz completeness for an appropriate description of
an approach to criticality is already obvious from our
parametrization. The chiral-symmetry breaking channel
$(S)^2$ in the Fierz-transformed Lagrangian in Eq.~\eqref{eq:LPsiInt2}
which, when becoming dominant, generates
a mass term $\sim i m \bar{\psi}^a\psi^a$, is associated
with a superposition of both four-fermion channels $\tilde g (P)^2$ and
$g (V)^2$
used in this work (see dashed line in Fig.~\ref{fig:fpnf3}). Ignoring
one of the channels {may}
lead to strong deviations from the
Fierz-complete result. 

In order to quantify the importance of Fierz completeness, we study
the dependence of our result for the \confcrit flavor number
$\Nfpc$ on a one-parameter family of Fierz-incomplete approximations.
To be specific, we first introduce a Fierz-complete reparametrization
of the couplings as follows:
\be
s_{\varphi}&=&g\sin\varphi+\tilde{g}\cos\varphi\,,\\
\tilde{s}_{\varphi}&=&g\cos\varphi-\tilde{g}\sin\varphi\,,
\ee
where the angle $\varphi$ parametrizes a family of couplings
$s_\varphi$, $\tilde{s}_\varphi$. From here, we arrive at a
Fierz-incomplete set by truncating
$\partial_t\tilde{s}_{\varphi} \equiv 0 \equiv \tilde{s}_{\varphi} $. The
angle~$\varphi$ can now be used to select a specific interaction channel.
For example for~$\varphi=\pi/4$, we have $\tilde{g}=g$, such that we
are left with the \chiral channel only, see also
Eq.~\eqref{eq:coupFierz} and the dashed line in Fig.~\ref{fig:fpnf3}. 
\begin{figure}
\includegraphics[clip=true,width=\columnwidth]{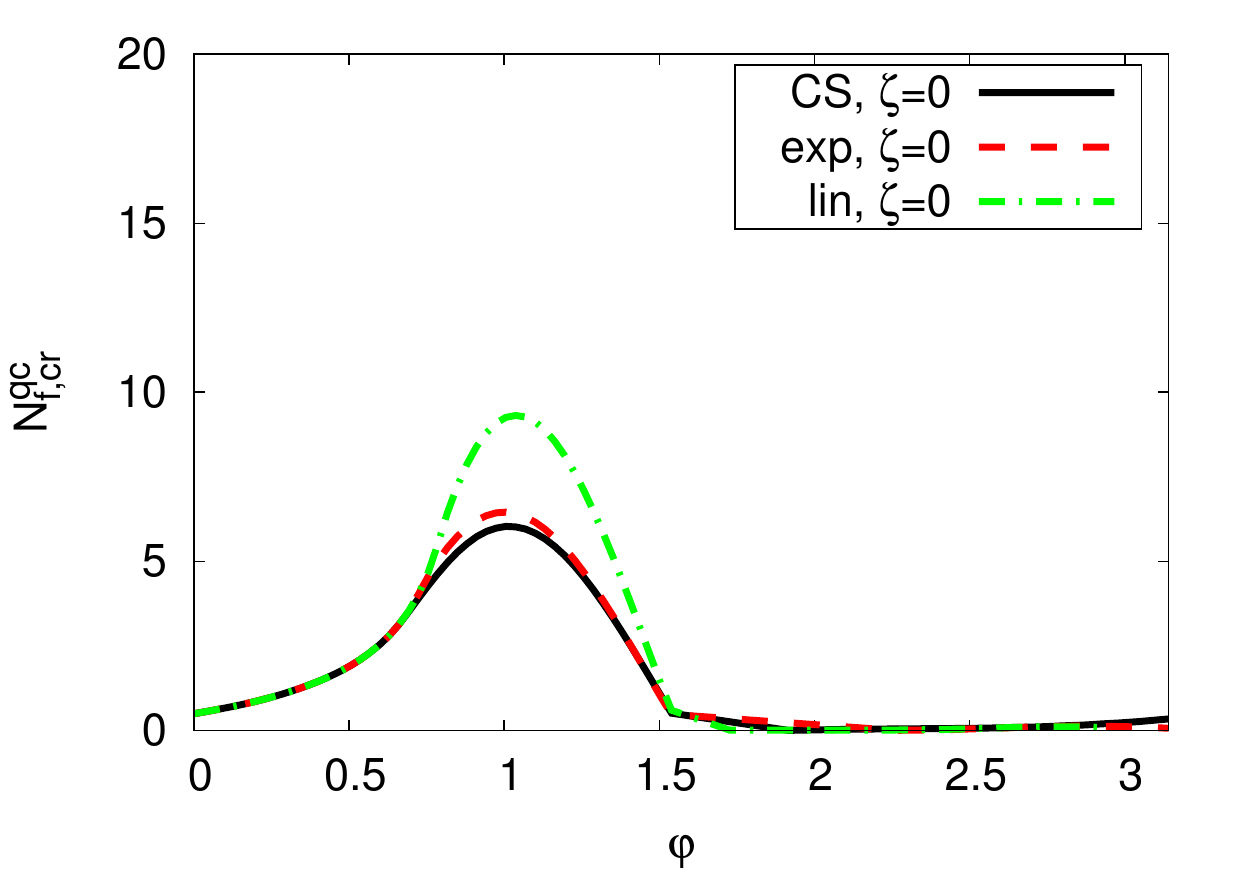}
\includegraphics[clip=true,width=\columnwidth]{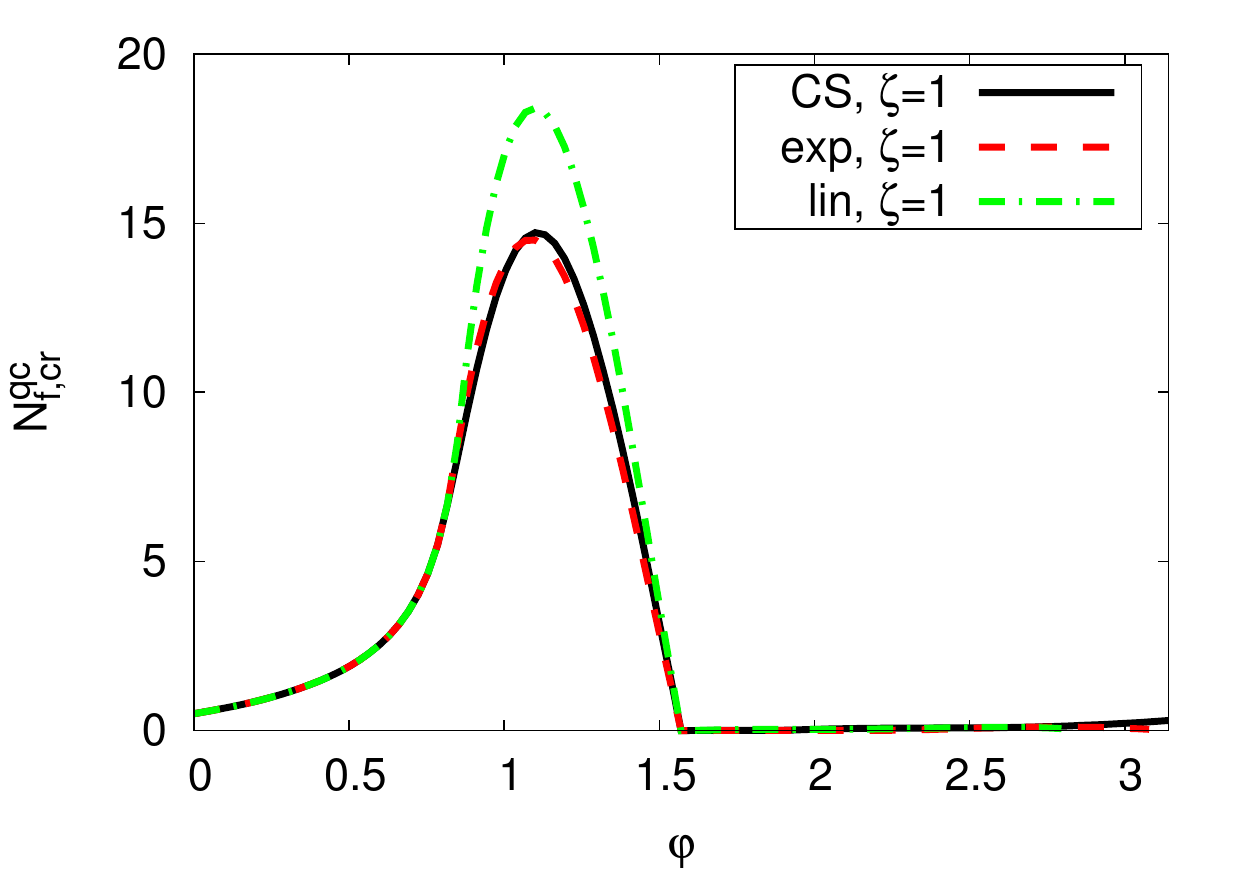}
\caption{\Confcrit flavor number~$\Nfpc$ as a function of the
  angle~$\varphi$ parametrizing an artificial
 Fierz incompleteness for~$\zeta = 0$ (top panel) and~$\zeta = 1$ (bottom panel).}
\label{fig:fi}
\end{figure}

With the $\varphi$-dependent Fierz-incomplete approximation at hand,
we can now compute the \confcrit flavor number again. In
Fig.~\ref{fig:fi}, we present our results for~$\Nfpc$ as a function of
the angle~$\varphi$ {for $\zeta=0$ (upper panel) and $\zeta=1$
  (lower panel)}. {We observe that the predictions for the conformal-critical
  flavor number strongly vary within this family of Fierz-incomplete
  approximations.}  Moreover, we find that a finite range of values
for~$\varphi$ exists for which we have~$\Nfpc=0$.  This was to be
expected, since for $\pi/2 \lesssim \varphi \lesssim \pi$ we project
onto a channel orthogonal to the chiral channel. There is no
annihilation of fixed points in this channel for any $\Nf$, since the
fixed points $\mathcal A$ and $\mathcal B$ do not approach the
Gau{\ss}ian fixed point $\mathcal O$ for any value of $e^2$, see
blue/bold lines in Fig.~\ref{fig:fpnf3}. This may be interpreted as a
consequence of the Vafa-Witten argument \cite{Vafa:1984xh},
prohibiting the spontaneous breaking of parity symmetry in QED$_3$. As
another specific example, let us consider a projection onto the chiral
channel corresponding to $\varphi = \pi/4$: here we find $\Nfpc
\approx 5$ even for all studied regulator functions and $\zeta$
values. However, this is still significantly different, for instance,
from the Fierz-complete result for $\zeta=1$.

Our analysis clearly demonstrates the necessity of a Fierz-complete
treatment as one may significantly overestimate {by almost a factor of 2} 
or underestimate~$({\Nfpc}=0)$ the {conformal-}critical flavor
number within a
Fierz-incomplete setup, see Fig.~\ref{fig:fi}. This strong
ambiguity of~${\Nfpc}$ within a Fierz-incomplete study represents the
second important result of our work.
{Moreover, any Fierz-incomplete study that is only sensitive to
  the chiral channel will inevitably identify $\Nfpc$ with
  $\Nfcr^\chi$. In this case, any information about a possibly
  existing intermediate phase will not be accessible because of
  Fierz incompleteness.}

While Fierz completeness is simple to implement in the present
approximation scheme of the exact RG flow, it is less obvious how this
issue might affect other methods. Mean-field methods are certainly
strongly affected, as the choice of a mean field immediately breaks
Fierz completeness \cite{Baier:2000yc}. 

By contrast, lattice simulations are by construction not affected, as
no choice of channels is required. Still, our results on Fierz
completeness can also be interpreted as a mandate to implement the
flavor symmetries exactly. Hence, lattice formulations should be given
preference that feature an exact (lattice version of) the U($2\Nf$)
flavor symmetry.

The largest body of literature on chiral-symmetry breaking in
QED${}_3$ relies on solutions of Dyson-Schwinger equations for the
photon and fermion propagators amended with suitable vertex
constructions. For the solution of the equation for the fermion
propagator $S_\psi(p)$, an ansatz of the following form is typically used,
\begin{equation}
S_\psi(p)^{-1} = i\slashed{p} A(p^2) + B(p^2), \label{eq:DSEansatzA}
\end{equation}
where $A(p^2)$ is related to the (inverse) wave-function
renormalization, and $B(p^2)$ parametrizes the mass function. In
particular, $\lim_{p\to 0} B(p^2) \neq 0$ signals fermion mass
generation and chiral symmetry breaking. This ansatz is also commonly
and successfully used for investigations of the strong-coupling regime of
QCD in $d=4$. 
Here, we note that the ansatz \eqref{eq:DSEansatzA} does not exhaust
all possible terms permitted by the special Dirac structure and flavor
symmetry of QED in $d=3$. As suggested by our results, the inclusion
of all terms permitted by the symmetries might be an essential
ingredient. On the level of the fermion propagator, a complete ansatz
would read
\begin{equation}
S_\psi(p)^{-1} = i\slashed{p} A(p^2) + B(p^2)+ \gamma_{45} C(p^2) + i\slashed{p} \gamma_{45} D(p^2), \label{eq:DSEansatzB}
\end{equation}
involving two further scalar functions $C$ and $D$. The case of
$\lim_{p\to 0} C(p^2) \neq 0$ would signal the generation of a
parity-breaking mass term. However, even in the parity-symmetric phase
where $\lim_{p\to 0} C(p^2) = 0$, the two further functions might
develop a nontrivial momentum dependence at intermediate scales,
potentially taking influence on the $B(p^2)$ function and thus on the
onset of chiral symmetry breaking.

Let us finally emphasize that there certainly is no one-to-one
correspondence between our results for Fierz-incomplete approximations
and flavor-symmetry-incomplete DSE ans\"{a}tze of the type of
\Eqref{eq:DSEansatzA}. It may well be that \Eqref{eq:DSEansatzA} is
perfectly sufficient to obtain quantitatively reliable results. Our
results, however, suggest that an ansatz of the type
\eqref{eq:DSEansatzB} exhausting the full symmetry 
could be worthwhile to be studied.

\section{Phase structure}
\label{sec:PS}

As our truncation based on pointlike fermion interaction channels is
not capable of entering the symmetry-broken regime, the scenario
developed in this section is founded only on limited
information which we can extract from the RG flow in the symmetric
regime. With these reservations in mind, we recall that we have
identified a \confcrit flavor number $\Nfpc$ above which
we found QED${}_3$ to be in the quasi-conformal phase. 

So far, we have carefully distinguished between $\Nfpc$ and a possible
critical flavor number $\Nfcr^{\chi}$, indicating the onset of a
chirally broken phase. From our results, we can primarily conclude
that $\Nfcr^{\chi}\leq\Nfpc$.
For a first attempt to estimate the possible value
of $\Nfcr^{\chi}$ within our approach, let us take a look at the RG flow
trajectories in
the plane of fermionic couplings for various flavor numbers below
$\Nfpc$. For illustrative purposes, we consider the flows obtained
with the linear regulator and a control parameter value $\zeta=1$,
which yielded the estimate $\Nfpc\simeq 10$. Also, we fix the gauge
coupling slightly above the critical value $e_{\text{cr}}^2$ where the
fixed points $\mathcal{O}$ and $\mathcal{C}$ annihilate,
$0< (e^2-e_{\text{cr}}^2) \ll 1$.

\begin{figure*}
\includegraphics[clip=true,width=\columnwidth]{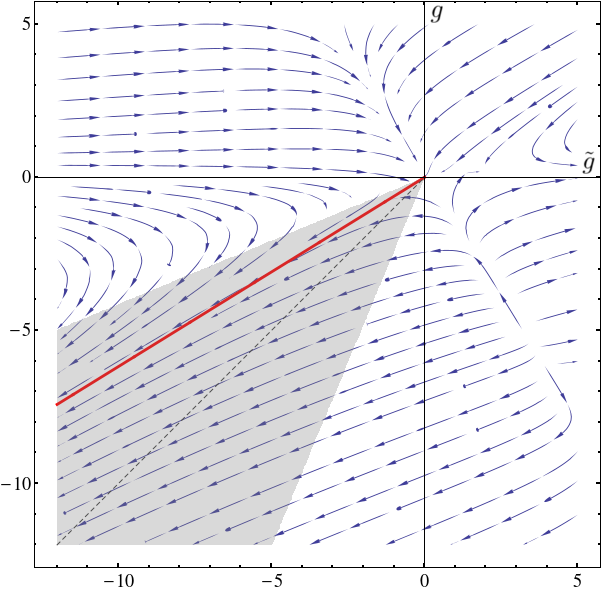}
\includegraphics[clip=true,width=\columnwidth]{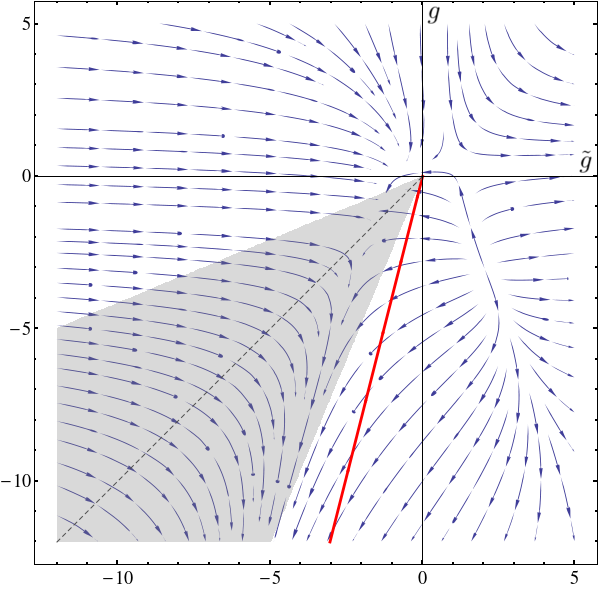}
\caption{RG flow of the four-fermion interactions in the plane spanned
  by the coulings~$\tilde g$ and~$g$ for $0<(e^2-e^2_{\rm cr})\ll 1$
  and~$\Nf=1$ (left panel) and~$\Nf=9$ (right panel), as obtained from
  the linear regulator function with~$\zeta=1$.  Recall
  that~$\Nfcr=10.0$ in this case.  The dashed line corresponds to the
  \chiral channel ($\tilde{g}=g$).  The solid red line
  represents the asymptotes of the RG trajectories. The gray-shaded
  area indicates a tentative measure for the chiral symmetry-breaking
  region, see main text for details.}
\label{fig:flowNf1Nf9}
\end{figure*}

The resulting fermionic flows in the $(\tilde g, g)$ plane are shown
in Fig.~\ref{fig:flowNf1Nf9} for the case of $\Nf=1$ (left panel) and
$\Nf=9$ (right panel). As before, the dashed line ($g=\tilde{g}$)
corresponds to the \chiral channel {$(S)^2$}, potentially
associated with chiral symmetry breaking
{when becoming dominant}. The solid red line marks the
direction of the asymptote of the RG trajectories for large
$\tilde g, g$. 
Starting the flow for vanishing fermionic interactions $\tilde g =
g = 0$, in general both $\tilde g$ and $g$ are generated and will approach this
asymptote in the course of the RG flow. The slope of the RG asymptote thus
determines the relative weight of the different possible channels in the IR.
For $\Nf = 1$ (left panel of Fig.~\ref{fig:flowNf1Nf9}), it is fairly close to
the dashed line associated with symmetry breaking in the chiral channel; in
fact, for $\Nf = 1.75$ (not shown)
the RG asymptote would lie exactly on top of the chiral channel. By 
contrast,
the $\Nf=9$ asymptote is closer to the pure vector channel $\sim
g {(V)^2}$. The fact that this asymptote rotates with increasing $\Nf$
towards
the vector channel is already known from studies of the Thirring model
\cite{Gies:2010st, Janssen:2012pq}. In fact, the depicted flows agree with those
of the Thirring model for asymptotically large $g$ and $\tilde{g}$, as we
have kept the gauge coupling at a fixed finite value. 
For any $\Nf < \Nfpc$, the RG asymptote in QED$_3$ thus coincides with
the Thirring-model asymptote within our approximation.

{On the basis of our pointlike fermionic truncation it is hard}
to judge which channel ultimately dominates as a
function of $\Nf$. This is because we do not have a metric in theory
(coupling) space available that could provide for a quantitative
measure of absolute distance from a certain channel. As a tentative
measure for the chiral symmetry-breaking region, we have depicted a
gray-shaded region between the angle bisectrix between the
\chiral axis and the $\tilde{g}$ axis and the one between the
\chiral axis and the $g$ axis.

For small $\Nf$ such as $\Nf=1$, the asymptote lies inside this region
where we expect chiral symmetry-breaking to occur,
cf. Fig.~\ref{fig:flowNf1Nf9} (left panel). For larger $\Nf$ such as
$\Nf=9$, the asymptote lies outside this region, cf.
Fig.~\ref{fig:flowNf1Nf9} (right panel). Taking this rough measure
seriously, we find that the asymptote of the four-fermion flows lies
within this suspected domain of attraction of the chiral channel for
$1\lesssim \Nf \lesssim 4$. 
{As a rough estimate,} {this suggests to} 
identify the maximal value of $\Nf$, for which the system is inside
this region with a dominant chiral channel, with the critical flavor
number for chiral-symmetry breaking $\Nfcr^\chi$. Independent of our
choice for the regulator function, we find the estimate
$\Nfcr^\chi\simeq 4$, which is in the ballpark of the most advanced
DSE studies~\cite{Fischer:2004nq,Bashir:2009aa,Bonnet:2011hh,Bonnet:2011ds,Bonnet:2012az}.

{For the linear regulator in the point-like limit $\zeta = 0$ and for all regulators with $\zeta = 1$, we find 
that the chiral-critical flavor number can in fact be smaller than the conformal-critical flavor number, $\Nfcr^\chi < \Nfpc$. }
This leaves us with the interesting conclusion that the many-flavor phase diagram of
QED${}_3$ could be more involved than
previously anticipated: in addition to the chiral symmetry-broken
phase for $\Nf<\Nfcr^\chi$ and the quasi-conformal phase for
$\Nf>\Nfpc$ there could be another phase in-between for
$\Nfcr^\chi<\Nf<\Nfpc$ characterized by different low-energy
properties.

At this point, it is instructive to compare our results
with those from the $3$d
Thirring model which shares with QED$_3$ both its
U($2\Nf$) chiral symmetry as well as
the corresponding possible symmetry-breaking patterns. In the Thirring
model, defined in terms of the non-Gau\ss{}ian {UV fixed point $\mathcal{C}$} (for
$e^2=0$), the long-range chiral properties in the pointlike language
are also determined by the competition between the chiral and the
vector channel. In \cite{Janssen:2012pq} the Thirring model was
studied in detail using dynamical bosonization techniques that allow
to enter the symmetry-broken regime and give direct access to the
order-parameter potentials, condensation phenomena and massive
excitations. The critical flavor number below which the system is in
the chiral symmetry broken phase was determined to be 
\begin{equation}
\Nfcr^{\chi, \text{Thirring}} \approx 5.1\,,
\label{eq:NfcrThirring}
\end{equation}
which is similar to our rough estimate for $\Nfcr^\chi$ for QED${}_3$
given above. In fact the mere quantitative difference between our
QED${}_3$ flows and those of the Thirring model within the same
approximation in the fermion sector are the gauge-coupling terms
{in the $\beta$ functions}. As
the approach to criticality is primarily indicated by diverging
four-fermion interactions, the following scenario is possible: if the
gauge contributions to the fermion self-interactions stay subdominant
for the approach to criticality, we conjecture that the critical
flavor number of QED${}_3$ and the $3$d Thirring model are identical.

For this conjecture to hold, the chiral critical flavor number of the
Thirring model must not lie in the quasi-conformal regime of
QED${}_3$. With our result for the \confcrit flavor number, $\Nfpc >
\Nfcr^{\chi, \text{Thirring}}$, this criterion appears to be satisfied
within our approximation for the linear regulator in the
  pointlike truncation with $\zeta=0$ and for all regulators with
  $\zeta=1$. Otherwise the QED${}_3$ system could still be trapped by
the IR attractive {fixed point $\mathcal{O}$} while the analogous
Thirring system would {already} be in the chirally broken phase,
such that the conjecture would fail. Whether the
gauge-contributions indeed stay subdominant during the approach to
criticality is a quantitative question that we cannot resolve within
our present simple truncation. For instance, using the simplified
$\beta$ function for the gauge coupling \eqref{eq:eFlowsimple}, the
gauge coupling {remains} bounded {by its fixed-point value, $e^2 \leq
  e^2_\ast$}, and the criterion is satisfied. In the more general
case, e.g., using \Eqref{eq:betaeq}, the situation is less clear and
requires a full numerical integration of the flow. Most likely a
definite answer requires a dynamically bosonized flow.  {However,}
even if the gauge contributions {do} not stay subdominant, it appears
plausible that the chiral-critical flavor numbers for QED${}_3$ and
the $3$d Thirring model {would still be} similar.

Let us now try to address the new possible phase in-between
$\Nfcr^\chi$ and $\Nfpc$, {assuming that $\Nfcr^\chi < \Nfpc$}. 
Again, the Thirring model may provide a
guideline: in \cite{Janssen:2012pq}, it was observed that for
$\Nf>\Nfcr^\chi$, the system 
not only is 
dominated by the vector
channel, but moreover the mass term of the vector channel $m_V^2$
approaches zero at a finite scale $k$. This indicates the possibility
of the appearance of a Lorentz symmetry {breaking}
condensate $\langle V_\mu
\rangle\neq 0$ for $\Nfcr^\chi<\Nf<\Nfpc$, going along with two 
massless Goldstone bosons and a massive ``radial'' mode.

\begin{figure}
\includegraphics[clip=true,width=\columnwidth]{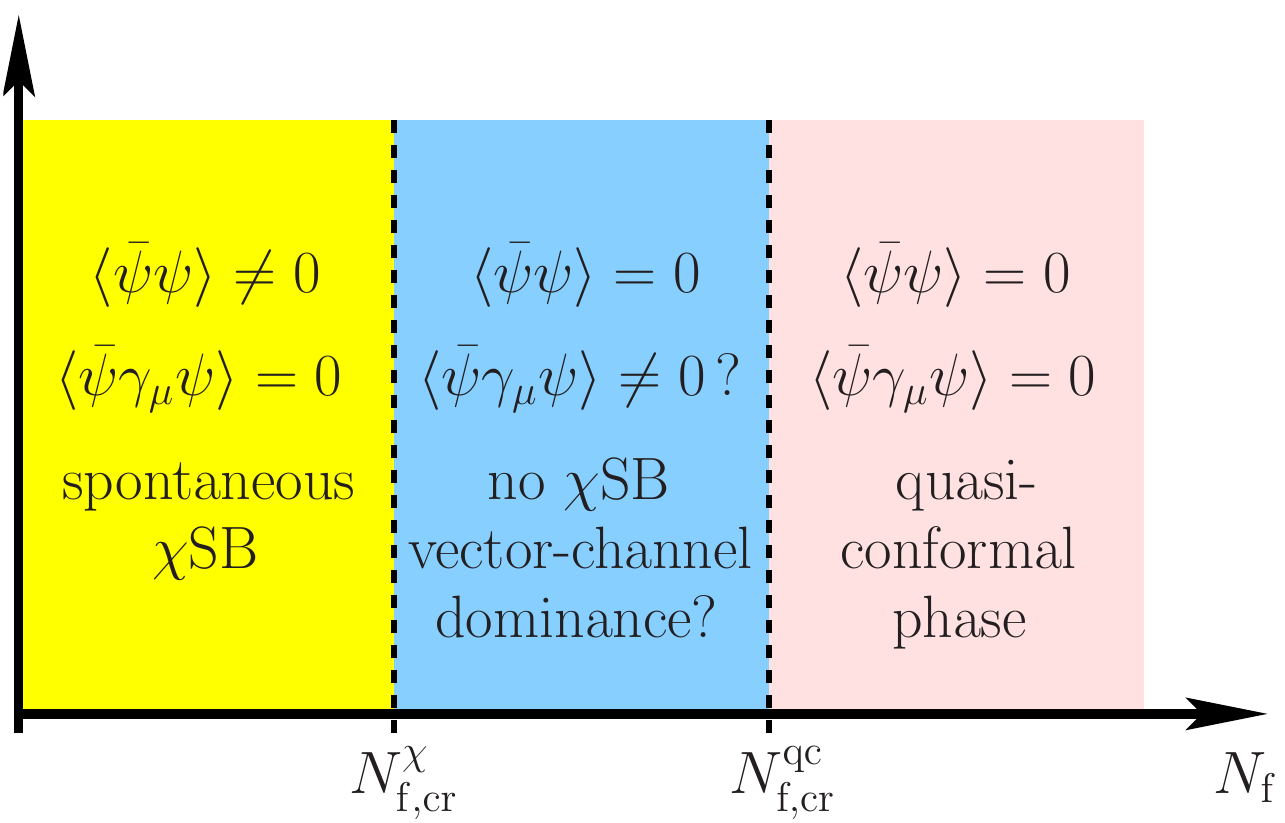}
\caption{Sketch of the conjectured many-flavor phase diagram of
  QED${}_3$. In addition to the phase governed by spontaneous chiral
  symmetry breaking ($\chi$SB) for small values of~$\Nf$, an
  intermediate phase driven by the vector-channel may exist, possibly
  exhibiting (spontaneous) breaking of Lorentz symmetry, see text for
  a discussion of the transition lines.}
\label{fig:cpd}
\end{figure}

These considerations suggest a many-flavor phase diagram of QED${}_3$
as schematically drawn in Fig.~\ref{fig:cpd} with a chirally broken
small-$\Nf$ phase, 
{possibly a} phase with {spontaneously} 
broken Lorentz symmetry at intermediate
$\Nf$, and a quasi-conformal massless phase at large $\Nf$ extending
to $\Nf\to\infty$. The nature of the phase transitions at $\Nfcr^\chi$
and $\Nfpc$ cannot be determined 
within our present approximation. 
For the Thirring model,
the dynamically bosonized study revealed that the
chiral phase transition at $\Nfcr^\chi$ is of second order
\cite{Janssen:2012pq}. In particular the chirally-broken and Lorentz-broken phases do not
overlap, but inhibit one another. This suggests the possibility of a second-order
phase transition at $\Nfcr^\chi$ also in QED${}_3$, if
the gauge coupling does not take a too strong influence on the
approach to criticality.

The nature of the transition at $\Nfpc$ is less {clear}. On the one
hand, the quasi-conformal mode vanishes because of the annihilation of
fixed points. This is similar to 
{Berezinsky-Kosterlitz-Thouless (BKT)-type}
phase transitions, such that
one might expect corresponding essential (or Miransky) scaling of
observables near {the phase transition}
\cite{Berezinskii,Berezinskii2,Kosterlitz:1973xp,Appelquist:1996dq,Chivukula:1996kg,Miransky:1988gk,Appelquist:1998xf,Kaplan:2009kr}
{with universal powerlaw corrections \cite{Braun:2010qs}, see
  also \cite{Braun:2005uj,Braun:2006jd,Bonnet:2012az}}.  On the other
hand, the spectra on {the two}
sides of the phase transition share some
similarities: 
{on both sides, the fermion and the photon fields are massless;}
there is a massive (but presumably unstable) vector excitation
on the quasi-conformal side, while there are a massive ``radial''
excitation and massless Goldstone bosons on the Lorentz
symmetry-breaking side. Near the transition at $\Nfpc$ all these
vector-like degrees of freedom can possibly mix nontrivially which
might influence the nature of the transition.

In order to check the scenario suggested above, it appears highly
worthwhile to search for vector condensates $\langle \bar\psi
\gamma_\mu \psi\rangle$ also with other nonperturbative methods in the
region above the chiral phase transition $\Nf\gtrsim \Nfcr^\chi$. If a
vector condensate {is}
found, our work suggests the existence of a further
transition to the quasi-conformal phase at $\Nfpc>\Nfcr^\chi$.
\section{Conclusions}
\label{sec:conc}
In the present work we have studied the many-flavor phase diagram of
QED${}_3$ by analyzing the RG fixed-point structure of the theory. In
addition to the asymptotically free Gau\ss{}ian fixed point, the fixed-point
structure of QED${}_3$ shares similarities with that of the
3-dimensional Thirring model which has the same global chiral/flavor
symmetries. 

For large flavor numbers $\Nf>\Nfpc$, the screening property of
fermionic fluctuations induces an IR attractive,
{quasi-conformal,} fixed point 
in the gauge 
{sector, which in the fermionic sector corresponds to a slightly shifted
Gau{\ss}ian fixed point,}
implying that the fermionic system remains
attracted by {this} fixed point. 
{For large $\Nf$,}  the system is 
in a quasi-conformal phase and remains massless in complete agreement
with expectations and literature results.
If this large-$\Nf$ phase described a condensed-matter system, the
existence of the quasi-conformal fixed point would indicate a so-called
algebraic-Fermi-liquid phase~\cite{Franz:2002qy}, with striking
consequences to the electronic, optical, and thermodynamic experimental
observables. Such a material would be one of the very rare examples above
$1+1$ dimensions and without disorder or magnetic field, which exhibit
genuine non-Fermi liquid behavior. If QED$_3$ is indeed an effective
theory for the superconductor-insulator transition in the cuprates, our result
of a large $\Nfpc > 2$, however, supports the scenario that cuprates at
$T=0$ are not in the quasi-conformal phase, and there is no
algebraic-Fermi-liquid behavior for any doping of the cuprates.

Lowering $\Nf$, the system approaches the lower end of the
``quasi-conformal window'' at $\Nfpc$ which is characterized by a
merger of the Gau\ss{}ian and the ``Thirring'' fixed point in the
fermionic interactions. This mechanism is similar to 
the one discovered in
4-dimensional many-flavor QCD~\cite{Gies:2005as,Braun:2005uj,Braun:2006jd},
{which gives} rise to
BKT-type scaling behavior
\cite{Miransky:1988gk,Kaplan:2009kr,Braun:2010qs}. As an important
difference, we observe {the possibility} in QED${}_3$ that the RG flow {can remain} 
dominated by the vector channel for $\Nf$ slightly below $\Nfpc$. Only for even
smaller $\Nf$, the chiral channel eventually takes over such that the
theory can definitely be expected to be in the chirally-broken phase
with massive fermions.

If these findings persist beyond the approximations underlying our analysis, the
phase diagram of QED${}_3$ along the many-flavor direction can exhibit
more phases than previously anticipated. In between the
chirally-broken phase for $\Nf<\Nfcr^\chi$ and the quasi-conformal
phase for $\Nf>\Nfpc$, there can exist a vector-channel dominated
phase provided that $\Nfcr^\chi<\Nfpc$. If the vector channel becomes
critical, this phase could be characterized by a Lorentz-breaking
vector condensate and a corresponding excitation spectrum with
photonlike Goldstone bosons as well as a massive radial-type mode.

From a technical perspective, we have discovered that a Fierz-complete set
of fermionic interactions is a mandatory ingredient for reliably
estimating quantities such as $\Nfpc$. Simple projections onto
seemingly physically relevant channels can imply a complete loss of
quantitative control. This result may inspire corresponding
improvements in other analytic approximation schemes used in the
literature. A similar word of caution applies to lattice approaches:
as Fierz completeness is a statement about the exact realization of
the U($2\Nf$) flavor symmetry of the model, a lattice formulation that
is not guaranteed to preserve the {full} continuum flavor symmetry
may simply simulate a different continuum model with possibly very
different values of $\Nfpc$. 
Indeed, a previous RG approach to such a QED$_3$ theory in the presence
of $\mathrm U(2\Nf)$-symmetry breaking interactions revealed that those
perturbatively irrelevant interactions may become relevant for strong gauge
coupling, significantly affecting the corresponding predictions for $\Nfpc$
\cite{Kaveh:2004qa}.
Also, while certainly tempting, it is thus premature to
speculate on possible consequences of the new vector-channel-dominated phase,
which we predict for $\Nfcr^\chi < \Nf < \Nfpc$, on the cuprate phase diagram:
Even if this new phase
reached all the way down to the physical flavor number $\Nf = 2$ (i.e., if
$\Nfcr^\chi$ was smaller than $2$, in contrast to most
of the previous findings, and also to our estimate), the actual cuprate
system does not have the full $\mathrm U(2\Nf)$ symmetry and it is momentarily
unclear how the presence of the symmetry-breaking short-range interactions will
affect the many-flavor phase diagram in QED$_3$ and the existence of the
vector-channel-dominated intermediate phase. This deserves further
investigation.

From a quantitative viewpoint, our result for $\Nfpc$ is still rather
strongly affected by artificial regularization-scheme
dependencies. This may hint to the insufficient resolution of momentum
dependencies of the vertices which in our work is only 
{estimated} by an
overall RG scale. We consider \Eqref{eq:NfpcEst} to represent our best
estimate: $\Nfpc \approx 4.1 \dots 10.0$.

For the chiral-critical flavor number, our results are compatible with
those of the most advanced DSE studies, suggesting $\Nfcr^\chi\simeq
4$. Hence, the window of theories in the vector-channel-dominance
phase could be finite and include theories with integer~$\Nf$.

However, under the assumption that the gauge
contributions to the approach to criticality stay subdominant, we
conjecture the chiral-critical flavor number of QED${}_3$ and the $3$d
Thirring model to be identical.
A recent study of the $3$d Thirring model suggests
that $\Nfcr^{\chi,\text{Thirring}} \approx 5.1$, see Ref.~\cite{Janssen:2012pq}. 
In the light of our QED$_3$-Thirring conjecture and the approximation
involved in our computation, we can therefore not exclude the possibility
that $\Nfcr^\chi$ and $\Nfpc$ are so close to each other that the
vector-dominance phase does not include a system with integer~$\Nf$.  
While it is certainly not inconceivable that $\Nfcr^\chi$ and $\Nfpc$ are
in fact identical, we see no natural reason for this coincidence to hold.  
Of course, a verification and exact determination of the phase boundaries
of the many-flavor phase diagram requires more elaborate studies in
the future, ideally by using
various different theoretical approaches. In
any case, the present work points to a so far overlooked new
intermediate phase and may therefore help to better our
understanding of the dynamics underlying low-dimensional fermionic
field theories and the corresponding strongly-correlated condensed-matter
systems.

\acknowledgments{The authors thank J.\ Berges, C.\ S.\ Fischer, and
  I.\ F.\ Herbut for useful discussions.  JB and DR acknowledge
  support by the Deutsche Forschungsgemeinschaft (DFG) under grant BR 4005/2-1 and by HIC for FAIR
  within the LOEWE program of the State of Hesse. Moreover, JB acknowledges 
  support by the DFG under grant SFB 634. HG acknowledges
  support by the DFG under grants Gi~328/5-2 (Heisenberg program), and
  Gi 328/6-2 (FOR 723). LJ is supported by the DFG under grants
  JA~2306/1-1, FOR~723, and GRK~1523.}

\appendix
\begin{table}[t]
\centering
\begin{tabularx}{\columnwidth}{@{\extracolsep{\fill}} c|c|cccc}
\hline\hline
 & & $R_{\rm{CS}}$ & $R_{\rm{exp}}$ & $R_{\rm{lin}}$ & $R_{\rm{SC}}$\\
\hline
\multirow{2}{*}{$l_1^{\rm (F)}$} & $\mathcal{N}$ & $\frac{\pi}{2}$ & $\frac{\sqrt{\pi}}{2}$ & $\frac{2}{3}$ & 1 \\
 & $\sim\eta_\psi$ & -0.858407 & -0.306377 & $-\frac{1}{6}$ & -- \\
\hline
\multirow{3}{*}{$l^{\rm (F,B)}_{1,1}$ } & $\mathcal{N}$ & $\frac{\pi}{4}$ & 1.03828 & $\frac{4}{3}$ & 1 \\
 & $\sim\eta_\psi$ & -0.237463 & -0.208436 & $-\frac{1}{6}$ & -- \\
 & $\sim\eta_A$ & $-\frac{\pi}{16}$ & -0.170823 & $-\frac{2}{15}$ & -- \\
\hline
\multirow{3}{*}{$l^{\rm (F,B)}_{2,1}$} & $\mathcal{N}$ & $\frac{3\pi}{16}$ & 1.02494 & 2 & 1 \\
 & $\sim\eta_\psi$ & -0.126032 & -0.153062 & $-\frac{1}{6}$ & -- \\
 & $\sim\eta_A$ & $-\frac{\pi}{16}$ & -0.243833 & $-\frac{4}{15}$ & -- \\
\hline
\multirow{3}{*}{$m^{\rm (F,B)}_{2,1}$} & $\mathcal{N}$ & $\frac{2}{3}$ & 0.821746 & 1 & $\frac{2}{3}$ \\
 & $\sim\eta_\psi$ & -0.077618 & -0.043037 & 0 & -- \\
 & $\sim\eta_A$ & $-\frac{4}{15}$ & -0.26131 & $-\frac{1}{4}$ & -- \\
\hline
\multirow{3}{*}{$\tilde{m}^{\rm (F,B)}_{1,1}$} & $\mathcal{N}$ & 1 & 1.23262 & $\frac{3}{2}$ & $1^{\ast}$ \\
 & $\sim\eta_\psi$ & -0.214602 & -0.19434 & $-\frac{1}{6}$  & -- \\
 & $\sim\eta_A$ & -$\frac{1}{3}$ & -0.298558 & $-\frac{1}{4}$ & -- \\
\hline\hline
\end{tabularx}
\caption{Numerical values for the threshold functions as obtained from
  the various regulators employed in this work and listed in
  App.~\ref{app:reg}. Depending on the type of internal lines in the
  1PI diagram underlying the different threshold functions, these
  functions can be written as sum of three terms: a pure (real-valued)
  number~(${\mathcal N}$), a number times $\eta_{\psi}$ (2nd row), and
  a number times~$\eta_{\rm A}$ (3rd row). Values with an
  asterisk ${}^{\ast}$ depend on the details of the definition of
  the non-analytic sharp cutoff.
}
\label{tab:reg}
\end{table}

\section{Irreducible representation}\label{app:irr}

Though the reducible representation using 4-component Dirac spinors $\psi^a$,
$a=1, \dots,\Nf$ has its merits from the viewpoint of applications in
condensed-matter systems, some aspects become more transparent in the
irreducible representation using 2-component spinors $\chi^i$, $i=1, \dots,
2\Nf$. In our conventions, the transition between these representations can
be defined using the projector
\begin{equation}
P_\mathrm{L,R}^{(45)} = \frac{1}{2} (1\pm \gamma_{45}).
\label{eq:P45}
\end{equation}
Decomposing $\chi^i$ into $(\chi^a, \chi^{a+\Nf})$, for $a=1,\dots,\Nf$, we
introduce the $\chi$ subcomponents by
\begin{equation}
P_\mathrm{L}\psi^a= \frac{1}{\sqrt{2}} \chi^a \otimes 
\left(\begin{array}{c} 1 \\ i  \end{array}\right), \bar\psi^a P_\mathrm{L} =
\frac{1}{\sqrt{2}} \bar{\chi}^a \otimes (1,-i),
\end{equation}
and
\begin{equation}
P_\mathrm{R}\psi^a= \frac{1}{\sqrt{2}} \chi^{a+\Nf} \otimes 
\left(\begin{array}{c} 1 \\ -i  \end{array}\right), \bar\psi^a P_\mathrm{R} =
-\frac{1}{\sqrt{2}} \bar{\chi}^{a+\Nf} \otimes (1,i).
\end{equation}
In {the irreducible representation}, the enhanced U($2\Nf$) symmetry of
QED${}_3$ becomes obvious, since
\begin{equation}
\bar\psi^a \gamma_\mu \psi^a = \bar\chi^i \sigma_\mu \chi^i, \quad
i=1,\dots,2\Nf,
\end{equation}
and $\sigma_\mu$ denote the Pauli matrices. Similarly, it is straightforward
to show that $\bar\psi^a\psi^a= \bar\chi^a \chi^a- \bar\chi^{a+\Nf}
\chi^{a+\Nf}$ and $\bar\psi^a \gamma_{45} \psi^a= \bar\chi^i \chi^i$.
The latter implies that a mass term of the form $i\tilde{m}\bar\psi^a \gamma_{45}
\psi^a$ actually preserves the U$(2\Nf)$ symmetry. Also, the interaction term
$(P)^2$ introduced in the main text in \Eqref{eq:SingChan} {in this
notation} indeed {becomes the standard}
Gross-Neveu interaction for two-component spinors.

In the same spirit the nonsinglet interaction channel $(S)^2$ as used in
\Eqref{eq:LPsiInt2} can be shown to read
\begin{equation}
(S)^2 =2 (\bar\chi^i\chi^j)^2 \equiv 2 \bar\chi^i \chi^j \bar\chi^j \chi^i,
\end{equation}
where the factor of two on the right-hand side motivates the
different coupling normalization between the $(V)^2$ and the $(S)^2$ term in
\Eqref{eq:LPsiInt2}.

\section{Regulator functions}\label{app:reg}
In this {appendix}, we summarize the regulator functions
employed in the present work. For the definition of the regulator functions, it
is convenient to introduce
so-called regulator shape functions~$r_{\rm F,B}$ for the fermions (F) and bosons (B), respectively:
\be
R_{\rm F}(p)=-\slashed{p}r_{\rm F}(y)\quad\text{and}\quad
R_{\rm B}(p^2)=p^2 r_{\rm B}(y)\,,
\ee
where $y=p^2/k^2$. Overall, we have used four different regulator functions, namely the Callan-Symanzik regulator $R_{\rm CS}$ with
\be
r_{\rm F}(y)= \sqrt{\frac{y+1}{y}}-1\,,\quad r_{\rm B}(y) = \frac{1}{y}\,,
\ee
the exponential regulator~$R_{\rm exp}$ with
\be
r_{\rm F}(y) = \frac{1}{\sqrt{1-e^{-y}}}-1\,,\quad r_{\rm B}(y) = \frac{1}{e^y-1}\,,
\ee
the linear regulator~$R_{\rm lin}$, see Refs.~\cite{Litim:2000ci,Litim:2001up,Litim:2001fd}, with
\be
&& r_{\rm F}(y) = \left(\frac{1}{\sqrt{y}}-1\right)\theta(1-y)\,,\\
&& \qquad r_{\rm B}(y) = \left(\frac{1}{y}-1\right)\theta(1-y)\,,
\ee
and the so-called sharp-cutoff regulator with
\be
r_{\rm F}(y)= \lim\limits_{b\rightarrow\infty} \sqrt{1+\frac{1}{y^b}}-1\,,\quad r_{\rm B}(y) = \lim\limits_{b\rightarrow\infty} \frac{1}{y^b}\,.
\label{eq:defscreg}
\ee
Note that the sharp-cutoff regulator has to be handled with
care as it requires a definite prescription of the order of the various
limiting
processes involved, in order to avoid ambiguities in the evaluation of the
loop integrals.
In particular, this is the case for the threshold function~$\tilde{m}_{1,1}^{\rm (F,B)}$, 
cf.\ also 
the RG equations in Ref.~\cite{Kaveh:2004qa}. These artifacts
of the sharp-cutoff scheme are well known, see, e.g.,
the discussion of the BKT-phase transition in \cite[Chapter
6.4]{herbut2007modern}.
In Tab.~\ref{tab:reg}, we list the numerical values for the threshold functions
as obtained from the various employed regulators.

\section{RG flow of $Z_A$}\label{app:za}
We briefly summarize the derivation of the equation for the anomalous dimension of the photon, $\eta_{A}=-\partial_t \ln Z_{A}$. We begin
by rewriting the Wetterich equation~\eqref{eq:wetterich} as follows:
\be
\partial _t\,\Gamma _k= \frac{1}{2}\,\STr \, \tilde{\partial}_t \ln \left( \Gamma ^{(2)} _k  + R_k \right)\,,\label{eq:trlog}
\ee
where $\tilde{\partial}_{t}$ denotes a formal derivative acting only on the 
{of the regulator function $R_k$}. The representation~\eqref{eq:trlog} of the Wetterich equation is a convenient starting point
for the computation of both the fermionic RG flows (see, e.g., Ref.~\cite{Braun:2011pp} for a detailed introduction) as well as for the anomalous dimensions.
In order to calculate the flow equation for~$Z_{A}$, 
we decompose the inverse regularized propagator $\Gamma ^{(2)} _k$ on the right-hand side of the flow equation into 
a field-independent (${\mathcal P}_k$) and a field-dependent (${\mathcal F}_k$) part,
\be
\Gamma _{k} ^{(2)}+R_k ={\mathcal P}_k + {\mathcal F}_k\,.
\ee
The flow equation can then be decomposed in powers of the fields:
\be
\label{eq:flowexp}
\partial_{t}\Gamma_{k}
=\frac{1}{2}\STr\bigg\{\tilde{\partial}_{t}\sum_{n=1}^{\infty}\frac{(-1)^{n+1}}{n} \left({{\mathcal{P}}_{k}^{\,-1}}\mathcal{F}_{k}\right)^n\bigg\}\,.
\ee
On the right-hand side we have dropped a field-independent term which is of no relevance for our present study.
The powers of ${{\mathcal{P}}_{k}^{\,-1}}\mathcal{F}_{k}$ can be {calculated by straightforward matrix multiplications.} It is then straightforward to 
project the various terms from the expansion
appearing on the right-hand side of Eq.~\eqref{eq:flowexp} onto our ansatz for the effective action. To the flow of~$Z_{A}$ only the second term
of the expansion contributes and we find 
\begin{widetext}
\be
\eta_A &=&
-\frac{1}{2Z_A}
\left\{ \frac{P^{\rm{T}}_{\mu\nu}(p)}{p^2} 
\left(
\int\frac{d^3q}{(2\pi)^3}
\lfdif{}{A_\mu(-p)}\frac{1}{2}\mbox{STr}\left[\tilde{\partial}_t\frac{
(-1)}{2}\left(\mathcal{P}_k^{-1}\mathcal{F}_k\right)^2\right]\rfdif{}{
A_\nu(q)}
\Bigg|_{\bar{\psi}=\psi=0,\,A_\mu=0}\right.\right.\nn\\ 
 &&\left.\left.\qquad\quad\; 
 - 
\int\frac{d^3q}{(2\pi)^3}
\lfdif{}{A_\mu(-p')}\frac{1}{2}\mbox{STr}\left[\tilde{
\partial}_t\frac{(-1)}{2}\left(\mathcal{P}_k^{-1}\mathcal{F}_k\right)^2\right]
\rfdif{}{A_\nu(q)}\Bigg|_{\bar{\psi}=\psi=0,\,
A_\mu=0,\, p'=0}
\right)\,\right\}_{p^2=\zeta^2 k^2},
\label{eq:C4}
 \ee
\end{widetext}
where we have used the transversal projector $P^\mathrm{T}_{\mu\nu}(p)
= \delta_{\mu\nu} - \frac{p_\mu p_\nu}{p^2}$. The second term
  corresponds to the subtraction of the zero-momentum limit of the 
  regularized flow which is constrained by the regulator-modified Ward
  identity. In this way, the transversal projection entering the
  definition of $\eta_A$ satisfies the standard Ward identity at all
  scales. This construction is based on the implicit assumption that
  the longitudinal and the transversal part of the photon propagator
  do not differ by non-analyticities at small momenta.
 From this expression, we then obtain
\be
\eta_A &=& 8 v_3 N_{\rm{f}} e^2 {\mathcal L}^{\rm (F)}_1\,,
\ee
where~$v_3=1/(8\pi)^2$ and 
\begin{widetext}
\be
{\mathcal L}^{(F)}_1 (\eta_{\psi};\zeta)\equiv {\mathcal L}^{(F)}_1&=&
\frac{1}{\zeta^2}
\int\limits_{0}^\infty\dif{y}\left\{\frac{2}{3}\frac{\partial_t
r_\psi(y)-\eta_\psi r_\psi(y)}{\sqrt{y}[1+r_\psi(y)]^3}-\right.
\frac{1}{2}\int\limits_{-1}^1\dif{x}\frac{\sqrt{y}x^2-\zeta x}{y-2\zeta
x\sqrt{y}+\zeta^2}\left[\frac{[\partial_t r_\psi](y)-\eta_\psi
r_\psi(y)}{[1+r_\psi(y)]^2[1+r_\psi(y-2\zeta x\sqrt{y}+\zeta^2)]}\right.\nn\\
&&\qquad +\left.\left.\frac{[\partial_t r_\psi](y-2\zeta x\sqrt{y}+ \zeta^2)-\eta_\psi r_\psi(y-2\zeta x\sqrt{y}+\zeta^2)}{[1+r_\psi(y)][1+r_\psi(y-2\zeta x\sqrt{y}+\zeta^2)]^2}\right]\right\}. \label{eq:AppL}
\ee
\end{widetext}
Here, we have introduced~$y=q^2/k^2$ for convenience and~$x=\cos\vartheta$. 

\bibliographystyle{apsrev4-1}
\bibliography{bibliography}

\begin{thebibliography}{119}%
\makeatletter
\providecommand \@ifxundefined [1]{%
 \@ifx{#1\undefined}
}%
\providecommand \@ifnum [1]{%
 \ifnum #1\expandafter \@firstoftwo
 \else \expandafter \@secondoftwo
 \fi
}%
\providecommand \@ifx [1]{%
 \ifx #1\expandafter \@firstoftwo
 \else \expandafter \@secondoftwo
 \fi
}%
\providecommand \natexlab [1]{#1}%
\providecommand \enquote  [1]{``#1''}%
\providecommand \bibnamefont  [1]{#1}%
\providecommand \bibfnamefont [1]{#1}%
\providecommand \citenamefont [1]{#1}%
\providecommand \href@noop [0]{\@secondoftwo}%
\providecommand \href [0]{\begingroup \@sanitize@url \@href}%
\providecommand \@href[1]{\@@startlink{#1}\@@href}%
\providecommand \@@href[1]{\endgroup#1\@@endlink}%
\providecommand \@sanitize@url [0]{\catcode `\\12\catcode `\$12\catcode
  `\&12\catcode `\#12\catcode `\^12\catcode `\_12\catcode `\%12\relax}%
\providecommand \@@startlink[1]{}%
\providecommand \@@endlink[0]{}%
\providecommand \url  [0]{\begingroup\@sanitize@url \@url }%
\providecommand \@url [1]{\endgroup\@href {#1}{\urlprefix }}%
\providecommand \urlprefix  [0]{URL }%
\providecommand \Eprint [0]{\href }%
\providecommand \doibase [0]{http://dx.doi.org/}%
\providecommand \selectlanguage [0]{\@gobble}%
\providecommand \bibinfo  [0]{\@secondoftwo}%
\providecommand \bibfield  [0]{\@secondoftwo}%
\providecommand \translation [1]{[#1]}%
\providecommand \BibitemOpen [0]{}%
\providecommand \bibitemStop [0]{}%
\providecommand \bibitemNoStop [0]{.\EOS\space}%
\providecommand \EOS [0]{\spacefactor3000\relax}%
\providecommand \BibitemShut  [1]{\csname bibitem#1\endcsname}%
\let\auto@bib@innerbib\@empty
\bibitem [{\citenamefont {Weinberg}(1979)}]{Weinberg:1979bn}%
  \BibitemOpen
  \bibfield  {author} {\bibinfo {author} {\bibfnamefont {S.}~\bibnamefont
  {Weinberg}},\ }\href {\doibase 10.1103/PhysRevD.19.1277} {\bibfield
  {journal} {\bibinfo  {journal} {Phys. Rev.}\ }\textbf {\bibinfo {volume}
  {D19}},\ \bibinfo {pages} {1277} (\bibinfo {year} {1979})}\BibitemShut
  {NoStop}%
\bibitem [{\citenamefont {Holdom}(1981)}]{Holdom:1981rm}%
  \BibitemOpen
  \bibfield  {author} {\bibinfo {author} {\bibfnamefont {B.}~\bibnamefont
  {Holdom}},\ }\href {\doibase 10.1103/PhysRevD.24.1441} {\bibfield  {journal}
  {\bibinfo  {journal} {Phys. Rev.}\ }\textbf {\bibinfo {volume} {D24}},\
  \bibinfo {pages} {1441} (\bibinfo {year} {1981})}\BibitemShut {NoStop}%
\bibitem [{\citenamefont {Hong}\ \emph {et~al.}(2004)\citenamefont {Hong},
  \citenamefont {Hsu},\ and\ \citenamefont {Sannino}}]{Hong:2004td}%
  \BibitemOpen
  \bibfield  {author} {\bibinfo {author} {\bibfnamefont {D.~K.}\ \bibnamefont
  {Hong}}, \bibinfo {author} {\bibfnamefont {S.~D.~H.}\ \bibnamefont {Hsu}}, \
  and\ \bibinfo {author} {\bibfnamefont {F.}~\bibnamefont {Sannino}},\ }\href
  {\doibase 10.1016/j.physletb.2004.07.007} {\bibfield  {journal} {\bibinfo
  {journal} {Phys. Lett.}\ }\textbf {\bibinfo {volume} {B597}},\ \bibinfo
  {pages} {89} (\bibinfo {year} {2004})},\ \Eprint
  {http://arxiv.org/abs/hep-ph/0406200} {arXiv:hep-ph/0406200} \BibitemShut
  {NoStop}%
\bibitem [{\citenamefont {Sannino}\ and\ \citenamefont
  {Tuominen}(2005)}]{Sannino:2004qp}%
  \BibitemOpen
  \bibfield  {author} {\bibinfo {author} {\bibfnamefont {F.}~\bibnamefont
  {Sannino}}\ and\ \bibinfo {author} {\bibfnamefont {K.}~\bibnamefont
  {Tuominen}},\ }\href {\doibase 10.1103/PhysRevD.71.051901} {\bibfield
  {journal} {\bibinfo  {journal} {Phys. Rev.}\ }\textbf {\bibinfo {volume}
  {D71}},\ \bibinfo {pages} {051901} (\bibinfo {year} {2005})},\ \Eprint
  {http://arxiv.org/abs/hep-ph/0405209} {arXiv:hep-ph/0405209} \BibitemShut
  {NoStop}%
\bibitem [{\citenamefont {Dietrich}\ \emph {et~al.}(2005)\citenamefont
  {Dietrich}, \citenamefont {Sannino},\ and\ \citenamefont
  {Tuominen}}]{Dietrich:2005jn}%
  \BibitemOpen
  \bibfield  {author} {\bibinfo {author} {\bibfnamefont {D.~D.}\ \bibnamefont
  {Dietrich}}, \bibinfo {author} {\bibfnamefont {F.}~\bibnamefont {Sannino}}, \
  and\ \bibinfo {author} {\bibfnamefont {K.}~\bibnamefont {Tuominen}},\ }\href
  {\doibase 10.1103/PhysRevD.72.055001} {\bibfield  {journal} {\bibinfo
  {journal} {Phys. Rev.}\ }\textbf {\bibinfo {volume} {D72}},\ \bibinfo {pages}
  {055001} (\bibinfo {year} {2005})},\ \Eprint
  {http://arxiv.org/abs/hep-ph/0505059} {arXiv:hep-ph/0505059} \BibitemShut
  {NoStop}%
\bibitem [{\citenamefont {Dietrich}\ and\ \citenamefont
  {Sannino}(2007)}]{Dietrich:2006cm}%
  \BibitemOpen
  \bibfield  {author} {\bibinfo {author} {\bibfnamefont {D.~D.}\ \bibnamefont
  {Dietrich}}\ and\ \bibinfo {author} {\bibfnamefont {F.}~\bibnamefont
  {Sannino}},\ }\href {\doibase 10.1103/PhysRevD.75.085018} {\bibfield
  {journal} {\bibinfo  {journal} {Phys. Rev.}\ }\textbf {\bibinfo {volume}
  {D75}},\ \bibinfo {pages} {085018} (\bibinfo {year} {2007})},\ \Eprint
  {http://arxiv.org/abs/hep-ph/0611341} {arXiv:hep-ph/0611341} \BibitemShut
  {NoStop}%
\bibitem [{\citenamefont {Sannino}(2009)}]{Sannino:2009za}%
  \BibitemOpen
  \bibfield  {author} {\bibinfo {author} {\bibfnamefont {F.}~\bibnamefont
  {Sannino}},\ }\href@noop {} {\bibfield  {journal} {\bibinfo  {journal} {Acta
  Phys. Polon.}\ }\textbf {\bibinfo {volume} {B40}},\ \bibinfo {pages} {3533}
  (\bibinfo {year} {2009})},\ \Eprint {http://arxiv.org/abs/0911.0931}
  {arXiv:0911.0931 [hep-ph]} \BibitemShut {NoStop}%
\bibitem [{\citenamefont {Cortijo}\ \emph {et~al.}(2012)\citenamefont
  {Cortijo}, \citenamefont {Guinea},\ and\ \citenamefont
  {Vozmediano}}]{Cortijo:2011aa}%
  \BibitemOpen
  \bibfield  {author} {\bibinfo {author} {\bibfnamefont {A.}~\bibnamefont
  {Cortijo}}, \bibinfo {author} {\bibfnamefont {F.}~\bibnamefont {Guinea}}, \
  and\ \bibinfo {author} {\bibfnamefont {M.~A.}\ \bibnamefont {Vozmediano}},\
  }\href {\doibase 10.1088/1751-8113/45/38/383001} {\bibfield  {journal}
  {\bibinfo  {journal} {J.Phys.}\ }\textbf {\bibinfo {volume} {A45}},\ \bibinfo
  {pages} {383001} (\bibinfo {year} {2012})},\ \Eprint
  {http://arxiv.org/abs/1112.2054} {arXiv:1112.2054 [cond-mat.mes-hall]}
  \BibitemShut {NoStop}%
\bibitem [{\citenamefont {Vafek}\ and\ \citenamefont
  {Vishwanath}(2014)}]{Vafek:2013mpa}%
  \BibitemOpen
  \bibfield  {author} {\bibinfo {author} {\bibfnamefont {O.}~\bibnamefont
  {Vafek}}\ and\ \bibinfo {author} {\bibfnamefont {A.}~\bibnamefont
  {Vishwanath}},\ }\href {\doibase 10.1146/annurev-conmatphys-031113-133841}
  {\bibfield  {journal} {\bibinfo  {journal} {Ann.\ Rev.\ Cond.\ Mat.\ Phys.}\
  }\textbf {\bibinfo {volume} {5}},\ \bibinfo {pages} {83} (\bibinfo {year}
  {2014})},\ \Eprint {http://arxiv.org/abs/1306.2272} {arXiv:1306.2272
  [cond-mat.mes-hall]} \BibitemShut {NoStop}%
\bibitem [{\citenamefont {Franz}\ and\ \citenamefont
  {Tesanovic}(2001)}]{Franz:2001zz}%
  \BibitemOpen
  \bibfield  {author} {\bibinfo {author} {\bibfnamefont {M.}~\bibnamefont
  {Franz}}\ and\ \bibinfo {author} {\bibfnamefont {Z.}~\bibnamefont
  {Tesanovic}},\ }\href {\doibase 10.1103/PhysRevLett.87.257003} {\bibfield
  {journal} {\bibinfo  {journal} {Phys.Rev.Lett.}\ }\textbf {\bibinfo {volume}
  {87}},\ \bibinfo {pages} {257003} (\bibinfo {year} {2001})}\BibitemShut
  {NoStop}%
\bibitem [{\citenamefont {Franz}\ \emph {et~al.}(2002)\citenamefont {Franz},
  \citenamefont {Tesanovic},\ and\ \citenamefont {Vafek}}]{Franz:2002qy}%
  \BibitemOpen
  \bibfield  {author} {\bibinfo {author} {\bibfnamefont {M.}~\bibnamefont
  {Franz}}, \bibinfo {author} {\bibfnamefont {Z.}~\bibnamefont {Tesanovic}}, \
  and\ \bibinfo {author} {\bibfnamefont {O.}~\bibnamefont {Vafek}},\ }\href
  {\doibase 10.1103/PhysRevB.66.054535} {\bibfield  {journal} {\bibinfo
  {journal} {Phys.Rev.}\ }\textbf {\bibinfo {volume} {B66}},\ \bibinfo {pages}
  {054535} (\bibinfo {year} {2002})},\ \Eprint
  {http://arxiv.org/abs/cond-mat/0203333} {arXiv:cond-mat/0203333 [cond-mat]}
  \BibitemShut {NoStop}%
\bibitem [{\citenamefont {Herbut}(2002{\natexlab{a}})}]{Herbut:2002wd}%
  \BibitemOpen
  \bibfield  {author} {\bibinfo {author} {\bibfnamefont {I.~F.}\ \bibnamefont
  {Herbut}},\ }\href {\doibase 10.1103/PhysRevLett.88.047006} {\bibfield
  {journal} {\bibinfo  {journal} {Phys.Rev.Lett.}\ }\textbf {\bibinfo {volume}
  {88}},\ \bibinfo {pages} {047006} (\bibinfo {year} {2002}{\natexlab{a}})},\
  \Eprint {http://arxiv.org/abs/cond-mat/0110188} {arXiv:cond-mat/0110188
  [cond-mat]} \BibitemShut {NoStop}%
\bibitem [{\citenamefont {Herbut}(2002{\natexlab{b}})}]{Herbut:2002yq}%
  \BibitemOpen
  \bibfield  {author} {\bibinfo {author} {\bibfnamefont {I.~F.}\ \bibnamefont
  {Herbut}},\ }\href {\doibase 10.1103/PhysRevB.66.094504} {\bibfield
  {journal} {\bibinfo  {journal} {Phys.Rev.}\ }\textbf {\bibinfo {volume}
  {B66}},\ \bibinfo {pages} {094504} (\bibinfo {year} {2002}{\natexlab{b}})},\
  \Eprint {http://arxiv.org/abs/cond-mat/0202491} {arXiv:cond-mat/0202491
  [cond-mat]} \BibitemShut {NoStop}%
\bibitem [{\citenamefont {Tesanovic}\ \emph {et~al.}(2002)\citenamefont
  {Tesanovic}, \citenamefont {Vafek},\ and\ \citenamefont
  {Franz}}]{Tesanovic:2002zz}%
  \BibitemOpen
  \bibfield  {author} {\bibinfo {author} {\bibfnamefont {Z.}~\bibnamefont
  {Tesanovic}}, \bibinfo {author} {\bibfnamefont {O.}~\bibnamefont {Vafek}}, \
  and\ \bibinfo {author} {\bibfnamefont {M.}~\bibnamefont {Franz}},\ }\href
  {\doibase 10.1103/PhysRevB.65.180511} {\bibfield  {journal} {\bibinfo
  {journal} {Phys.Rev.}\ }\textbf {\bibinfo {volume} {B65}},\ \bibinfo {pages}
  {180511} (\bibinfo {year} {2002})}\BibitemShut {NoStop}%
\bibitem [{\citenamefont {Mavromatos}\ and\ \citenamefont
  {Papavassiliou}(2004)}]{Mavromatos:2003ss}%
  \BibitemOpen
  \bibfield  {author} {\bibinfo {author} {\bibfnamefont {N.}~\bibnamefont
  {Mavromatos}}\ and\ \bibinfo {author} {\bibfnamefont {J.}~\bibnamefont
  {Papavassiliou}},\ }\href@noop {} {\bibfield  {journal} {\bibinfo  {journal}
  {Recent Res.Devel.Phys.}\ }\textbf {\bibinfo {volume} {5}},\ \bibinfo {pages}
  {369} (\bibinfo {year} {2004})},\ \Eprint
  {http://arxiv.org/abs/cond-mat/0311421} {arXiv:cond-mat/0311421 [cond-mat]}
  \BibitemShut {NoStop}%
\bibitem [{\citenamefont {Herbut}(2005)}]{Herbut:2004ue}%
  \BibitemOpen
  \bibfield  {author} {\bibinfo {author} {\bibfnamefont {I.~F.}\ \bibnamefont
  {Herbut}},\ }\href {\doibase 10.1103/PhysRevLett.94.237001} {\bibfield
  {journal} {\bibinfo  {journal} {Phys.Rev.Lett.}\ }\textbf {\bibinfo {volume}
  {94}},\ \bibinfo {pages} {237001} (\bibinfo {year} {2005})},\ \Eprint
  {http://arxiv.org/abs/cond-mat/0410557} {arXiv:cond-mat/0410557 [cond-mat]}
  \BibitemShut {NoStop}%
\bibitem [{\citenamefont {Appelquist}\ \emph {et~al.}(1988)\citenamefont
  {Appelquist}, \citenamefont {Nash},\ and\ \citenamefont
  {Wijewardhana}}]{Appelquist:1988sr}%
  \BibitemOpen
  \bibfield  {author} {\bibinfo {author} {\bibfnamefont {T.}~\bibnamefont
  {Appelquist}}, \bibinfo {author} {\bibfnamefont {D.}~\bibnamefont {Nash}}, \
  and\ \bibinfo {author} {\bibfnamefont {L.~C.~R.}\ \bibnamefont
  {Wijewardhana}},\ }\href {\doibase 10.1103/PhysRevLett.60.2575} {\bibfield
  {journal} {\bibinfo  {journal} {Phys. Rev. Lett.}\ }\textbf {\bibinfo
  {volume} {60}},\ \bibinfo {pages} {2575} (\bibinfo {year}
  {1988})}\BibitemShut {NoStop}%
\bibitem [{\citenamefont {Nash}(1989)}]{Nash:1989xx}%
  \BibitemOpen
  \bibfield  {author} {\bibinfo {author} {\bibfnamefont {D.}~\bibnamefont
  {Nash}},\ }\href {\doibase 10.1103/PhysRevLett.62.3024} {\bibfield  {journal}
  {\bibinfo  {journal} {Phys.Rev.Lett.}\ }\textbf {\bibinfo {volume} {62}},\
  \bibinfo {pages} {3024} (\bibinfo {year} {1989})}\BibitemShut {NoStop}%
\bibitem [{\citenamefont {Pennington}\ and\ \citenamefont
  {Walsh}(1991)}]{Pennington:1990bx}%
  \BibitemOpen
  \bibfield  {author} {\bibinfo {author} {\bibfnamefont {M.~R.}\ \bibnamefont
  {Pennington}}\ and\ \bibinfo {author} {\bibfnamefont {D.}~\bibnamefont
  {Walsh}},\ }\href {\doibase 10.1016/0370-2693(91)91392-9} {\bibfield
  {journal} {\bibinfo  {journal} {Phys. Lett.}\ }\textbf {\bibinfo {volume}
  {B253}},\ \bibinfo {pages} {246} (\bibinfo {year} {1991})}\BibitemShut
  {NoStop}%
\bibitem [{\citenamefont {Atkinson}\ \emph {et~al.}(1990)\citenamefont
  {Atkinson}, \citenamefont {Johnson},\ and\ \citenamefont
  {Maris}}]{Atkinson:1989fp}%
  \BibitemOpen
  \bibfield  {author} {\bibinfo {author} {\bibfnamefont {D.}~\bibnamefont
  {Atkinson}}, \bibinfo {author} {\bibfnamefont {P.}~\bibnamefont {Johnson}}, \
  and\ \bibinfo {author} {\bibfnamefont {P.}~\bibnamefont {Maris}},\ }\href
  {\doibase 10.1103/PhysRevD.42.602} {\bibfield  {journal} {\bibinfo  {journal}
  {Phys.Rev.}\ }\textbf {\bibinfo {volume} {D42}},\ \bibinfo {pages} {602}
  (\bibinfo {year} {1990})}\BibitemShut {NoStop}%
\bibitem [{\citenamefont {Curtis}\ \emph {et~al.}(1992)\citenamefont {Curtis},
  \citenamefont {Pennington},\ and\ \citenamefont {Walsh}}]{Curtis:1992gm}%
  \BibitemOpen
  \bibfield  {author} {\bibinfo {author} {\bibfnamefont {D.}~\bibnamefont
  {Curtis}}, \bibinfo {author} {\bibfnamefont {M.}~\bibnamefont {Pennington}},
  \ and\ \bibinfo {author} {\bibfnamefont {D.}~\bibnamefont {Walsh}},\ }\href
  {\doibase 10.1016/0370-2693(92)91572-Q} {\bibfield  {journal} {\bibinfo
  {journal} {Phys.Lett.}\ }\textbf {\bibinfo {volume} {B295}},\ \bibinfo
  {pages} {313} (\bibinfo {year} {1992})}\BibitemShut {NoStop}%
\bibitem [{\citenamefont {Maris}(1995)}]{Maris:1995ns}%
  \BibitemOpen
  \bibfield  {author} {\bibinfo {author} {\bibfnamefont {P.}~\bibnamefont
  {Maris}},\ }\href {\doibase 10.1103/PhysRevD.52.6087} {\bibfield  {journal}
  {\bibinfo  {journal} {Phys.Rev.}\ }\textbf {\bibinfo {volume} {D52}},\
  \bibinfo {pages} {6087} (\bibinfo {year} {1995})},\ \Eprint
  {http://arxiv.org/abs/hep-ph/9508323} {arXiv:hep-ph/9508323 [hep-ph]}
  \BibitemShut {NoStop}%
\bibitem [{\citenamefont {Ebihara}\ \emph {et~al.}(1995)\citenamefont
  {Ebihara}, \citenamefont {Iizuka}, \citenamefont {Kondo},\ and\ \citenamefont
  {Tanaka}}]{Ebihara:1995aa}%
  \BibitemOpen
  \bibfield  {author} {\bibinfo {author} {\bibfnamefont {T.}~\bibnamefont
  {Ebihara}}, \bibinfo {author} {\bibfnamefont {T.}~\bibnamefont {Iizuka}},
  \bibinfo {author} {\bibfnamefont {K.-I.}\ \bibnamefont {Kondo}}, \ and\
  \bibinfo {author} {\bibfnamefont {E.}~\bibnamefont {Tanaka}},\ }\href
  {\doibase 10.1016/0550-3213(94)00457-P} {\bibfield  {journal} {\bibinfo
  {journal} {Nucl.Phys.}\ }\textbf {\bibinfo {volume} {B434}},\ \bibinfo
  {pages} {85} (\bibinfo {year} {1995})},\ \Eprint
  {http://arxiv.org/abs/hep-ph/9404361} {arXiv:hep-ph/9404361 [hep-ph]}
  \BibitemShut {NoStop}%
\bibitem [{\citenamefont {Maris}(1996)}]{Maris:1996zg}%
  \BibitemOpen
  \bibfield  {author} {\bibinfo {author} {\bibfnamefont {P.}~\bibnamefont
  {Maris}},\ }\href {\doibase 10.1103/PhysRevD.54.4049} {\bibfield  {journal}
  {\bibinfo  {journal} {Phys.Rev.}\ }\textbf {\bibinfo {volume} {D54}},\
  \bibinfo {pages} {4049} (\bibinfo {year} {1996})},\ \Eprint
  {http://arxiv.org/abs/hep-ph/9606214} {arXiv:hep-ph/9606214 [hep-ph]}
  \BibitemShut {NoStop}%
\bibitem [{\citenamefont {Aitchison}\ \emph {et~al.}(1997)\citenamefont
  {Aitchison}, \citenamefont {Mavromatos},\ and\ \citenamefont
  {McNeill}}]{Aitchison:1997aa}%
  \BibitemOpen
  \bibfield  {author} {\bibinfo {author} {\bibfnamefont {I.}~\bibnamefont
  {Aitchison}}, \bibinfo {author} {\bibfnamefont {N.}~\bibnamefont
  {Mavromatos}}, \ and\ \bibinfo {author} {\bibfnamefont {D.}~\bibnamefont
  {McNeill}},\ }\href {\doibase 10.1016/S0370-2693(97)00447-4} {\bibfield
  {journal} {\bibinfo  {journal} {Phys.Lett.}\ }\textbf {\bibinfo {volume}
  {B402}},\ \bibinfo {pages} {154} (\bibinfo {year} {1997})},\ \Eprint
  {http://arxiv.org/abs/hep-th/9701087} {arXiv:hep-th/9701087 [hep-th]}
  \BibitemShut {NoStop}%
\bibitem [{\citenamefont {Appelquist}\ \emph {et~al.}(1999)\citenamefont
  {Appelquist}, \citenamefont {Cohen},\ and\ \citenamefont
  {Schmaltz}}]{Appelquist:1999hr}%
  \BibitemOpen
  \bibfield  {author} {\bibinfo {author} {\bibfnamefont {T.}~\bibnamefont
  {Appelquist}}, \bibinfo {author} {\bibfnamefont {A.~G.}\ \bibnamefont
  {Cohen}}, \ and\ \bibinfo {author} {\bibfnamefont {M.}~\bibnamefont
  {Schmaltz}},\ }\href {\doibase 10.1103/PhysRevD.60.045003} {\bibfield
  {journal} {\bibinfo  {journal} {Phys.Rev.}\ }\textbf {\bibinfo {volume}
  {D60}},\ \bibinfo {pages} {045003} (\bibinfo {year} {1999})},\ \Eprint
  {http://arxiv.org/abs/hep-th/9901109} {arXiv:hep-th/9901109 [hep-th]}
  \BibitemShut {NoStop}%
\bibitem [{\citenamefont {Gusynin}\ and\ \citenamefont
  {Reenders}(2003)}]{Gusynin:2003ww}%
  \BibitemOpen
  \bibfield  {author} {\bibinfo {author} {\bibfnamefont {V.}~\bibnamefont
  {Gusynin}}\ and\ \bibinfo {author} {\bibfnamefont {M.}~\bibnamefont
  {Reenders}},\ }\href {\doibase 10.1103/PhysRevD.68.025017} {\bibfield
  {journal} {\bibinfo  {journal} {Phys.Rev.}\ }\textbf {\bibinfo {volume}
  {D68}},\ \bibinfo {pages} {025017} (\bibinfo {year} {2003})},\ \Eprint
  {http://arxiv.org/abs/hep-ph/0304302} {arXiv:hep-ph/0304302 [hep-ph]}
  \BibitemShut {NoStop}%
\bibitem [{\citenamefont {Fischer}\ \emph {et~al.}(2004)\citenamefont
  {Fischer}, \citenamefont {Alkofer}, \citenamefont {Dahm},\ and\ \citenamefont
  {Maris}}]{Fischer:2004nq}%
  \BibitemOpen
  \bibfield  {author} {\bibinfo {author} {\bibfnamefont {C.~S.}\ \bibnamefont
  {Fischer}}, \bibinfo {author} {\bibfnamefont {R.}~\bibnamefont {Alkofer}},
  \bibinfo {author} {\bibfnamefont {T.}~\bibnamefont {Dahm}}, \ and\ \bibinfo
  {author} {\bibfnamefont {P.}~\bibnamefont {Maris}},\ }\href {\doibase
  10.1103/PhysRevD.70.073007} {\bibfield  {journal} {\bibinfo  {journal} {Phys.
  Rev.}\ }\textbf {\bibinfo {volume} {D70}},\ \bibinfo {pages} {073007}
  (\bibinfo {year} {2004})},\ \Eprint {http://arxiv.org/abs/hep-ph/0407104}
  {arXiv:hep-ph/0407104} \BibitemShut {NoStop}%
\bibitem [{\citenamefont {Kaveh}\ and\ \citenamefont
  {Herbut}(2005)}]{Kaveh:2004qa}%
  \BibitemOpen
  \bibfield  {author} {\bibinfo {author} {\bibfnamefont {K.}~\bibnamefont
  {Kaveh}}\ and\ \bibinfo {author} {\bibfnamefont {I.~F.}\ \bibnamefont
  {Herbut}},\ }\href {\doibase 10.1103/PhysRevB.71.184519} {\bibfield
  {journal} {\bibinfo  {journal} {Phys.Rev.}\ }\textbf {\bibinfo {volume}
  {B71}},\ \bibinfo {pages} {184519} (\bibinfo {year} {2005})},\ \Eprint
  {http://arxiv.org/abs/cond-mat/0411594} {arXiv:cond-mat/0411594 [cond-mat]}
  \BibitemShut {NoStop}%
\bibitem [{\citenamefont {Kubota}\ and\ \citenamefont
  {Terao}(2001)}]{Kubota:2001kk}%
  \BibitemOpen
  \bibfield  {author} {\bibinfo {author} {\bibfnamefont {K.-I.}\ \bibnamefont
  {Kubota}}\ and\ \bibinfo {author} {\bibfnamefont {H.}~\bibnamefont {Terao}},\
  }\href {\doibase 10.1143/PTP.105.809} {\bibfield  {journal} {\bibinfo
  {journal} {Prog.Theor.Phys.}\ }\textbf {\bibinfo {volume} {105}},\ \bibinfo
  {pages} {809} (\bibinfo {year} {2001})},\ \Eprint
  {http://arxiv.org/abs/hep-ph/0101073} {arXiv:hep-ph/0101073 [hep-ph]}
  \BibitemShut {NoStop}%
\bibitem [{\citenamefont {Hands}\ \emph {et~al.}(2002)\citenamefont {Hands},
  \citenamefont {Kogut},\ and\ \citenamefont {Strouthos}}]{Hands:2002dv}%
  \BibitemOpen
  \bibfield  {author} {\bibinfo {author} {\bibfnamefont {S.}~\bibnamefont
  {Hands}}, \bibinfo {author} {\bibfnamefont {J.}~\bibnamefont {Kogut}}, \ and\
  \bibinfo {author} {\bibfnamefont {C.}~\bibnamefont {Strouthos}},\ }\href
  {\doibase 10.1016/S0550-3213(02)00869-6} {\bibfield  {journal} {\bibinfo
  {journal} {Nucl.Phys.}\ }\textbf {\bibinfo {volume} {B645}},\ \bibinfo
  {pages} {321} (\bibinfo {year} {2002})},\ \Eprint
  {http://arxiv.org/abs/hep-lat/0208030} {arXiv:hep-lat/0208030 [hep-lat]}
  \BibitemShut {NoStop}%
\bibitem [{\citenamefont {Hands}\ \emph {et~al.}(2004)\citenamefont {Hands},
  \citenamefont {Kogut}, \citenamefont {Scorzato},\ and\ \citenamefont
  {Strouthos}}]{Hands:2004bh}%
  \BibitemOpen
  \bibfield  {author} {\bibinfo {author} {\bibfnamefont {S.}~\bibnamefont
  {Hands}}, \bibinfo {author} {\bibfnamefont {J.}~\bibnamefont {Kogut}},
  \bibinfo {author} {\bibfnamefont {L.}~\bibnamefont {Scorzato}}, \ and\
  \bibinfo {author} {\bibfnamefont {C.}~\bibnamefont {Strouthos}},\ }\href
  {\doibase 10.1103/PhysRevB.70.104501} {\bibfield  {journal} {\bibinfo
  {journal} {Phys.Rev.}\ }\textbf {\bibinfo {volume} {B70}},\ \bibinfo {pages}
  {104501} (\bibinfo {year} {2004})},\ \Eprint
  {http://arxiv.org/abs/hep-lat/0404013} {arXiv:hep-lat/0404013 [hep-lat]}
  \BibitemShut {NoStop}%
\bibitem [{\citenamefont {Mitra}\ \emph {et~al.}(2007)\citenamefont {Mitra},
  \citenamefont {Ratabole},\ and\ \citenamefont
  {Sharatchandra}}]{Mitra:2007aa}%
  \BibitemOpen
  \bibfield  {author} {\bibinfo {author} {\bibfnamefont {I.}~\bibnamefont
  {Mitra}}, \bibinfo {author} {\bibfnamefont {R.}~\bibnamefont {Ratabole}}, \
  and\ \bibinfo {author} {\bibfnamefont {H.}~\bibnamefont {Sharatchandra}},\
  }\href {\doibase 10.1142/S0217732307020981} {\bibfield  {journal} {\bibinfo
  {journal} {Mod.Phys.Lett.}\ }\textbf {\bibinfo {volume} {A22}},\ \bibinfo
  {pages} {297} (\bibinfo {year} {2007})},\ \Eprint
  {http://arxiv.org/abs/hep-th/0601058} {arXiv:hep-th/0601058 [hep-th]}
  \BibitemShut {NoStop}%
\bibitem [{\citenamefont {Strouthos}\ and\ \citenamefont
  {Kogut}(2009)}]{Strouthos:2008hs}%
  \BibitemOpen
  \bibfield  {author} {\bibinfo {author} {\bibfnamefont {C.}~\bibnamefont
  {Strouthos}}\ and\ \bibinfo {author} {\bibfnamefont {J.~B.}\ \bibnamefont
  {Kogut}},\ }\href {\doibase 10.1088/1742-6596/150/5/052247} {\bibfield
  {journal} {\bibinfo  {journal} {J.Phys.Conf.Ser.}\ }\textbf {\bibinfo
  {volume} {150}},\ \bibinfo {pages} {052247} (\bibinfo {year} {2009})},\
  \Eprint {http://arxiv.org/abs/0808.2714} {arXiv:0808.2714
  [cond-mat.supr-con]} \BibitemShut {NoStop}%
\bibitem [{\citenamefont {Bashir}\ \emph {et~al.}(2008)\citenamefont {Bashir},
  \citenamefont {Raya}, \citenamefont {Cloet},\ and\ \citenamefont
  {Roberts}}]{Bashir:2008aa}%
  \BibitemOpen
  \bibfield  {author} {\bibinfo {author} {\bibfnamefont {A.}~\bibnamefont
  {Bashir}}, \bibinfo {author} {\bibfnamefont {A.}~\bibnamefont {Raya}},
  \bibinfo {author} {\bibfnamefont {I.}~\bibnamefont {Cloet}}, \ and\ \bibinfo
  {author} {\bibfnamefont {C.}~\bibnamefont {Roberts}},\ }\href {\doibase
  10.1103/PhysRevC.78.055201} {\bibfield  {journal} {\bibinfo  {journal}
  {Phys.Rev.}\ }\textbf {\bibinfo {volume} {C78}},\ \bibinfo {pages} {055201}
  (\bibinfo {year} {2008})},\ \Eprint {http://arxiv.org/abs/0806.3305}
  {arXiv:0806.3305 [hep-ph]} \BibitemShut {NoStop}%
\bibitem [{\citenamefont {Bashir}\ \emph {et~al.}(2009)\citenamefont {Bashir},
  \citenamefont {Raya}, \citenamefont {Sanchez-Madrigal},\ and\ \citenamefont
  {Roberts}}]{Bashir:2009aa}%
  \BibitemOpen
  \bibfield  {author} {\bibinfo {author} {\bibfnamefont {A.}~\bibnamefont
  {Bashir}}, \bibinfo {author} {\bibfnamefont {A.}~\bibnamefont {Raya}},
  \bibinfo {author} {\bibfnamefont {S.}~\bibnamefont {Sanchez-Madrigal}}, \
  and\ \bibinfo {author} {\bibfnamefont {C.}~\bibnamefont {Roberts}},\ }\href
  {\doibase 10.1007/s00601-009-0069-9} {\bibfield  {journal} {\bibinfo
  {journal} {Few Body Syst.}\ }\textbf {\bibinfo {volume} {46}},\ \bibinfo
  {pages} {229} (\bibinfo {year} {2009})},\ \Eprint
  {http://arxiv.org/abs/0905.1337} {arXiv:0905.1337 [hep-ph]} \BibitemShut
  {NoStop}%
\bibitem [{\citenamefont {Feng}\ \emph {et~al.}(2012)\citenamefont {Feng},
  \citenamefont {Wang}, \citenamefont {Sun},\ and\ \citenamefont
  {Zong}}]{Feng:2012aa}%
  \BibitemOpen
  \bibfield  {author} {\bibinfo {author} {\bibfnamefont {H.-t.}\ \bibnamefont
  {Feng}}, \bibinfo {author} {\bibfnamefont {B.}~\bibnamefont {Wang}}, \bibinfo
  {author} {\bibfnamefont {W.-m.}\ \bibnamefont {Sun}}, \ and\ \bibinfo
  {author} {\bibfnamefont {H.-s.}\ \bibnamefont {Zong}},\ }\href {\doibase
  10.1103/PhysRevD.86.105042} {\bibfield  {journal} {\bibinfo  {journal}
  {Phys.Rev.}\ }\textbf {\bibinfo {volume} {D86}},\ \bibinfo {pages} {105042}
  (\bibinfo {year} {2012})}\BibitemShut {NoStop}%
\bibitem [{\citenamefont {Grover}(2012)}]{Grover:2012aa}%
  \BibitemOpen
  \bibfield  {author} {\bibinfo {author} {\bibfnamefont {T.}~\bibnamefont
  {Grover}},\ }\href@noop {} {\  (\bibinfo {year} {2012})},\ \Eprint
  {http://arxiv.org/abs/1211.1392} {arXiv:1211.1392 [hep-th]} \BibitemShut
  {NoStop}%
\bibitem [{\citenamefont {Bonnet}\ \emph {et~al.}(2011)\citenamefont {Bonnet},
  \citenamefont {Fischer},\ and\ \citenamefont {Williams}}]{Bonnet:2011hh}%
  \BibitemOpen
  \bibfield  {author} {\bibinfo {author} {\bibfnamefont {J.~A.}\ \bibnamefont
  {Bonnet}}, \bibinfo {author} {\bibfnamefont {C.~S.}\ \bibnamefont {Fischer}},
  \ and\ \bibinfo {author} {\bibfnamefont {R.}~\bibnamefont {Williams}},\
  }\href {\doibase 10.1103/PhysRevB.84.024520} {\bibfield  {journal} {\bibinfo
  {journal} {Phys.Rev.}\ }\textbf {\bibinfo {volume} {B84}},\ \bibinfo {pages}
  {024520} (\bibinfo {year} {2011})},\ \Eprint {http://arxiv.org/abs/1103.1578}
  {arXiv:1103.1578 [hep-ph]} \BibitemShut {NoStop}%
\bibitem [{\citenamefont {Bonnet}\ \emph {et~al.}(2012)\citenamefont {Bonnet},
  \citenamefont {Fischer},\ and\ \citenamefont {Williams}}]{Bonnet:2011ds}%
  \BibitemOpen
  \bibfield  {author} {\bibinfo {author} {\bibfnamefont {J.~A.}\ \bibnamefont
  {Bonnet}}, \bibinfo {author} {\bibfnamefont {C.~S.}\ \bibnamefont {Fischer}},
  \ and\ \bibinfo {author} {\bibfnamefont {R.}~\bibnamefont {Williams}},\
  }\href {\doibase 10.1016/j.ppnp.2011.12.026} {\bibfield  {journal} {\bibinfo
  {journal} {Prog.Part.Nucl.Phys.}\ }\textbf {\bibinfo {volume} {67}},\
  \bibinfo {pages} {245} (\bibinfo {year} {2012})},\ \Eprint
  {http://arxiv.org/abs/1111.0182} {arXiv:1111.0182 [hep-ph]} \BibitemShut
  {NoStop}%
\bibitem [{\citenamefont {Bonnet}\ and\ \citenamefont
  {Fischer}(2012)}]{Bonnet:2012az}%
  \BibitemOpen
  \bibfield  {author} {\bibinfo {author} {\bibfnamefont {J.~A.}\ \bibnamefont
  {Bonnet}}\ and\ \bibinfo {author} {\bibfnamefont {C.~S.}\ \bibnamefont
  {Fischer}},\ }\href {\doibase 10.1016/j.physletb.2012.11.004} {\bibfield
  {journal} {\bibinfo  {journal} {Phys.Lett.}\ }\textbf {\bibinfo {volume}
  {B718}},\ \bibinfo {pages} {532} (\bibinfo {year} {2012})},\ \Eprint
  {http://arxiv.org/abs/1201.6139} {arXiv:1201.6139 [hep-ph]} \BibitemShut
  {NoStop}%
\bibitem [{\citenamefont {Popovici}\ \emph {et~al.}(2013)\citenamefont
  {Popovici}, \citenamefont {Fischer},\ and\ \citenamefont {von
  Smekal}}]{Popovici:2013wfa}%
  \BibitemOpen
  \bibfield  {author} {\bibinfo {author} {\bibfnamefont {C.}~\bibnamefont
  {Popovici}}, \bibinfo {author} {\bibfnamefont {C.}~\bibnamefont {Fischer}}, \
  and\ \bibinfo {author} {\bibfnamefont {L.}~\bibnamefont {von Smekal}},\
  }\href {\doibase 10.1103/PhysRevB.88.205429} {\bibfield  {journal} {\bibinfo
  {journal} {Phys.Rev.}\ }\textbf {\bibinfo {volume} {B88}},\ \bibinfo {pages}
  {205429} (\bibinfo {year} {2013})},\ \Eprint {http://arxiv.org/abs/1308.6199}
  {arXiv:1308.6199 [hep-ph]} \BibitemShut {NoStop}%
\bibitem [{\citenamefont {Gies}\ and\ \citenamefont
  {Janssen}(2010)}]{Gies:2010st}%
  \BibitemOpen
  \bibfield  {author} {\bibinfo {author} {\bibfnamefont {H.}~\bibnamefont
  {Gies}}\ and\ \bibinfo {author} {\bibfnamefont {L.}~\bibnamefont {Janssen}},\
  }\href {\doibase 10.1103/PhysRevD.82.085018} {\bibfield  {journal} {\bibinfo
  {journal} {Phys. Rev.}\ }\textbf {\bibinfo {volume} {D82}},\ \bibinfo {pages}
  {085018} (\bibinfo {year} {2010})},\ \Eprint {http://arxiv.org/abs/1006.3747}
  {arXiv:1006.3747 [hep-th]} \BibitemShut {NoStop}%
\bibitem [{\citenamefont {Janssen}(2012)}]{Janssen:2012oca}%
  \BibitemOpen
  \bibfield  {author} {\bibinfo {author} {\bibfnamefont {L.}~\bibnamefont
  {Janssen}},\ }\href@noop {} {\bibfield  {journal} {\bibinfo  {journal} {PhD
  thesis, Jena, http://www.db-thueringen.de/servlets/DocumentServlet?id=20856}\
  } (\bibinfo {year} {2012})}\BibitemShut {NoStop}%
\bibitem [{\citenamefont {Semenoff}(1984)}]{Semenoff:1984dq}%
  \BibitemOpen
  \bibfield  {author} {\bibinfo {author} {\bibfnamefont {G.~W.}\ \bibnamefont
  {Semenoff}},\ }\href {\doibase 10.1103/PhysRevLett.53.2449} {\bibfield
  {journal} {\bibinfo  {journal} {Phys.Rev.Lett.}\ }\textbf {\bibinfo {volume}
  {53}},\ \bibinfo {pages} {2449} (\bibinfo {year} {1984})}\BibitemShut
  {NoStop}%
\bibitem [{\citenamefont {Hands}\ and\ \citenamefont
  {Strouthos}(2008)}]{Hands:2008id}%
  \BibitemOpen
  \bibfield  {author} {\bibinfo {author} {\bibfnamefont {S.}~\bibnamefont
  {Hands}}\ and\ \bibinfo {author} {\bibfnamefont {C.}~\bibnamefont
  {Strouthos}},\ }\href {\doibase 10.1103/PhysRevB.78.165423} {\bibfield
  {journal} {\bibinfo  {journal} {Phys.Rev.}\ }\textbf {\bibinfo {volume}
  {B78}},\ \bibinfo {pages} {165423} (\bibinfo {year} {2008})},\ \Eprint
  {http://arxiv.org/abs/0806.4877} {arXiv:0806.4877 [cond-mat.str-el]}
  \BibitemShut {NoStop}%
\bibitem [{\citenamefont {Herbut}\ \emph {et~al.}(2009)\citenamefont {Herbut},
  \citenamefont {Juricic},\ and\ \citenamefont {Roy}}]{Herbut:2009qb}%
  \BibitemOpen
  \bibfield  {author} {\bibinfo {author} {\bibfnamefont {I.~F.}\ \bibnamefont
  {Herbut}}, \bibinfo {author} {\bibfnamefont {V.}~\bibnamefont {Juricic}}, \
  and\ \bibinfo {author} {\bibfnamefont {B.}~\bibnamefont {Roy}},\ }\href
  {\doibase 10.1103/PhysRevB.79.085116} {\bibfield  {journal} {\bibinfo
  {journal} {Phys.Rev.}\ }\textbf {\bibinfo {volume} {B79}},\ \bibinfo {pages}
  {085116} (\bibinfo {year} {2009})},\ \Eprint {http://arxiv.org/abs/0811.0610}
  {arXiv:0811.0610 [cond-mat.str-el]} \BibitemShut {NoStop}%
\bibitem [{\citenamefont {Drut}\ and\ \citenamefont
  {Lahde}(2009)}]{Drut:2009aj}%
  \BibitemOpen
  \bibfield  {author} {\bibinfo {author} {\bibfnamefont {J.~E.}\ \bibnamefont
  {Drut}}\ and\ \bibinfo {author} {\bibfnamefont {T.~A.}\ \bibnamefont
  {Lahde}},\ }\href {\doibase 10.1103/PhysRevB.79.165425} {\bibfield  {journal}
  {\bibinfo  {journal} {Phys.Rev.}\ }\textbf {\bibinfo {volume} {B79}},\
  \bibinfo {pages} {165425} (\bibinfo {year} {2009})},\ \Eprint
  {http://arxiv.org/abs/0901.0584} {arXiv:0901.0584 [cond-mat.str-el]}
  \BibitemShut {NoStop}%
\bibitem [{\citenamefont {Armour}\ \emph {et~al.}(2010)\citenamefont {Armour},
  \citenamefont {Hands},\ and\ \citenamefont {Strouthos}}]{Armour:2009vj}%
  \BibitemOpen
  \bibfield  {author} {\bibinfo {author} {\bibfnamefont {W.}~\bibnamefont
  {Armour}}, \bibinfo {author} {\bibfnamefont {S.}~\bibnamefont {Hands}}, \
  and\ \bibinfo {author} {\bibfnamefont {C.}~\bibnamefont {Strouthos}},\ }\href
  {\doibase 10.1103/PhysRevB.81.125105} {\bibfield  {journal} {\bibinfo
  {journal} {Phys.Rev.}\ }\textbf {\bibinfo {volume} {B81}},\ \bibinfo {pages}
  {125105} (\bibinfo {year} {2010})},\ \Eprint {http://arxiv.org/abs/0910.5646}
  {arXiv:0910.5646 [cond-mat.str-el]} \BibitemShut {NoStop}%
\bibitem [{\citenamefont {Gusynin}\ \emph {et~al.}(2007)\citenamefont
  {Gusynin}, \citenamefont {Sharapov},\ and\ \citenamefont
  {Carbotte}}]{Gusynin:2007ix}%
  \BibitemOpen
  \bibfield  {author} {\bibinfo {author} {\bibfnamefont {V.}~\bibnamefont
  {Gusynin}}, \bibinfo {author} {\bibfnamefont {S.}~\bibnamefont {Sharapov}}, \
  and\ \bibinfo {author} {\bibfnamefont {J.}~\bibnamefont {Carbotte}},\ }\href
  {\doibase 10.1142/S0217979207038022} {\bibfield  {journal} {\bibinfo
  {journal} {Int.J.Mod.Phys.}\ }\textbf {\bibinfo {volume} {B21}},\ \bibinfo
  {pages} {4611} (\bibinfo {year} {2007})},\ \Eprint
  {http://arxiv.org/abs/0706.3016} {arXiv:0706.3016 [cond-mat.mes-hall]}
  \BibitemShut {NoStop}%
\bibitem [{\citenamefont {Smith}\ and\ \citenamefont {von
  Smekal}(2014)}]{Smith:2014tha}%
  \BibitemOpen
  \bibfield  {author} {\bibinfo {author} {\bibfnamefont {D.}~\bibnamefont
  {Smith}}\ and\ \bibinfo {author} {\bibfnamefont {L.}~\bibnamefont {von
  Smekal}},\ }\href@noop {} {\  (\bibinfo {year} {2014})},\ \Eprint
  {http://arxiv.org/abs/1403.3620} {arXiv:1403.3620 [hep-lat]} \BibitemShut
  {NoStop}%
\bibitem [{\citenamefont {Gies}\ and\ \citenamefont
  {Jaeckel}(2006)}]{Gies:2005as}%
  \BibitemOpen
  \bibfield  {author} {\bibinfo {author} {\bibfnamefont {H.}~\bibnamefont
  {Gies}}\ and\ \bibinfo {author} {\bibfnamefont {J.}~\bibnamefont {Jaeckel}},\
  }\href@noop {} {\bibfield  {journal} {\bibinfo  {journal} {Eur. Phys. J.}\
  }\textbf {\bibinfo {volume} {C46}},\ \bibinfo {pages} {433} (\bibinfo {year}
  {2006})},\ \Eprint {http://arxiv.org/abs/hep-ph/0507171} {hep-ph/0507171}
  \BibitemShut {NoStop}%
\bibitem [{\citenamefont {Braun}\ and\ \citenamefont
  {Gies}(2007)}]{Braun:2005uj}%
  \BibitemOpen
  \bibfield  {author} {\bibinfo {author} {\bibfnamefont {J.}~\bibnamefont
  {Braun}}\ and\ \bibinfo {author} {\bibfnamefont {H.}~\bibnamefont {Gies}},\
  }\href {\doibase 10.1016/j.physletb.2006.11.059} {\bibfield  {journal}
  {\bibinfo  {journal} {Phys. Lett.}\ }\textbf {\bibinfo {volume} {B645}},\
  \bibinfo {pages} {53} (\bibinfo {year} {2007})},\ \Eprint
  {http://arxiv.org/abs/hep-ph/0512085} {arXiv:hep-ph/0512085} \BibitemShut
  {NoStop}%
\bibitem [{\citenamefont {Braun}\ and\ \citenamefont
  {Gies}(2006)}]{Braun:2006jd}%
  \BibitemOpen
  \bibfield  {author} {\bibinfo {author} {\bibfnamefont {J.}~\bibnamefont
  {Braun}}\ and\ \bibinfo {author} {\bibfnamefont {H.}~\bibnamefont {Gies}},\
  }\href {\doibase 10.1088/1126-6708/2006/06/024} {\bibfield  {journal}
  {\bibinfo  {journal} {JHEP}\ }\textbf {\bibinfo {volume} {0606}},\ \bibinfo
  {pages} {024} (\bibinfo {year} {2006})},\ \Eprint
  {http://arxiv.org/abs/hep-ph/0602226} {arXiv:hep-ph/0602226 [hep-ph]}
  \BibitemShut {NoStop}%
\bibitem [{\citenamefont {Hubbard}(1959)}]{Hubbard:1959ub}%
  \BibitemOpen
  \bibfield  {author} {\bibinfo {author} {\bibfnamefont {J.}~\bibnamefont
  {Hubbard}},\ }\href {\doibase 10.1103/PhysRevLett.3.77} {\bibfield  {journal}
  {\bibinfo  {journal} {Phys. Rev. Lett.}\ }\textbf {\bibinfo {volume} {3}},\
  \bibinfo {pages} {77} (\bibinfo {year} {1959})}\BibitemShut {NoStop}%
\bibitem [{\citenamefont {Stratonovich}(1957)}]{Stratonovich}%
  \BibitemOpen
  \bibfield  {author} {\bibinfo {author} {\bibfnamefont {R.}~\bibnamefont
  {Stratonovich}},\ }\href@noop {} {\bibfield  {journal} {\bibinfo  {journal}
  {Dokl. Akad. Nauk.}\ }\textbf {\bibinfo {volume} {115}},\ \bibinfo {pages}
  {1097} (\bibinfo {year} {1957})}\BibitemShut {NoStop}%
\bibitem [{\citenamefont {Baier}\ \emph {et~al.}(2000)\citenamefont {Baier},
  \citenamefont {Bick},\ and\ \citenamefont {Wetterich}}]{Baier:2000yc}%
  \BibitemOpen
  \bibfield  {author} {\bibinfo {author} {\bibfnamefont {T.}~\bibnamefont
  {Baier}}, \bibinfo {author} {\bibfnamefont {E.}~\bibnamefont {Bick}}, \ and\
  \bibinfo {author} {\bibfnamefont {C.}~\bibnamefont {Wetterich}},\ }\href
  {\doibase 10.1103/PhysRevB.62.15471} {\bibfield  {journal} {\bibinfo
  {journal} {Phys.Rev.}\ }\textbf {\bibinfo {volume} {B62}},\ \bibinfo {pages}
  {15471} (\bibinfo {year} {2000})},\ \Eprint
  {http://arxiv.org/abs/cond-mat/0005218} {arXiv:cond-mat/0005218 [cond-mat]}
  \BibitemShut {NoStop}%
\bibitem [{\citenamefont {Gies}\ and\ \citenamefont
  {Wetterich}(2002)}]{Gies:2001nw}%
  \BibitemOpen
  \bibfield  {author} {\bibinfo {author} {\bibfnamefont {H.}~\bibnamefont
  {Gies}}\ and\ \bibinfo {author} {\bibfnamefont {C.}~\bibnamefont
  {Wetterich}},\ }\href@noop {} {\bibfield  {journal} {\bibinfo  {journal}
  {Phys. Rev.}\ }\textbf {\bibinfo {volume} {D65}},\ \bibinfo {pages} {065001}
  (\bibinfo {year} {2002})},\ \Eprint {http://arxiv.org/abs/hep-th/0107221}
  {hep-th/0107221} \BibitemShut {NoStop}%
\bibitem [{\citenamefont {Jaeckel}\ and\ \citenamefont
  {Wetterich}(2003)}]{Jaeckel:2002rm}%
  \BibitemOpen
  \bibfield  {author} {\bibinfo {author} {\bibfnamefont {J.}~\bibnamefont
  {Jaeckel}}\ and\ \bibinfo {author} {\bibfnamefont {C.}~\bibnamefont
  {Wetterich}},\ }\href@noop {} {\bibfield  {journal} {\bibinfo  {journal}
  {Phys. Rev.}\ }\textbf {\bibinfo {volume} {D68}},\ \bibinfo {pages} {025020}
  (\bibinfo {year} {2003})},\ \Eprint {http://arxiv.org/abs/hep-ph/0207094}
  {hep-ph/0207094} \BibitemShut {NoStop}%
\bibitem [{\citenamefont {Janssen}\ and\ \citenamefont
  {Gies}(2012)}]{Janssen:2012pq}%
  \BibitemOpen
  \bibfield  {author} {\bibinfo {author} {\bibfnamefont {L.}~\bibnamefont
  {Janssen}}\ and\ \bibinfo {author} {\bibfnamefont {H.}~\bibnamefont {Gies}},\
  }\href {\doibase 10.1103/PhysRevD.86.105007} {\bibfield  {journal} {\bibinfo
  {journal} {Phys.Rev.}\ }\textbf {\bibinfo {volume} {D86}},\ \bibinfo {pages}
  {105007} (\bibinfo {year} {2012})},\ \Eprint {http://arxiv.org/abs/1208.3327}
  {arXiv:1208.3327 [hep-th]} \BibitemShut {NoStop}%
\bibitem [{\citenamefont {Bjorken}(1963)}]{Bjorken:1963vg}%
  \BibitemOpen
  \bibfield  {author} {\bibinfo {author} {\bibfnamefont {J.}~\bibnamefont
  {Bjorken}},\ }\href {\doibase 10.1016/0003-4916(63)90069-1} {\bibfield
  {journal} {\bibinfo  {journal} {Annals Phys.}\ }\textbf {\bibinfo {volume}
  {24}},\ \bibinfo {pages} {174} (\bibinfo {year} {1963})}\BibitemShut
  {NoStop}%
\bibitem [{\citenamefont {Bialynicki-Birula}(1963)}]{BialynickiBirula:1963zz}%
  \BibitemOpen
  \bibfield  {author} {\bibinfo {author} {\bibfnamefont {I.}~\bibnamefont
  {Bialynicki-Birula}},\ }\href {\doibase 10.1103/PhysRev.130.465} {\bibfield
  {journal} {\bibinfo  {journal} {Phys.Rev.}\ }\textbf {\bibinfo {volume}
  {130}},\ \bibinfo {pages} {465} (\bibinfo {year} {1963})}\BibitemShut
  {NoStop}%
\bibitem [{\citenamefont {Guralnik}(1964)}]{Guralnik:1964zz}%
  \BibitemOpen
  \bibfield  {author} {\bibinfo {author} {\bibfnamefont {G.}~\bibnamefont
  {Guralnik}},\ }\href {\doibase 10.1103/PhysRev.136.B1404} {\bibfield
  {journal} {\bibinfo  {journal} {Phys.Rev.}\ }\textbf {\bibinfo {volume}
  {136}},\ \bibinfo {pages} {B1404} (\bibinfo {year} {1964})}\BibitemShut
  {NoStop}%
\bibitem [{\citenamefont {Banks}\ and\ \citenamefont
  {Zaks}(1981)}]{Banks:1980rh}%
  \BibitemOpen
  \bibfield  {author} {\bibinfo {author} {\bibfnamefont {T.}~\bibnamefont
  {Banks}}\ and\ \bibinfo {author} {\bibfnamefont {A.}~\bibnamefont {Zaks}},\
  }\href {\doibase 10.1016/0550-3213(81)90220-0} {\bibfield  {journal}
  {\bibinfo  {journal} {Nucl.Phys.}\ }\textbf {\bibinfo {volume} {B184}},\
  \bibinfo {pages} {303} (\bibinfo {year} {1981})}\BibitemShut {NoStop}%
\bibitem [{\citenamefont {Kraus}\ and\ \citenamefont
  {Tomboulis}(2002)}]{oai:arXiv.org:hep-th/0203221}%
  \BibitemOpen
  \bibfield  {author} {\bibinfo {author} {\bibfnamefont {P.}~\bibnamefont
  {Kraus}}\ and\ \bibinfo {author} {\bibfnamefont {E.}~\bibnamefont
  {Tomboulis}},\ }\href {\doibase 10.1103/PhysRevD.66.045015} {\bibfield
  {journal} {\bibinfo  {journal} {Phys.Rev.}\ }\textbf {\bibinfo {volume}
  {D66}},\ \bibinfo {pages} {045015} (\bibinfo {year} {2002})},\ \Eprint
  {http://arxiv.org/abs/hep-th/0203221} {arXiv:hep-th/0203221 [hep-th]}
  \BibitemShut {NoStop}%
\bibitem [{\citenamefont {Wetterich}(1993)}]{Wetterich:1992yh}%
  \BibitemOpen
  \bibfield  {author} {\bibinfo {author} {\bibfnamefont {C.}~\bibnamefont
  {Wetterich}},\ }\href@noop {} {\bibfield  {journal} {\bibinfo  {journal}
  {Phys. Lett.}\ }\textbf {\bibinfo {volume} {B301}},\ \bibinfo {pages} {90}
  (\bibinfo {year} {1993})}\BibitemShut {NoStop}%
\bibitem [{\citenamefont {Strack}\ \emph {et~al.}(2010)\citenamefont {Strack},
  \citenamefont {Takei},\ and\ \citenamefont {Metzner}}]{Strack:2009ia}%
  \BibitemOpen
  \bibfield  {author} {\bibinfo {author} {\bibfnamefont {P.}~\bibnamefont
  {Strack}}, \bibinfo {author} {\bibfnamefont {S.}~\bibnamefont {Takei}}, \
  and\ \bibinfo {author} {\bibfnamefont {W.}~\bibnamefont {Metzner}},\ }\href
  {\doibase 10.1103/PhysRevB.81.125103} {\bibfield  {journal} {\bibinfo
  {journal} {Phys.Rev.}\ }\textbf {\bibinfo {volume} {B81}},\ \bibinfo {pages}
  {125103} (\bibinfo {year} {2010})},\ \Eprint {http://arxiv.org/abs/0905.3894}
  {arXiv:0905.3894 [cond-mat.str-el]} \BibitemShut {NoStop}%
\bibitem [{\citenamefont {Abbott}(1981)}]{Abbott:1980hw}%
  \BibitemOpen
  \bibfield  {author} {\bibinfo {author} {\bibfnamefont {L.~F.}\ \bibnamefont
  {Abbott}},\ }\href@noop {} {\bibfield  {journal} {\bibinfo  {journal} {Nucl.
  Phys.}\ }\textbf {\bibinfo {volume} {B185}},\ \bibinfo {pages} {189}
  (\bibinfo {year} {1981})}\BibitemShut {NoStop}%
\bibitem [{\citenamefont {Abbott}(1982)}]{Abbott:1981ke}%
  \BibitemOpen
  \bibfield  {author} {\bibinfo {author} {\bibfnamefont {L.}~\bibnamefont
  {Abbott}},\ }\href@noop {} {\bibfield  {journal} {\bibinfo  {journal} {Acta
  Phys.Polon.}\ }\textbf {\bibinfo {volume} {B13}},\ \bibinfo {pages} {33}
  (\bibinfo {year} {1982})}\BibitemShut {NoStop}%
\bibitem [{\citenamefont {Dittrich}\ and\ \citenamefont
  {Reuter}(1986)}]{Dittrich:1985tr}%
  \BibitemOpen
  \bibfield  {author} {\bibinfo {author} {\bibfnamefont {W.}~\bibnamefont
  {Dittrich}}\ and\ \bibinfo {author} {\bibfnamefont {M.}~\bibnamefont
  {Reuter}},\ }\href@noop {} {\bibfield  {journal} {\bibinfo  {journal}
  {Lect.Notes Phys.}\ }\textbf {\bibinfo {volume} {244}},\ \bibinfo {pages} {1}
  (\bibinfo {year} {1986})}\BibitemShut {NoStop}%
\bibitem [{\citenamefont {Reuter}\ and\ \citenamefont
  {Wetterich}(1994)}]{Reuter:1993kw}%
  \BibitemOpen
  \bibfield  {author} {\bibinfo {author} {\bibfnamefont {M.}~\bibnamefont
  {Reuter}}\ and\ \bibinfo {author} {\bibfnamefont {C.}~\bibnamefont
  {Wetterich}},\ }\href {\doibase 10.1016/0550-3213(94)90543-6} {\bibfield
  {journal} {\bibinfo  {journal} {Nucl.Phys.}\ }\textbf {\bibinfo {volume}
  {B417}},\ \bibinfo {pages} {181} (\bibinfo {year} {1994})}\BibitemShut
  {NoStop}%
\bibitem [{\citenamefont {Reuter}\ and\ \citenamefont
  {Wetterich}(1997)}]{Reuter:1997gx}%
  \BibitemOpen
  \bibfield  {author} {\bibinfo {author} {\bibfnamefont {M.}~\bibnamefont
  {Reuter}}\ and\ \bibinfo {author} {\bibfnamefont {C.}~\bibnamefont
  {Wetterich}},\ }\href@noop {} {\bibfield  {journal} {\bibinfo  {journal}
  {Phys. Rev.}\ }\textbf {\bibinfo {volume} {D56}},\ \bibinfo {pages} {7893}
  (\bibinfo {year} {1997})},\ \Eprint {http://arxiv.org/abs/hep-th/9708051}
  {hep-th/9708051} \BibitemShut {NoStop}%
\bibitem [{\citenamefont {Gies}(2002)}]{Gies:2002af}%
  \BibitemOpen
  \bibfield  {author} {\bibinfo {author} {\bibfnamefont {H.}~\bibnamefont
  {Gies}},\ }\href {\doibase 10.1103/PhysRevD.66.025006} {\bibfield  {journal}
  {\bibinfo  {journal} {Phys. Rev.}\ }\textbf {\bibinfo {volume} {D66}},\
  \bibinfo {pages} {025006} (\bibinfo {year} {2002})},\ \Eprint
  {http://arxiv.org/abs/hep-th/0202207} {arXiv:hep-th/0202207} \BibitemShut
  {NoStop}%
\bibitem [{\citenamefont {Gies}\ and\ \citenamefont
  {Jaeckel}(2004)}]{Gies:2004hy}%
  \BibitemOpen
  \bibfield  {author} {\bibinfo {author} {\bibfnamefont {H.}~\bibnamefont
  {Gies}}\ and\ \bibinfo {author} {\bibfnamefont {J.}~\bibnamefont {Jaeckel}},\
  }\href@noop {} {\bibfield  {journal} {\bibinfo  {journal} {Phys. Rev. Lett.}\
  }\textbf {\bibinfo {volume} {93}},\ \bibinfo {pages} {110405} (\bibinfo
  {year} {2004})},\ \Eprint {http://arxiv.org/abs/hep-ph/0405183}
  {hep-ph/0405183} \BibitemShut {NoStop}%
\bibitem [{\citenamefont {Gies}(2012)}]{Gies:2006wv}%
  \BibitemOpen
  \bibfield  {author} {\bibinfo {author} {\bibfnamefont {H.}~\bibnamefont
  {Gies}},\ }\href {\doibase 10.1007/978-3-642-27320-9_6} {\bibfield  {journal}
  {\bibinfo  {journal} {Lect.Notes Phys.}\ }\textbf {\bibinfo {volume} {852}},\
  \bibinfo {pages} {287} (\bibinfo {year} {2012})},\ \Eprint
  {http://arxiv.org/abs/hep-ph/0611146} {arXiv:hep-ph/0611146 [hep-ph]}
  \BibitemShut {NoStop}%
\bibitem [{\citenamefont {Pisarski}(1984)}]{Pisarski:1984dj}%
  \BibitemOpen
  \bibfield  {author} {\bibinfo {author} {\bibfnamefont {R.~D.}\ \bibnamefont
  {Pisarski}},\ }\href {\doibase 10.1103/PhysRevD.29.2423} {\bibfield
  {journal} {\bibinfo  {journal} {Phys.Rev.}\ }\textbf {\bibinfo {volume}
  {D29}},\ \bibinfo {pages} {2423} (\bibinfo {year} {1984})}\BibitemShut
  {NoStop}%
\bibitem [{\citenamefont {Ellwanger}\ \emph {et~al.}(1996)\citenamefont
  {Ellwanger}, \citenamefont {Hirsch},\ and\ \citenamefont
  {Weber}}]{Ellwanger:1995qf}%
  \BibitemOpen
  \bibfield  {author} {\bibinfo {author} {\bibfnamefont {U.}~\bibnamefont
  {Ellwanger}}, \bibinfo {author} {\bibfnamefont {M.}~\bibnamefont {Hirsch}}, \
  and\ \bibinfo {author} {\bibfnamefont {A.}~\bibnamefont {Weber}},\
  }\href@noop {} {\bibfield  {journal} {\bibinfo  {journal} {Z. Phys.}\
  }\textbf {\bibinfo {volume} {C69}},\ \bibinfo {pages} {687} (\bibinfo {year}
  {1996})},\ \Eprint {http://arxiv.org/abs/hep-th/9506019} {hep-th/9506019}
  \BibitemShut {NoStop}%
\bibitem [{\citenamefont {Ellwanger}\ \emph {et~al.}(1998)\citenamefont
  {Ellwanger}, \citenamefont {Hirsch},\ and\ \citenamefont
  {Weber}}]{Ellwanger:1996wy}%
  \BibitemOpen
  \bibfield  {author} {\bibinfo {author} {\bibfnamefont {U.}~\bibnamefont
  {Ellwanger}}, \bibinfo {author} {\bibfnamefont {M.}~\bibnamefont {Hirsch}}, \
  and\ \bibinfo {author} {\bibfnamefont {A.}~\bibnamefont {Weber}},\ }\href
  {\doibase 10.1007/s100520050105} {\bibfield  {journal} {\bibinfo  {journal}
  {Eur.Phys.J.}\ }\textbf {\bibinfo {volume} {C1}},\ \bibinfo {pages} {563}
  (\bibinfo {year} {1998})},\ \Eprint {http://arxiv.org/abs/hep-ph/9606468}
  {arXiv:hep-ph/9606468 [hep-ph]} \BibitemShut {NoStop}%
\bibitem [{\citenamefont {Litim}\ and\ \citenamefont
  {Pawlowski}(1998)}]{Litim:1998qi}%
  \BibitemOpen
  \bibfield  {author} {\bibinfo {author} {\bibfnamefont {D.~F.}\ \bibnamefont
  {Litim}}\ and\ \bibinfo {author} {\bibfnamefont {J.~M.}\ \bibnamefont
  {Pawlowski}},\ }\href {\doibase 10.1016/S0370-2693(98)00761-8} {\bibfield
  {journal} {\bibinfo  {journal} {Phys.Lett.}\ }\textbf {\bibinfo {volume}
  {B435}},\ \bibinfo {pages} {181} (\bibinfo {year} {1998})},\ \Eprint
  {http://arxiv.org/abs/hep-th/9802064} {arXiv:hep-th/9802064 [hep-th]}
  \BibitemShut {NoStop}%
\bibitem [{\citenamefont {Lauscher}\ and\ \citenamefont
  {Reuter}(2002)}]{Lauscher:2001ya}%
  \BibitemOpen
  \bibfield  {author} {\bibinfo {author} {\bibfnamefont {O.}~\bibnamefont
  {Lauscher}}\ and\ \bibinfo {author} {\bibfnamefont {M.}~\bibnamefont
  {Reuter}},\ }\href {\doibase 10.1103/PhysRevD.65.025013} {\bibfield
  {journal} {\bibinfo  {journal} {Phys. Rev.}\ }\textbf {\bibinfo {volume}
  {D65}},\ \bibinfo {pages} {025013} (\bibinfo {year} {2002})},\ \Eprint
  {http://arxiv.org/abs/hep-th/0108040} {arXiv:hep-th/0108040} \BibitemShut
  {NoStop}%
\bibitem [{\citenamefont {Braun}(2012)}]{Braun:2011pp}%
  \BibitemOpen
  \bibfield  {author} {\bibinfo {author} {\bibfnamefont {J.}~\bibnamefont
  {Braun}},\ }\href {\doibase 10.1088/0954-3899/39/3/033001} {\bibfield
  {journal} {\bibinfo  {journal} {J.Phys.}\ }\textbf {\bibinfo {volume}
  {G39}},\ \bibinfo {pages} {033001} (\bibinfo {year} {2012})},\ \Eprint
  {http://arxiv.org/abs/1108.4449} {arXiv:1108.4449 [hep-ph]} \BibitemShut
  {NoStop}%
\bibitem [{\citenamefont {Pokorski}(1987)}]{Pokorski:1987ed}%
  \BibitemOpen
  \bibfield  {author} {\bibinfo {author} {\bibfnamefont {S.}~\bibnamefont
  {Pokorski}},\ }\href@noop {} {\emph {\bibinfo {title} {Gauge field
  theories}}}\ (\bibinfo  {publisher} {Cambridge University Press},\ \bibinfo
  {year} {1987})\BibitemShut {NoStop}%
\bibitem [{\citenamefont {Ellwanger}(1994)}]{Ellwanger:1994iz}%
  \BibitemOpen
  \bibfield  {author} {\bibinfo {author} {\bibfnamefont {U.}~\bibnamefont
  {Ellwanger}},\ }\href@noop {} {\bibfield  {journal} {\bibinfo  {journal}
  {Phys. Lett.}\ }\textbf {\bibinfo {volume} {B335}},\ \bibinfo {pages} {364}
  (\bibinfo {year} {1994})},\ \Eprint {http://arxiv.org/abs/hep-th/9402077}
  {hep-th/9402077} \BibitemShut {NoStop}%
\bibitem [{\citenamefont {Litim}\ and\ \citenamefont
  {Pawlowski}()}]{Litim:1998nf}%
  \BibitemOpen
  \bibfield  {author} {\bibinfo {author} {\bibfnamefont {D.~F.}\ \bibnamefont
  {Litim}}\ and\ \bibinfo {author} {\bibfnamefont {J.~M.}\ \bibnamefont
  {Pawlowski}},\ }\href@noop {} {\ }\Eprint
  {http://arxiv.org/abs/hep-th/9901063} {hep-th/9901063} \BibitemShut {NoStop}%
\bibitem [{\citenamefont {Litim}\ and\ \citenamefont
  {Pawlowski}(2002)}]{Litim:2002ce}%
  \BibitemOpen
  \bibfield  {author} {\bibinfo {author} {\bibfnamefont {D.~F.}\ \bibnamefont
  {Litim}}\ and\ \bibinfo {author} {\bibfnamefont {J.~M.}\ \bibnamefont
  {Pawlowski}},\ }\href@noop {} {\bibfield  {journal} {\bibinfo  {journal}
  {JHEP}\ }\textbf {\bibinfo {volume} {09}},\ \bibinfo {pages} {049} (\bibinfo
  {year} {2002})},\ \Eprint {http://arxiv.org/abs/hep-th/0203005}
  {hep-th/0203005} \BibitemShut {NoStop}%
\bibitem [{\citenamefont {Pawlowski}(2007)}]{Pawlowski:2005xe}%
  \BibitemOpen
  \bibfield  {author} {\bibinfo {author} {\bibfnamefont {J.~M.}\ \bibnamefont
  {Pawlowski}},\ }\href {\doibase 10.1016/j.aop.2007.01.007} {\bibfield
  {journal} {\bibinfo  {journal} {Annals Phys.}\ }\textbf {\bibinfo {volume}
  {322}},\ \bibinfo {pages} {2831} (\bibinfo {year} {2007})},\ \Eprint
  {http://arxiv.org/abs/hep-th/0512261} {arXiv:hep-th/0512261} \BibitemShut
  {NoStop}%
\bibitem [{\citenamefont {Gies}\ \emph {et~al.}(2004)\citenamefont {Gies},
  \citenamefont {Jaeckel},\ and\ \citenamefont {Wetterich}}]{Gies:2003dp}%
  \BibitemOpen
  \bibfield  {author} {\bibinfo {author} {\bibfnamefont {H.}~\bibnamefont
  {Gies}}, \bibinfo {author} {\bibfnamefont {J.}~\bibnamefont {Jaeckel}}, \
  and\ \bibinfo {author} {\bibfnamefont {C.}~\bibnamefont {Wetterich}},\ }\href
  {\doibase 10.1103/PhysRevD.69.105008} {\bibfield  {journal} {\bibinfo
  {journal} {Phys.Rev.}\ }\textbf {\bibinfo {volume} {D69}},\ \bibinfo {pages}
  {105008} (\bibinfo {year} {2004})},\ \Eprint
  {http://arxiv.org/abs/hep-ph/0312034} {arXiv:hep-ph/0312034 [hep-ph]}
  \BibitemShut {NoStop}%
\bibitem [{\citenamefont {Freire}\ \emph {et~al.}(2000)\citenamefont {Freire},
  \citenamefont {Litim},\ and\ \citenamefont {Pawlowski}}]{Freire:2000bq}%
  \BibitemOpen
  \bibfield  {author} {\bibinfo {author} {\bibfnamefont {F.}~\bibnamefont
  {Freire}}, \bibinfo {author} {\bibfnamefont {D.~F.}\ \bibnamefont {Litim}}, \
  and\ \bibinfo {author} {\bibfnamefont {J.~M.}\ \bibnamefont {Pawlowski}},\
  }\href@noop {} {\bibfield  {journal} {\bibinfo  {journal} {Phys. Lett.}\
  }\textbf {\bibinfo {volume} {B495}},\ \bibinfo {pages} {256} (\bibinfo {year}
  {2000})},\ \Eprint {http://arxiv.org/abs/hep-th/0009110} {hep-th/0009110}
  \BibitemShut {NoStop}%
\bibitem [{\citenamefont {Fischer}\ and\ \citenamefont
  {Gies}(2004)}]{Fischer:2004uk}%
  \BibitemOpen
  \bibfield  {author} {\bibinfo {author} {\bibfnamefont {C.~S.}\ \bibnamefont
  {Fischer}}\ and\ \bibinfo {author} {\bibfnamefont {H.}~\bibnamefont {Gies}},\
  }\href@noop {} {\bibfield  {journal} {\bibinfo  {journal} {JHEP}\ }\textbf
  {\bibinfo {volume} {10}},\ \bibinfo {pages} {048} (\bibinfo {year} {2004})},\
  \Eprint {http://arxiv.org/abs/hep-ph/0408089} {hep-ph/0408089} \BibitemShut
  {NoStop}%
\bibitem [{\citenamefont {Herbut}(2007)}]{herbut2007modern}%
  \BibitemOpen
  \bibfield  {author} {\bibinfo {author} {\bibfnamefont {I.~F.}\ \bibnamefont
  {Herbut}},\ }\href@noop {} {\emph {\bibinfo {title} {A modern approach to
  critical phenomena}}}\ (\bibinfo  {publisher} {Cambridge University Press},\
  \bibinfo {year} {2007})\BibitemShut {NoStop}%
\bibitem [{\citenamefont {Berges}\ \emph {et~al.}(2002)\citenamefont {Berges},
  \citenamefont {Tetradis},\ and\ \citenamefont {Wetterich}}]{Berges:2000ew}%
  \BibitemOpen
  \bibfield  {author} {\bibinfo {author} {\bibfnamefont {J.}~\bibnamefont
  {Berges}}, \bibinfo {author} {\bibfnamefont {N.}~\bibnamefont {Tetradis}}, \
  and\ \bibinfo {author} {\bibfnamefont {C.}~\bibnamefont {Wetterich}},\
  }\href@noop {} {\bibfield  {journal} {\bibinfo  {journal} {Phys. Rept.}\
  }\textbf {\bibinfo {volume} {363}},\ \bibinfo {pages} {223} (\bibinfo {year}
  {2002})},\ \Eprint {http://arxiv.org/abs/hep-ph/0005122} {hep-ph/0005122}
  \BibitemShut {NoStop}%
\bibitem [{\citenamefont {Gies}\ and\ \citenamefont
  {Wetterich}(2004)}]{Gies:2002hq}%
  \BibitemOpen
  \bibfield  {author} {\bibinfo {author} {\bibfnamefont {H.}~\bibnamefont
  {Gies}}\ and\ \bibinfo {author} {\bibfnamefont {C.}~\bibnamefont
  {Wetterich}},\ }\href@noop {} {\bibfield  {journal} {\bibinfo  {journal}
  {Phys. Rev.}\ }\textbf {\bibinfo {volume} {D69}},\ \bibinfo {pages} {025001}
  (\bibinfo {year} {2004})},\ \Eprint {http://arxiv.org/abs/hep-th/0209183}
  {hep-th/0209183} \BibitemShut {NoStop}%
\bibitem [{\citenamefont {Mesterhazy}\ \emph {et~al.}(2012)\citenamefont
  {Mesterhazy}, \citenamefont {Berges},\ and\ \citenamefont {von
  Smekal}}]{oai:arXiv.org:1207.4054}%
  \BibitemOpen
  \bibfield  {author} {\bibinfo {author} {\bibfnamefont {D.}~\bibnamefont
  {Mesterhazy}}, \bibinfo {author} {\bibfnamefont {J.}~\bibnamefont {Berges}},
  \ and\ \bibinfo {author} {\bibfnamefont {L.}~\bibnamefont {von Smekal}},\
  }\href {\doibase 10.1103/PhysRevB.86.245431} {\bibfield  {journal} {\bibinfo
  {journal} {Phys.Rev.}\ }\textbf {\bibinfo {volume} {B86}},\ \bibinfo {pages}
  {245431} (\bibinfo {year} {2012})},\ \Eprint {http://arxiv.org/abs/1207.4054}
  {arXiv:1207.4054 [cond-mat.str-el]} \BibitemShut {NoStop}%
\bibitem [{\citenamefont {Christofi}\ \emph {et~al.}(2007)\citenamefont
  {Christofi}, \citenamefont {Hands},\ and\ \citenamefont
  {Strouthos}}]{oai:arXiv.org:hep-lat/0701016}%
  \BibitemOpen
  \bibfield  {author} {\bibinfo {author} {\bibfnamefont {S.}~\bibnamefont
  {Christofi}}, \bibinfo {author} {\bibfnamefont {S.}~\bibnamefont {Hands}}, \
  and\ \bibinfo {author} {\bibfnamefont {C.}~\bibnamefont {Strouthos}},\ }\href
  {\doibase 10.1103/PhysRevD.75.101701} {\bibfield  {journal} {\bibinfo
  {journal} {Phys.Rev.}\ }\textbf {\bibinfo {volume} {D75}},\ \bibinfo {pages}
  {101701} (\bibinfo {year} {2007})},\ \Eprint
  {http://arxiv.org/abs/hep-lat/0701016} {arXiv:hep-lat/0701016 [hep-lat]}
  \BibitemShut {NoStop}%
\bibitem [{\citenamefont {Gomes}\ \emph {et~al.}(1991)\citenamefont {Gomes},
  \citenamefont {Mendes}, \citenamefont {Ribeiro},\ and\ \citenamefont
  {da~Silva}}]{Gomes:1991aa}%
  \BibitemOpen
  \bibfield  {author} {\bibinfo {author} {\bibfnamefont {M.}~\bibnamefont
  {Gomes}}, \bibinfo {author} {\bibfnamefont {R.}~\bibnamefont {Mendes}},
  \bibinfo {author} {\bibfnamefont {R.}~\bibnamefont {Ribeiro}}, \ and\
  \bibinfo {author} {\bibfnamefont {A.}~\bibnamefont {da~Silva}},\ }\href
  {\doibase 10.1103/PhysRevD.43.3516} {\bibfield  {journal} {\bibinfo
  {journal} {Phys.Rev.}\ }\textbf {\bibinfo {volume} {D43}},\ \bibinfo {pages}
  {3516} (\bibinfo {year} {1991})}\BibitemShut {NoStop}%
\bibitem [{\citenamefont {Hong}\ and\ \citenamefont
  {Park}(1994)}]{Hong:1993qk}%
  \BibitemOpen
  \bibfield  {author} {\bibinfo {author} {\bibfnamefont {D.~K.}\ \bibnamefont
  {Hong}}\ and\ \bibinfo {author} {\bibfnamefont {S.~H.}\ \bibnamefont
  {Park}},\ }\href {\doibase 10.1103/PhysRevD.49.5507} {\bibfield  {journal}
  {\bibinfo  {journal} {Phys.Rev.}\ }\textbf {\bibinfo {volume} {D49}},\
  \bibinfo {pages} {5507} (\bibinfo {year} {1994})},\ \Eprint
  {http://arxiv.org/abs/hep-th/9307186} {arXiv:hep-th/9307186 [hep-th]}
  \BibitemShut {NoStop}%
\bibitem [{\citenamefont {Itoh}\ \emph {et~al.}(1995)\citenamefont {Itoh},
  \citenamefont {Kim}, \citenamefont {Sugiura},\ and\ \citenamefont
  {Yamawaki}}]{oai:arXiv.org:hep-th/9411201}%
  \BibitemOpen
  \bibfield  {author} {\bibinfo {author} {\bibfnamefont {T.}~\bibnamefont
  {Itoh}}, \bibinfo {author} {\bibfnamefont {Y.}~\bibnamefont {Kim}}, \bibinfo
  {author} {\bibfnamefont {M.}~\bibnamefont {Sugiura}}, \ and\ \bibinfo
  {author} {\bibfnamefont {K.}~\bibnamefont {Yamawaki}},\ }\href {\doibase
  10.1143/PTP.93.417} {\bibfield  {journal} {\bibinfo  {journal}
  {Prog.Theor.Phys.}\ }\textbf {\bibinfo {volume} {93}},\ \bibinfo {pages}
  {417} (\bibinfo {year} {1995})},\ \Eprint
  {http://arxiv.org/abs/hep-th/9411201} {arXiv:hep-th/9411201 [hep-th]}
  \BibitemShut {NoStop}%
\bibitem [{\citenamefont {Sugiura}(1997)}]{oai:arXiv.org:hep-th/9611198}%
  \BibitemOpen
  \bibfield  {author} {\bibinfo {author} {\bibfnamefont {M.}~\bibnamefont
  {Sugiura}},\ }\href {\doibase 10.1143/PTP.97.311} {\bibfield  {journal}
  {\bibinfo  {journal} {Prog.Theor.Phys.}\ }\textbf {\bibinfo {volume} {97}},\
  \bibinfo {pages} {311} (\bibinfo {year} {1997})},\ \Eprint
  {http://arxiv.org/abs/hep-th/9611198} {arXiv:hep-th/9611198 [hep-th]}
  \BibitemShut {NoStop}%
\bibitem [{\citenamefont {Kondo}(1995)}]{oai:arXiv.org:hep-th/9502070}%
  \BibitemOpen
  \bibfield  {author} {\bibinfo {author} {\bibfnamefont {K.-I.}\ \bibnamefont
  {Kondo}},\ }\href {\doibase 10.1016/0550-3213(95)00316-K} {\bibfield
  {journal} {\bibinfo  {journal} {Nucl.Phys.}\ }\textbf {\bibinfo {volume}
  {B450}},\ \bibinfo {pages} {251} (\bibinfo {year} {1995})},\ \Eprint
  {http://arxiv.org/abs/hep-th/9502070} {arXiv:hep-th/9502070 [hep-th]}
  \BibitemShut {NoStop}%
\bibitem [{\citenamefont {Kim}\ and\ \citenamefont
  {Kim}()}]{oai:arXiv.org:hep-lat/9605021}%
  \BibitemOpen
  \bibfield  {author} {\bibinfo {author} {\bibfnamefont {S.}~\bibnamefont
  {Kim}}\ and\ \bibinfo {author} {\bibfnamefont {Y.}~\bibnamefont {Kim}},\
  }\href@noop {} {\ }\Eprint {http://arxiv.org/abs/hep-lat/9605021}
  {arXiv:hep-lat/9605021 [hep-lat]} \BibitemShut {NoStop}%
\bibitem [{\citenamefont {Del~Debbio}\ \emph {et~al.}(1997)\citenamefont
  {Del~Debbio}, \citenamefont {Hands},\ and\ \citenamefont
  {Mehegan}}]{oai:arXiv.org:hep-lat/9701016}%
  \BibitemOpen
  \bibfield  {author} {\bibinfo {author} {\bibfnamefont {L.}~\bibnamefont
  {Del~Debbio}}, \bibinfo {author} {\bibfnamefont {S.}~\bibnamefont {Hands}}, \
  and\ \bibinfo {author} {\bibfnamefont {J.}~\bibnamefont {Mehegan}} (\bibinfo
  {collaboration} {UKQCD Collaboration}),\ }\href {\doibase
  10.1016/S0550-3213(97)00435-5} {\bibfield  {journal} {\bibinfo  {journal}
  {Nucl.Phys.}\ }\textbf {\bibinfo {volume} {B502}},\ \bibinfo {pages} {269}
  (\bibinfo {year} {1997})},\ \Eprint {http://arxiv.org/abs/hep-lat/9701016}
  {arXiv:hep-lat/9701016 [hep-lat]} \BibitemShut {NoStop}%
\bibitem [{\citenamefont {Hands}\ and\ \citenamefont
  {Lucini}(1999)}]{oai:arXiv.org:hep-lat/9906008}%
  \BibitemOpen
  \bibfield  {author} {\bibinfo {author} {\bibfnamefont {S.}~\bibnamefont
  {Hands}}\ and\ \bibinfo {author} {\bibfnamefont {B.}~\bibnamefont {Lucini}},\
  }\href {\doibase 10.1016/S0370-2693(99)00843-6} {\bibfield  {journal}
  {\bibinfo  {journal} {Phys.Lett.}\ }\textbf {\bibinfo {volume} {B461}},\
  \bibinfo {pages} {263} (\bibinfo {year} {1999})},\ \Eprint
  {http://arxiv.org/abs/hep-lat/9906008} {arXiv:hep-lat/9906008 [hep-lat]}
  \BibitemShut {NoStop}%
\bibitem [{\citenamefont {Rosa}\ \emph {et~al.}(2001)\citenamefont {Rosa},
  \citenamefont {Vitale},\ and\ \citenamefont {Wetterich}}]{Rosa:2000ju}%
  \BibitemOpen
  \bibfield  {author} {\bibinfo {author} {\bibfnamefont {L.}~\bibnamefont
  {Rosa}}, \bibinfo {author} {\bibfnamefont {P.}~\bibnamefont {Vitale}}, \ and\
  \bibinfo {author} {\bibfnamefont {C.}~\bibnamefont {Wetterich}},\ }\href
  {\doibase 10.1103/PhysRevLett.86.958} {\bibfield  {journal} {\bibinfo
  {journal} {Phys.Rev.Lett.}\ }\textbf {\bibinfo {volume} {86}},\ \bibinfo
  {pages} {958} (\bibinfo {year} {2001})},\ \Eprint
  {http://arxiv.org/abs/hep-th/0007093} {arXiv:hep-th/0007093 [hep-th]}
  \BibitemShut {NoStop}%
\bibitem [{\citenamefont {Hofling}\ \emph {et~al.}(2002)\citenamefont
  {Hofling}, \citenamefont {Nowak},\ and\ \citenamefont
  {Wetterich}}]{Hofling:2002hj}%
  \BibitemOpen
  \bibfield  {author} {\bibinfo {author} {\bibfnamefont {F.}~\bibnamefont
  {Hofling}}, \bibinfo {author} {\bibfnamefont {C.}~\bibnamefont {Nowak}}, \
  and\ \bibinfo {author} {\bibfnamefont {C.}~\bibnamefont {Wetterich}},\ }\href
  {\doibase 10.1103/PhysRevB.66.205111} {\bibfield  {journal} {\bibinfo
  {journal} {Phys. Rev.}\ }\textbf {\bibinfo {volume} {B66}},\ \bibinfo {pages}
  {205111} (\bibinfo {year} {2002})},\ \Eprint
  {http://arxiv.org/abs/cond-mat/0203588} {arXiv:cond-mat/0203588} \BibitemShut
  {NoStop}%
\bibitem [{\citenamefont {Braun}\ \emph
  {et~al.}(2011{\natexlab{a}})\citenamefont {Braun}, \citenamefont {Gies},\
  and\ \citenamefont {Scherer}}]{Braun:2010tt}%
  \BibitemOpen
  \bibfield  {author} {\bibinfo {author} {\bibfnamefont {J.}~\bibnamefont
  {Braun}}, \bibinfo {author} {\bibfnamefont {H.}~\bibnamefont {Gies}}, \ and\
  \bibinfo {author} {\bibfnamefont {D.~D.}\ \bibnamefont {Scherer}},\ }\href
  {\doibase 10.1103/PhysRevD.83.085012} {\bibfield  {journal} {\bibinfo
  {journal} {Phys. Rev.}\ }\textbf {\bibinfo {volume} {D83}},\ \bibinfo {pages}
  {085012} (\bibinfo {year} {2011}{\natexlab{a}})},\ \Eprint
  {http://arxiv.org/abs/1011.1456} {arXiv:1011.1456 [hep-th]} \BibitemShut
  {NoStop}%
\bibitem [{\citenamefont {Litim}(2000)}]{Litim:2000ci}%
  \BibitemOpen
  \bibfield  {author} {\bibinfo {author} {\bibfnamefont {D.~F.}\ \bibnamefont
  {Litim}},\ }\href@noop {} {\bibfield  {journal} {\bibinfo  {journal} {Phys.
  Lett.}\ }\textbf {\bibinfo {volume} {B486}},\ \bibinfo {pages} {92} (\bibinfo
  {year} {2000})},\ \Eprint {http://arxiv.org/abs/hep-th/0005245}
  {hep-th/0005245} \BibitemShut {NoStop}%
\bibitem [{\citenamefont {Litim}(2001{\natexlab{a}})}]{Litim:2001up}%
  \BibitemOpen
  \bibfield  {author} {\bibinfo {author} {\bibfnamefont {D.~F.}\ \bibnamefont
  {Litim}},\ }\href@noop {} {\bibfield  {journal} {\bibinfo  {journal} {Phys.
  Rev.}\ }\textbf {\bibinfo {volume} {D64}},\ \bibinfo {pages} {105007}
  (\bibinfo {year} {2001}{\natexlab{a}})},\ \Eprint
  {http://arxiv.org/abs/hep-th/0103195} {hep-th/0103195} \BibitemShut {NoStop}%
\bibitem [{\citenamefont {Litim}(2001{\natexlab{b}})}]{Litim:2001fd}%
  \BibitemOpen
  \bibfield  {author} {\bibinfo {author} {\bibfnamefont {D.~F.}\ \bibnamefont
  {Litim}},\ }\href {\doibase 10.1142/S0217751X01004748} {\bibfield  {journal}
  {\bibinfo  {journal} {Int. J. Mod. Phys.}\ }\textbf {\bibinfo {volume}
  {A16}},\ \bibinfo {pages} {2081} (\bibinfo {year} {2001}{\natexlab{b}})},\
  \Eprint {http://arxiv.org/abs/hep-th/0104221} {arXiv:hep-th/0104221}
  \BibitemShut {NoStop}%
\bibitem [{\citenamefont {Goecke}\ \emph {et~al.}(2009)\citenamefont {Goecke},
  \citenamefont {Fischer},\ and\ \citenamefont {Williams}}]{Goecke:2008zh}%
  \BibitemOpen
  \bibfield  {author} {\bibinfo {author} {\bibfnamefont {T.}~\bibnamefont
  {Goecke}}, \bibinfo {author} {\bibfnamefont {C.~S.}\ \bibnamefont {Fischer}},
  \ and\ \bibinfo {author} {\bibfnamefont {R.}~\bibnamefont {Williams}},\
  }\href {\doibase 10.1103/PhysRevB.79.064513} {\bibfield  {journal} {\bibinfo
  {journal} {Phys.Rev.}\ }\textbf {\bibinfo {volume} {B79}},\ \bibinfo {pages}
  {064513} (\bibinfo {year} {2009})},\ \Eprint {http://arxiv.org/abs/0811.1887}
  {arXiv:0811.1887 [hep-ph]} \BibitemShut {NoStop}%
\bibitem [{\citenamefont {Vafa}\ and\ \citenamefont
  {Witten}(1984)}]{Vafa:1984xh}%
  \BibitemOpen
  \bibfield  {author} {\bibinfo {author} {\bibfnamefont {C.}~\bibnamefont
  {Vafa}}\ and\ \bibinfo {author} {\bibfnamefont {E.}~\bibnamefont {Witten}},\
  }\href {\doibase 10.1007/BF01212397} {\bibfield  {journal} {\bibinfo
  {journal} {Commun.Math.Phys.}\ }\textbf {\bibinfo {volume} {95}},\ \bibinfo
  {pages} {257} (\bibinfo {year} {1984})}\BibitemShut {NoStop}%
\bibitem [{\citenamefont {Berezinskii}(1971)}]{Berezinskii}%
  \BibitemOpen
  \bibfield  {author} {\bibinfo {author} {\bibfnamefont {V.~L.}\ \bibnamefont
  {Berezinskii}},\ }\href@noop {} {\bibfield  {journal} {\bibinfo  {journal}
  {Sov. Phys. JETP}\ }\textbf {\bibinfo {volume} {32}},\ \bibinfo {pages} {493}
  (\bibinfo {year} {1971})}\BibitemShut {NoStop}%
\bibitem [{\citenamefont {Berezinskii}(1972)}]{Berezinskii2}%
  \BibitemOpen
  \bibfield  {author} {\bibinfo {author} {\bibfnamefont {V.~L.}\ \bibnamefont
  {Berezinskii}},\ }\href@noop {} {\bibfield  {journal} {\bibinfo  {journal}
  {Sov. Phys. JETP}\ }\textbf {\bibinfo {volume} {34}},\ \bibinfo {pages} {610}
  (\bibinfo {year} {1972})}\BibitemShut {NoStop}%
\bibitem [{\citenamefont {Kosterlitz}\ and\ \citenamefont
  {Thouless}(1973)}]{Kosterlitz:1973xp}%
  \BibitemOpen
  \bibfield  {author} {\bibinfo {author} {\bibfnamefont {J.~M.}\ \bibnamefont
  {Kosterlitz}}\ and\ \bibinfo {author} {\bibfnamefont {D.~J.}\ \bibnamefont
  {Thouless}},\ }\href@noop {} {\bibfield  {journal} {\bibinfo  {journal} {J.
  Phys.}\ }\textbf {\bibinfo {volume} {C6}},\ \bibinfo {pages} {1181} (\bibinfo
  {year} {1973})}\BibitemShut {NoStop}%
\bibitem [{\citenamefont {Appelquist}\ \emph {et~al.}(1996)\citenamefont
  {Appelquist}, \citenamefont {Terning},\ and\ \citenamefont
  {Wijewardhana}}]{Appelquist:1996dq}%
  \BibitemOpen
  \bibfield  {author} {\bibinfo {author} {\bibfnamefont {T.}~\bibnamefont
  {Appelquist}}, \bibinfo {author} {\bibfnamefont {J.}~\bibnamefont {Terning}},
  \ and\ \bibinfo {author} {\bibfnamefont {L.}~\bibnamefont {Wijewardhana}},\
  }\href {\doibase 10.1103/PhysRevLett.77.1214} {\bibfield  {journal} {\bibinfo
   {journal} {Phys.Rev.Lett.}\ }\textbf {\bibinfo {volume} {77}},\ \bibinfo
  {pages} {1214} (\bibinfo {year} {1996})},\ \Eprint
  {http://arxiv.org/abs/hep-ph/9602385} {arXiv:hep-ph/9602385 [hep-ph]}
  \BibitemShut {NoStop}%
\bibitem [{\citenamefont {Chivukula}(1997)}]{Chivukula:1996kg}%
  \BibitemOpen
  \bibfield  {author} {\bibinfo {author} {\bibfnamefont {R.~S.}\ \bibnamefont
  {Chivukula}},\ }\href {\doibase 10.1103/PhysRevD.55.5238} {\bibfield
  {journal} {\bibinfo  {journal} {Phys. Rev.}\ }\textbf {\bibinfo {volume}
  {D55}},\ \bibinfo {pages} {5238} (\bibinfo {year} {1997})},\ \Eprint
  {http://arxiv.org/abs/hep-ph/9612267} {arXiv:hep-ph/9612267} \BibitemShut
  {NoStop}%
\bibitem [{\citenamefont {Miransky}\ and\ \citenamefont
  {Yamawaki}(1989)}]{Miransky:1988gk}%
  \BibitemOpen
  \bibfield  {author} {\bibinfo {author} {\bibfnamefont {V.~A.}\ \bibnamefont
  {Miransky}}\ and\ \bibinfo {author} {\bibfnamefont {K.}~\bibnamefont
  {Yamawaki}},\ }\href {\doibase 10.1142/S0217732389000186} {\bibfield
  {journal} {\bibinfo  {journal} {Mod. Phys. Lett.}\ }\textbf {\bibinfo
  {volume} {A4}},\ \bibinfo {pages} {129} (\bibinfo {year} {1989})}\BibitemShut
  {NoStop}%
\bibitem [{\citenamefont {Appelquist}\ and\ \citenamefont
  {Sannino}(1999)}]{Appelquist:1998xf}%
  \BibitemOpen
  \bibfield  {author} {\bibinfo {author} {\bibfnamefont {T.}~\bibnamefont
  {Appelquist}}\ and\ \bibinfo {author} {\bibfnamefont {F.}~\bibnamefont
  {Sannino}},\ }\href {\doibase 10.1103/PhysRevD.59.067702} {\bibfield
  {journal} {\bibinfo  {journal} {Phys.Rev.}\ }\textbf {\bibinfo {volume}
  {D59}},\ \bibinfo {pages} {067702} (\bibinfo {year} {1999})},\ \Eprint
  {http://arxiv.org/abs/hep-ph/9806409} {arXiv:hep-ph/9806409 [hep-ph]}
  \BibitemShut {NoStop}%
\bibitem [{\citenamefont {Kaplan}\ \emph {et~al.}(2009)\citenamefont {Kaplan},
  \citenamefont {Lee}, \citenamefont {Son},\ and\ \citenamefont
  {Stephanov}}]{Kaplan:2009kr}%
  \BibitemOpen
  \bibfield  {author} {\bibinfo {author} {\bibfnamefont {D.~B.}\ \bibnamefont
  {Kaplan}}, \bibinfo {author} {\bibfnamefont {J.-W.}\ \bibnamefont {Lee}},
  \bibinfo {author} {\bibfnamefont {D.~T.}\ \bibnamefont {Son}}, \ and\
  \bibinfo {author} {\bibfnamefont {M.~A.}\ \bibnamefont {Stephanov}},\ }\href
  {\doibase 10.1103/PhysRevD.80.125005} {\bibfield  {journal} {\bibinfo
  {journal} {Phys.Rev.}\ }\textbf {\bibinfo {volume} {D80}},\ \bibinfo {pages}
  {125005} (\bibinfo {year} {2009})},\ \Eprint {http://arxiv.org/abs/0905.4752}
  {arXiv:0905.4752 [hep-th]} \BibitemShut {NoStop}%
\bibitem [{\citenamefont {Braun}\ \emph
  {et~al.}(2011{\natexlab{b}})\citenamefont {Braun}, \citenamefont {Fischer},\
  and\ \citenamefont {Gies}}]{Braun:2010qs}%
  \BibitemOpen
  \bibfield  {author} {\bibinfo {author} {\bibfnamefont {J.}~\bibnamefont
  {Braun}}, \bibinfo {author} {\bibfnamefont {C.~S.}\ \bibnamefont {Fischer}},
  \ and\ \bibinfo {author} {\bibfnamefont {H.}~\bibnamefont {Gies}},\ }\href
  {\doibase 10.1103/PhysRevD.84.034045} {\bibfield  {journal} {\bibinfo
  {journal} {Phys.Rev.}\ }\textbf {\bibinfo {volume} {D84}},\ \bibinfo {pages}
  {034045} (\bibinfo {year} {2011}{\natexlab{b}})},\ \Eprint
  {http://arxiv.org/abs/1012.4279} {arXiv:1012.4279 [hep-ph]} \BibitemShut
  {NoStop}%
\end{thebibliography}%

\end{document}